\documentclass[a4paper,11pt]{article}

\usepackage{jheppub}
\usepackage[T1]{fontenc}
\usepackage{empheq}
\usepackage{amsmath}
\usepackage{bbm}
\usepackage{amssymb}
\usepackage{dsfont}
\usepackage{graphicx}
\usepackage{hyperref}
\usepackage{stackrel}
\usepackage{mathrsfs}
\usepackage{comment}
\usepackage[export]{adjustbox}
\usepackage{float}
\usepackage{enumitem}
\usepackage{epsfig,euscript}
\usepackage{amsfonts,mathtools}
\usepackage{subcaption}
\allowdisplaybreaks[4]        
\usepackage[dvipsnames]{xcolor}
\usepackage{fancyhdr}
\usepackage[header,title,page,titletoc]{appendix}  
\usepackage{relsize}
\usepackage[percent]{overpic}

\DeclareMathOperator{\tr}{Tr}

\renewcommand{\(}{\left(}
\renewcommand{\)}{\right)}
\renewcommand{\[}{\left[}
\renewcommand{\]}{\right]}
\newcommand{\eg}{{\it e.g.,}\ }
\newcommand{\ie}{{\it i.e.,}\ }

\def\be{\begin{equation}}
\def\ee{\end{equation}}
\def\bea{\begin{eqnarray}}
\def\eea{\end{eqnarray}}
\newcommand{\beq}{\begin{equation}}
\newcommand{\eeq}{\end{equation}}
\newcommand{\beqa}{\begin{eqnarray}}
\newcommand{\eeqa}{\end{eqnarray}}

\def\ba{\begin{eqnarray}}
\def\ea{\end{eqnarray}}

\def\a{\alpha}

\def\t{\tau}

\renewcommand{\(}{\left(}
\renewcommand{\)}{\right)}

\newcommand{\Tr}{\ {\rm Tr}\ }
\newcommand{\Fcal}{\mathcal{F}}

\newcommand{\nbox}{{\,\lower0.9pt\vbox{\hrule \hbox{\vrule height 0.2 cm \hskip 0.19 cm \vrule height 0.2 cm}\hrule}\,}}

\newcommand{\vareps}{\varepsilon}
\newcommand{\veps}{\varepsilon}
\newcommand{\eps}{\epsilon}

\newcommand{\bz}{\bar{z}}
\newcommand{\tsigma}{\tilde{\sigma}}

\newcommand{\bw}{\bar{w}}

\newcommand{\ft}{\mathfrak{t}}

\begin{document}

\title{Temporal Entanglement from Twist Correlators\\ in 2d Conformal Field Theory and Holography}

\author{%
  \resizebox{\textwidth}{!}{%
    Alice Bernamonti$^{a,b}$, Federico Galli$^{b}$,
    Michal P. Heller$^{c,d}$, Fabio Ori$^{c}$,
    Alexandre Serantes$^{c}$%
  }%
}

\affiliation[a]{Dipartimento di Fisica e Astronomia, Università di Firenze\\
Via G. Sansone 1, I-50019 Sesto Fiorentino, Italy}
\affiliation[b]{INFN, Sezione di Firenze\\
Via G. Sansone 1, I-50019 Sesto Fiorentino, Italy}
\affiliation[c]{Department of Physics and Astronomy, Ghent University \\
9000 Ghent, Belgium}
\affiliation[d]{Institute of Theoretical Physics and Mark Kac Center for Complex Systems Research \\
Jagiellonian University, 30-348 Cracow, Poland}

\emailAdd{alice.bernamonti@unifi.it}
\emailAdd{federico.galli@fi.infn.it}
\emailAdd{michal.p.heller@ugent.be}
\emailAdd{fabio.ori@ugent.be}
\emailAdd{alexandre.serantesrubianes@ugent.be}

\abstract{We formulate timelike entanglement entropy and its R{\'e}nyi extension in two-dimensional conformal field theory through the analytic continuation of replica twist correlators to time-ordered, timelike-separated insertions. This field-theoretic construction grounds and generalizes recent developments, and applies to temporal subregions of arbitrary extent. Within three-dimensional holography, the semiclassical boundary correlator identifies boundary-anchored complex geodesics as the relevant bulk saddles and selects the one with the smallest real part of the length. This provides a direct boundary derivation of the proposed complex extremal surface prescription and extends to R{\'e}nyi index $n>1$, for which we explicitly construct the corresponding complex cosmic brane geometry in the vacuum. 
We develop these ideas in several representative settings, including locally and globally excited states and quantum operator quenches, making manifest the precise agreement between boundary twist correlator and bulk complex geodesic calculations. For AdS-Vaidya, our approach predicts a different result from earlier piecewise geodesic constructions, while reproducing the field theory answer. Across these examples, the operator ordering uniquely determines the imaginary part of the complex-valued entropy, which is quantized in units of $c\pi/6$ and sensitive to the effective causal structure but not to the underlying dynamics.}

\maketitle

%%%%%%%%%%%%%%%%%%%%%%%%%%%%%%%%%%%%%%%%%%%%%
%%%%%%%%%%%%%%%%%%%%%%%%%%%%%%%%%%%%%%%%%%%%%
\newpage

%%%%%%%%%%%%%%%%%%%%%%%%%%%%%%
\section{Introduction}

In quantum many-body physics, tensor network methods are successful because many physically relevant states, such as the vacuum, low-lying excited states, and thermal states, contain only a limited amount of entanglement. This makes it possible to model them in a controllable way using contractions of local tensors with moderate bond dimension~\cite{Orus:2013kga,Cirac:2020obd}. The situation is much more challenging for time-dependent processes. During time evolution, entanglement is typically produced rapidly, and the evolving state soon becomes too costly to be represented within standard tensor network approaches. This well-known entanglement barrier has motivated alternative contraction schemes for tensor networks describing time evolution.

One such scheme, which is of relevance for the present paper, is the so-called transverse contraction~\cite{Banuls:2009jmn,Hastings:2014qqa}. In this approach, one obtains tensor network structures usually associated with quantum states, but now defined in an auxiliary Hilbert space emerging along the temporal direction. This observation underlies various notions of temporal entanglement~\cite{Giudice:2021smd,Lerose:2020fhd,Foligno:2023dih,Carignano:2023xbz,Carignano:2024jxb,Bou-Comas:2024pxf,Cerezo-Roquebrun:2026ocx}, and provides the conceptual backdrop for this article, as we will explore quantum field-theoretical and holographic realizations of temporal entanglement. Importantly, the natural analogue of the reduced density matrix in these quantum many-body settings is non-Hermitian. As a result, the analogues of Rényi entropies, including the entanglement entropy itself, are in general complex-valued. The non-Hermiticity of the underlying object places these quantities in the class of pseudoentropies, in the terminology introduced in~\cite{Nakata:2020luh}. Despite their unorthodox definition and unfamiliar properties, complex-valued entropies thus arise as natural quantities in the time evolution of quantum many-body systems.

Independently of these developments, structurally similar quantities began to appear more recently in the context of conformal field theory (CFT) and holography \cite{Doi:2022iyj, Doi:2023zaf}. The key idea of \cite{Doi:2022iyj, Doi:2023zaf} was to exploit the few cases in which the entanglement entropy of a CFT is known in closed form as a functional of the spatial subregion. Starting from these expressions, the authors defined a notion of timelike entanglement entropy (TEE) by analytically continuing them beyond the lightcone to timelike separations leading to complex-valued quantities.  One main motivation was to construct a new probe of holographic spacetimes. In the standard holographic prescription for entanglement entropy, the time direction is probed by using Poincar{\'e} transformations to move or boost the boundary entangling subregion, which anchors the corresponding bulk extremal surface to a different time slice~\cite{Rangamani:2016dms}. The vision of \cite{Doi:2022iyj, Doi:2023zaf} was that the analytically continued quantity would admit a bulk representation that probes the time direction more directly.

The resulting picture has proven more subtle than initial expectations. As shown by three of us in \cite{Heller:2024whi}, the natural bulk objects computing the analytically continued entanglement entropy in holography are complex extremal surfaces. These surfaces arise as genuine saddle points of the bulk gravitational path integral, can be identified systematically, and yield physically sensible holographic results. Other developments along these lines include~\cite{Nunez:2025gxq,Nunez:2025ppd,Nunez:2025puk,Zhao:2025zgm}. 
In parallel, a large body of work, including~\cite{
Doi:2022iyj,Doi:2023zaf,Li:2022tsv,Bohra:2025mhb,
Jiang:2023ffu,Basu:2024bal,Chang:2024voo,
Chu:2023zah,
Das:2023yyl,
He:2023ubi,
Basak:2023otu,Afrasiar:2024lsi,Jena:2024tly,Afrasiar:2024ldn,
Anegawa:2024kdj,Afrasiar:2025eam,Li:2026fcr,
Katoch:2025bnh,Katoch:2026dzs,
Wen:2024yny,Chu:2025sjv,
Goki:2026hpl,Prihadi:2026nua}, investigated geometric constructions for these quantities in which the real part is attributed to spacelike geodesic segments and the imaginary part to timelike ones, with the resulting object being a composite curve rather than a single extremal surface. In this work, we show that such piecewise constructions fail to capture the full structure of the TEE, already in the vacuum state of a CFT on a circle, dual to global AdS$_3$. Furthermore, in time-dependent settings such as global quenches that remain under full bulk-boundary control, these proxies of holographic TEE exhibit qualitatively incorrect features. 

Another key work motivating our present study is \cite{Heller:2025kvp}, which proposed a prescription for computing holographic TEE starting from holographic entanglement entropy. This consists of geometrically continuing all candidate bulk extremal surfaces beyond the lightcone, including subleading ones, and selecting as dominant the one that minimizes the real part of the area. Throughout this analytic continuation, the boundary subregion is transported while keeping both its shape and its coordinate size fixed. This prescription can always be implemented in the bulk for holographic CFTs in Minkowski space, reproduces all previously known closed-form expressions for holographic TEE, and has not led to any paradoxes.

Despite these strengths, the prescription of \cite{Heller:2025kvp} has important limitations. First, it was formulated and analyzed from a bulk perspective, leaving its CFT origin unexplored. Moreover, the extension beyond the lightcone into the timelike regime relies on a regularization procedure that is not derived from the dual CFT. As a consequence, the imaginary part of the TEE is not fixed unambiguously within the construction, and its physical interpretation remains unclear. Finally, on spacetimes with compact spatial directions, such as a 2d CFT on a Lorentzian cylinder, the prescription applies only to timelike subregions whose temporal extent is bounded by the size of a constant-time spatial slice.

These limitations motivate us to re-examine timelike entanglement in the present paper in the simplest and most tractable setting of 2d (holographic) CFTs and their gravity duals.
This setting is especially useful because the $n$-th R\'enyi entropy of a single interval can be computed as a two-point function of replica twist operators \cite{Calabrese:2009qy}. 
This formulation has long provided a powerful framework for computing entanglement entropy in a wide range of physical settings, exploiting the rich structure and analytic control of 2d CFT. It also played a central role in establishing the holographic entanglement entropy prescription from the boundary perspective.  In the semiclassical large-$c$ limit, the replica twist construction naturally reorganizes into a sum over saddles that are identified with appropriate spacelike extremal surfaces, establishing a direct correspondence between boundary replica computations and the Ryu-Takayanagi prescription \cite{Headrick:2010zt,Hartman:2013mia,Faulkner:2013yia,Asplund:2014coa}. 

\paragraph{Key idea.} The central idea of this paper is to formulate timelike entanglement and Rényi entropies in 2d CFT directly in terms of the time-ordered correlator of replica twist operators. As we describe in detail in the next section, the starting point is the Euclidean replica twist construction, in which the Rényi entropy of a spatial interval is encoded in the replica twist correlator
\be\nonumber
\Tr \rho_A^n =\langle\sigma_n(z_1,\bar z_1)\tilde\sigma_n(z_2,\bar z_2)\rangle_\rho \, .
\ee
Our proposal is to analytically continue the replica twist correlator to a time-ordered Lorentzian correlator with timelike-separated insertions,
\begin{equation}\nonumber
\langle \sigma_n(z_1,\bar z_1)\tilde\sigma_n(z_2,\bar z_2)\rangle_\rho
\;\longrightarrow\;
\langle T\!\left\{\sigma_n(t_1,x_1)\tilde\sigma_n(t_2,x_2)\right\}\rangle_\rho \, .
\end{equation}
Timelike Rényi entropies are then defined by applying the standard replica prescription to this Lorentzian correlator, and TEE is obtained in the replica limit,
\begin{equation}\nonumber
S^{\rm (T)}
=
\lim_{n\to1}
\frac{1}{1-n}
\log
\langle T\!\left\{\sigma_n(t_1,x_1)\tilde\sigma_n(t_2,x_2)\right\} \rangle_\rho \, .
\end{equation}
While the idea of introducing timelike-separated replica twist operators has already been explored in the context of pseudoentropy~\cite{Chen:2023gnh,Omidi:2023env,Guo:2023tjv,He:2023syy,Shinmyo:2023eci}, TEE~\cite{Gong:2025pnu,Guo:2025ase,Kanda:2026jyk}, and timelike Rényi entropies~\cite{Castro-Alvaredo:2026ohy}, here we develop a systematic field-theoretic framework for this construction in both 2d CFT and holography. 

This formulation naturally extends the standard replica twist construction to timelike-separated insertions and has several immediate consequences that address the central questions raised above:
\begin{enumerate}
\item[(1)] It defines TEE for single intervals of arbitrary temporal extent and enables explicit computations in a broad class of 2d CFT states.
\item[(2)] It uniquely fixes the imaginary part of the resulting complex-valued entropies through the operator ordering, thereby removing the continuation ambiguity and clarifying its physical origin.
\item[(3)] In holographic CFTs, the boundary computation selects the complex geodesics that minimize the real part of the length as the natural holographic dual of TEE.
\end{enumerate}

\paragraph{Outline and summary of results.} Our proposal is developed in detail in section \ref{sec:SITEE}, where we formulate the replica twist construction systematically, discuss its Lorentzian continuation, and illustrate it in the simplest example: the vacuum state on the line. On the CFT side, we show that time ordering and anti-time ordering select complex-conjugate branches of the twist correlator and thereby fix the sign of the imaginary part of the TEE, whose absolute value is $c \,\pi/6$. On the holographic side, the geodesic approximation to the time-ordered twist correlator provides a direct derivation of complex geodesics as the bulk carriers of TEE, with the choice of operator ordering selecting one of the two complex-conjugate solutions.

Importantly, we extend our construction beyond the von Neumann limit to R\'enyi index $n>1$. Holographically, timelike R\'enyi entropies are computed by the area of a codimension-two cosmic brane with $n$-dependent tension~\cite{Dong:2016fnf,Dong:2018lsk}. The backreacted geometry is constructed explicitly using the Bañados and Roberts maps~\cite{Banados:1998gg,Roberts:2012aq}. For timelike-separated twist insertions, the bulk spacetime, the brane profile, and its regularized area all become complex; see eqs.~\eqref{eq:uTlambda} and~\eqref{eq:RlengthFinal}. Integrating the refined entropy then reproduces the CFT timelike R\'enyi entropy~\eqref{eq:TRSvac}. This provides a timelike extension of the cosmic brane prescription, along the lines of the real-time replica analysis of~\cite{Colin-Ellerin:2021jev}.

Building on this framework, the remainder of the paper explores its consequences in a variety of settings, both in 2d CFT and in holography.

In section \ref{sec:Im_TEE}, we turn to the vacuum state on the Lorentzian cylinder. In our construction, the time-ordered twist correlator can be continued to temporal separations of arbitrary extent, thereby removing the restriction to a single causal diamond present in the geometric continuation of~\cite{Heller:2025kvp}. Each additional lightcone crossing moves the correlator onto a new sheet. As a result, the imaginary part of TEE is quantized in units of $c\,\pi/6$ and counts the number of causal diamonds separating the twist insertions. The corresponding complex geodesics in global AdS$_3$ reproduce both the real part of the TEE and the accumulated phase. This example makes manifest that the imaginary part of TEE, whose sign is fixed by the operator ordering, is also sensitive to the causal structure of the boundary correlator. 

In section \ref{sec.locallyexcited}, we extend the analysis to states prepared by heavy local primary operators. We begin by considering high-energy eigenstates on the circle, dual to conical defect and BTZ microstate geometries. For these states, the relevant twist correlators are described by semiclassical heavy-heavy-light-light four-point functions, whose dominant contributions come from competing Virasoro identity-block channels~\cite{Asplund:2014coa}. These channels are mapped holographically to complex geodesics with different winding numbers in conical defect and BTZ geometries. The competition between these channels shows directly that the dominant contribution is the one with the smallest real part of the geodesic length, thereby providing a boundary derivation of the saddle-selection rule proposed in~\cite{Heller:2025kvp}. 
In the conical-defect regime, the imaginary part counts light-cone crossings in a rescaled effective causal structure, whereas in the BTZ regime it reduces to the Minkowski-space value whenever the dominant semiclassical channel is timelike. Together with the results of section~\ref{sec:Im_TEE}, this shows that the imaginary part of the TEE probes the effective causal structure encoded by the dominant semiclassical channel.

We next go beyond static states and analyze a local operator quench. In this case, the CFT computation shows that the real part of the TEE departs from its vacuum value only while one of the two twist insertions has crossed the lightcone of the excitation and the other has not. This behavior naturally admits a quasiparticle picture interpretation. By contrast, the imaginary part of the TEE remains constant and equal to its vacuum value at all times. This boundary result is matched exactly by a bulk complex geodesic computation in a boosted conical defect geometry.

In section \ref{sec.excited}, we study globally excited states. We first revisit a thermal state on the line and its planar black hole dual, providing a further cross-check of our construction. We then analyze a global quench protocol whose bulk dual is the planar AdS$_3$-Vaidya geometry. In the thin-shell limit, the known equal-space CFT correlator~\cite{Anous:2016kss} gives the TEE directly. Its real part is again time dependent for insertions across the quench and is compatible with a quasiparticle interpretation, whereas its imaginary part remains time independent, coinciding with both its vacuum and thermal equilibrium values. On the gravity side, the time-dependent part of the CFT result is reproduced exactly by complex geodesics crossing the collapsing shell \cite{Balasubramanian:2012tu}.
This constitutes a sharp test of the holographic dictionary: the complex geodesic answer agrees with the explicit CFT calculation but differs from the piecewise-curve construction of~\cite{Katoch:2025bnh,Katoch:2026dzs}. 
We verify the same picture numerically for thick-shell Vaidya spacetimes. These results provide further evidence that complex geodesics are not merely one possible representation of holographic TEE, but are the saddles selected by its CFT definition. They do not exclude piecewise constructions as independent geometric probes of the bulk, but they do exclude them as the gravitational carriers of TEE.
\newpage
We conclude with a discussion of future directions in section \ref{sec:discussion}, including extending our construction to multiple intervals, further exploring the quasiparticle picture, deepening our understanding of the imaginary part of the TEE, developing higher-dimensional generalizations, understanding the role of the homology constraint in the timelike setting, and exploring connections with quantum many-body physics and dS/CFT. 

Technical details concerning the complex geodesics, the numerical treatment of thick-shell Vaidya geometries, and the Lorentzian ordering of the four-point functions are collected in appendices~\ref{App:geo}, \ref{app:Vaidya}, and~\ref{app:TO}, respectively.\\

\noindent \emph{Note on conventions.} Throughout the paper we set the AdS curvature radius to unity. Similarly, when working with CFTs on the Lorentzian cylinder, we also set the radius of the associated circle to unity. Finally, we choose to refer to the key quantity we are considering as TEE in general and holographic TEE when referring to its gravitational dual. Our main motivation was continuity of naming conventions with respect to earlier works, in particular the original papers proposing this kind of quantity~\cite{Doi:2022iyj,Doi:2023zaf}.

%%%%%%%%%%%%%%%%%%%%%%%%%%%%%%%%%%%%%%%%%%%%%
%%%%%%%%%%%%%%%%%%%%%%%%%%%%%%%%%%%%%%%%%%%%%
\section{Replica twist correlators and complex geodesics} \label{sec:SITEE}

The entanglement entropy of a spacelike interval $A$ in a 2d CFT is 
\be \label{eq:EE}
S_A = - \tr \rho_A \log \rho_A\,,
\ee
and can be obtained from the R\'enyi entropy
\be \label{eq:RE}
S_{A,n} = \frac{1}{1-n}\log \tr \rho_A^n
\ee
by analytic continuation from integer values $n\geq 2$. Here $\rho_A=\tr_{\bar A}\rho$ is the reduced density matrix obtained by tracing over the complement of $A$. We consider states $\rho$ that admit a Euclidean path integral representation, such that $\tr \rho_A^n$ and the R\'enyi entropies can be computed by gluing together $n$ copies of the Euclidean path integral along the region $A$. These include the vacuum state prepared by a path integral on the half-plane or disk, thermal states prepared by Euclidean path integrals on a cylinder or torus, and excited states obtained from these by acting with operator insertions.

The resulting replica partition function can equivalently be expressed as a two-point function of replica twist operators inserted at the endpoints of the interval $A$ (see \cite{Calabrese:2009qy} for a review),
\be \label{eq:replicacorr}
\Tr \rho_A^n = \langle \sigma_n(z_1,\bz_1)\tilde \sigma_n(z_2,\bz_2)\rangle_\rho .
\ee
The correlator is defined in the cyclic orbifold CFT$^n/\mathbb{Z}_n$, where $\sigma_n$ and $\tilde{\sigma}_n$ are the twist and anti-twist operators implementing cyclic permutations of the $n$ replicas in opposite orientations. Here $\langle \dots \rangle_\rho$ denotes the Euclidean path integral representation of the correlation function. For example, the vacuum and thermal states correspond respectively to correlators evaluated on the plane and on the thermal cylinder.

The twist operators are operators of the orbifold CFT that transform as primaries, with conformal dimensions
\be
h_n=\bar h_n=\frac{c}{24}\left(n-\frac{1}{n}\right)\, .
\ee
In writing the replica correlator \eqref{eq:replicacorr}, we have suppressed an overall UV-dependent normalization factor arising from regulating the twist operator insertions. This contribution will be reinstated below. 

The Euclidean correlator \eqref{eq:replicacorr} can then be analytically continued to Lorentzian signature. For spacelike-separated twist insertions, it reproduces the standard R\'enyi and entanglement entropies \eqref{eq:RE} and \eqref{eq:EE}. This suggests that the replica correlator itself provides a natural object to analytically continue beyond spacelike separations.

For timelike separations of the twist insertions, it is natural to define a timelike extension of these quantities by analytically continuing the replica correlator \eqref{eq:replicacorr} to the time-ordered expression
\be
 \langle T \left\{ \sigma_n (t_1,x_1) \tilde \sigma_n(t_2,x_2) \right\} \rangle_\rho \,,
 \label{eq:tord}
\ee
where $T\{ \dots\}$ denotes time ordering and the operator ordering in the Lorentzian regime can be fixed by an appropriate $i\varepsilon$ prescription (see, \eg \cite{Hartman:2015lfa,Kundu:2025jsm}).

From this Lorentzian correlator one obtains a timelike extension of the R\'enyi entropy
\be
S_{A,n}^{\rm (T)} \equiv \frac{1}{1-n}\log \,
\langle T \left\{ \sigma_n (t_1,x_1) \tilde \sigma_n(t_2,x_2) \right\} \rangle_\rho \,,
\ee
and of the entanglement entropy
\be \label{eq:TEEdef}
S_A^{\rm (T)} \equiv  \lim_{n\to 1}  S_{A,n}^{\rm (T)} = \lim_{n\to 1}\frac{1}{1-n}\log
\, \langle T \left\{ \sigma_n (t_1,x_1) \tilde \sigma_n(t_2,x_2) \right\}\rangle_\rho \,,
\ee
where $A$ now labels the ordered timelike endpoint pair defined by the continued twist correlator. 
The analytic continuation of the entanglement entropy to a timelike subsystem was dubbed TEE in \cite{Doi:2022iyj,Doi:2023zaf} and denoted $S_A^{\rm (T)}$, as above. Here we propose a replica-based definition of this quantity by directly analytically continuing the time-ordered replica twist correlator, rather than the final expression obtained for the spacelike entanglement entropy in \cite{Doi:2022iyj,Doi:2023zaf}.

The holographic implementation of this prescription in a holographic 2d CFT is then straightforward: one needs to evaluate the holographic correlator \eqref{eq:tord} and plug the result in the entropy formula \eqref{eq:TEEdef}. Given that the replica twists transform as heavy primary operators, their two-point functions in the large-$c$ limit can be evaluated in a geodesic approximation (see, \eg \cite{Balasubramanian:2010ce,Balasubramanian:2011ur} for a review). While real spacelike geodesics are the relevant saddles for spacelike-separated boundary points, no real geodesic saddle connects timelike-separated points on the AdS boundary, but complex geodesics can. These curves make excursions in a complexified AdS geometry before returning to real points on the boundary and were first studied in \cite{Kraus:2002iv,Fidkowski:2003nf,Festuccia:2005pi}, later used to evaluate holographic retarded Green functions in \cite{Balasubramanian:2012tu} and more recently the holographic TEE in \cite{Heller:2024whi,Heller:2025kvp,Nunez:2025gxq,Nunez:2025ppd}.

In this section, we illustrate these ideas in detail by revisiting the simplest example: a single interval in the vacuum state of a 2d CFT on a line dual to planar AdS$_3$, matching step by step the CFT and geometric computations, both for the TEE and the timelike R\'enyi entropy. While all final results are well known in the literature, this example allows us to clarify three main points: 
\begin{itemize}
    \item  The TEE is obtained by analytically continuing the replica twist correlator to a time-ordered Lorentzian correlator.
    \item Through the holographic dictionary, this becomes a two-point function of heavy replica twist insertions, whose semiclassical contribution is captured by a geodesic approximation.
    \item For timelike separations no real saddles exist and one needs to consider complex ones.
\end{itemize}

%%%%%%%%%%%%%%%%%%%%%%%%%%%%%%%%%%%%%%%%%%%%%
\subsection{TEE in the vacuum state of a CFT on a line}

The simplest case in which we can evaluate \eqref{eq:tord} is the vacuum state of a 2d CFT on the infinite line. In the Euclidean plane we then consider the two-point function
\be 
\langle 0 |  \sigma_n(z_1,\bz_1)   \tilde \sigma_n(z_2,\bz_2)  | 0\rangle    
= \frac{1}{[(z_1-z_2)(\bz_1-\bz_2)]^{2 h_n}}\, . 
\label{eq:2ptplane}
\ee
For spacelike-separated insertions, this correlator is uniquely fixed by conformal invariance, 
and for an interval $A$ of size $\Delta x$ yields the well-known R\'enyi and entanglement entropies
\bea
S_{A,n} &=& \frac{c}{6} \left(1+\frac{1}{n}\right) \log\[ \frac{\Delta x}{\delta} \]\label{eq:SAnSpacelike}\\[.5em]
S_A &=& \frac{c}{3}\log\[\frac{\Delta x}{\delta}\]\, .
\eea
Here we restored the UV cutoff $\delta$ that regulates the twist insertions, and in writing 
\eqref{eq:SAnSpacelike} we dropped an $n$-dependent non-universal constant 
\cite{Calabrese:2009qy}. 

When continuing to Lorentzian time, timelike separations introduce branch cuts associated with lightcone crossings. Assuming $\Delta t\geq0$, the time-ordered correlator can be obtained by continuing \eqref{eq:2ptplane} with the following $i\varepsilon$ prescription for $\varepsilon>0$
\be
\begin{aligned}\label{eq:eq:2ptplaneTO}
\langle 0 |  \sigma_n(\Delta t/2,\Delta x/2)  \tilde \sigma_n(-\Delta t/2,-\Delta x/2)  | 0\rangle
&= \lim_{\varepsilon \to 0 } \frac{1}{\left[-(\Delta t-i\varepsilon)^2+\Delta x^2\right]^{2 h_n}} \\
&= \frac{1}{\left(e^{i\pi}(\Delta t^2-\Delta x^2)\right)^{2 h_n}}\, .
\end{aligned}
\ee
For timelike separations $\Delta t^2>\Delta x^2$, the continuation crosses the branch cut of the
logarithm appearing in the twist correlator. The choice of the $i\varepsilon$ prescription fixes 
which side of the branch cut is selected and therefore determines the imaginary contribution to 
the entropy. Thus the timelike R\'enyi and entanglement entropies are
\bea
S_{A,n}^{\rm (T)} &=&\frac{c}{6}\left(1+\frac{1}{n}\right)
\left(\log \[ \frac{\sqrt{\Delta t^2-\Delta x^2}}{\delta}
\] +i\frac{\pi}{2}\right) \label{eq:TRSvac} \\[.5em]
S_A^{\rm (T)}&=& \frac{c}{3}\log\[\frac{\sqrt{\Delta t^2-\Delta x^2}}{\delta}\]+i\frac{c}{6}\pi \, . \label{eq:TSvac}
\eea
The latter coincides with the definition given in \cite{Doi:2023zaf} through the continuation of 
the closed-form expression for the entanglement entropy. 

The anti-time-ordered correlator is obtained analogously, with the opposite sign for the 
$i\varepsilon$ regularization,
\be
\begin{aligned}
\langle 0 | \tilde \sigma_n(-\Delta t/2,-\Delta x/2)  \sigma_n(\Delta t/2,\Delta x/2) | 0\rangle   
&= \lim_{\varepsilon \to 0 } \frac{1}{\left[-(\Delta t+i\varepsilon)^2+\Delta x^2\right]^{2 h_n}}\\
&=  \frac{1}{\left(e^{-i\pi}(\Delta t^2-\Delta x^2)\right)^{2 h_n}}\, .\label{eq:eq:2ptplaneATO}
\end{aligned}
\ee
This selects the opposite side of the branch cut and therefore gives the opposite sign for the 
imaginary contribution in \eqref{eq:TSvac}. Hence, the imaginary part of the TEE is fixed by the 
operator ordering chosen in the Lorentzian continuation.

Equivalently, one may describe the analytic continuation in terms of paths in the complex plane rather than the $i\varepsilon$ prescription \cite{Hartman:2015lfa}. Starting from the Euclidean correlator \eqref{eq:2ptplane} with $z_1=-z_2=i\tau/2$, $\bz_1=-\bz_2=-i\tau/2$ (for $\Delta x=0$), the time-ordered Lorentzian correlator \eqref{eq:eq:2ptplaneTO} is obtained by setting $\tau=\Delta t\,e^{i\theta}$ and following the path $\theta\in[0,\pi/2]$. 
The anti-time-ordered continuation instead corresponds to taking the path around the singularity at $z=0$ on the opposite side. This amounts to the replacement $z_1-z_2\rightarrow (z_1-z_2)e^{-i\pi}$, selecting the opposite branch and therefore yielding the opposite imaginary contribution.

%%%%%%%%%%%%%%%%%%%%%%%%%%%%%%%%%%%%%%%%%%%%%
\subsection{Holographic TEE in Poincar\'e AdS$_3$}

The dual geometry in this case is Poincar\'e AdS$_3$
\be \label{eq:PAdSmetric}
ds^2 = \frac{1}{y^2} \(-dt^2  + dy^2 + dx^2\) \,,
\ee
where $y$ denotes the holographic radial direction, with boundary at $y=0$. The result 
\eqref{eq:TSvac} can equivalently be obtained from the holographic two-point function of 
replica twists \eqref{eq:tord}. For such heavy insertions at timelike separations, the 
correlator can be evaluated geometrically in a geodesic approximation via the renormalized 
length $\delta {\mathcal L}$ of complex solutions to the spacelike geodesic equations 
\cite{Balasubramanian:2012tu,Heller:2024whi}
\be
\langle 0 |  \sigma_n(\Delta t/2, \Delta x/2) \tilde \sigma_n (- \Delta t/2, - \Delta x/2)  | 0\rangle  \sim e^{-2h_n \delta {\mathcal L}}\,. \label{eq:geodesicapprox}
\ee
A spacelike geodesic connecting the points
\be
(t_1,x_1,y_1)  = \(\frac{\Delta t}{2},\frac{\Delta x}{2},\eps\)\,, \quad  (t_2,x_2,y_2) = \(- \frac{\Delta t}{2}, -\frac{\Delta x}{2},\eps\)
\ee
on the regularized boundary $y=\epsilon$ of Poincar\'e AdS$_3$ can be written as (see appendix \ref{app:AdSPoincare})
\be \label{eq:tAdS}
\begin{aligned}
t(\lambda) &= \frac{\Delta t}{2}\tanh\lambda \\[.5em]
x(\lambda) &= \frac{\Delta x}{2}\tanh\lambda  \\[.5em]
y(\lambda) &= \frac{\sqrt{\Delta x^2-\Delta t^2}}{2}\,\operatorname{sech}\,\lambda  
\end{aligned}
\ee
in the limit of a small boundary regulator. The affine parameter $\lambda$ is in general 
complex. Translation invariance allows for a symmetric parametrization, and the boundary 
points are reached for
\be
\lambda \to \pm \lambda_* = \pm \frac 1 2  \log\!\left[ \frac{-\Delta t^2 + \Delta x^2}{\epsilon^2}\right] \,,
\ee
corresponding to a geodesic length
\be
{\mathcal L} \equiv 2 \lambda_* =  \log\!\left[ \frac{-\Delta t^2 + \Delta x^2}{\epsilon^2}\right].
\ee
For spacelike separations, one recovers the usual result with an affine parameter connecting 
the real extrema $\pm\lambda_*$. For timelike separations of the boundary endpoints one can 
follow the same procedure as in the 2d CFT and regulate the computation with the appropriate 
$i\varepsilon$ prescription. For time-ordered insertions, we obtain
\be
\lambda_*  = \lim_{\vareps \to 0} \frac 1 2 \log\!\left[ \frac{-(\Delta t - i \varepsilon)^2 + \Delta x^2}{\epsilon^2}\right] =  \frac 1 2 \log\!\left[ \frac{\Delta t^2 -  \Delta x^2}{\epsilon^2}\right] + i \frac{\pi}{2}\,.
\ee
Requiring that the usual spacelike result is recovered as $\Delta t$ is decreased resolves the 
ambiguity in the choice of sheet for the logarithm and gives the complex geodesic length
\be
{\mathcal L}\equiv 2 \lambda_*  = \log\!\left[ \frac{\Delta t^2 -  \Delta x^2}{\epsilon^2}\right] + i \pi \,. \label{eq:LAdS}
\ee
Notice that the solution \eqref{eq:tAdS} parametrizes a single complex geodesic 
with multiple sections and length \eqref{eq:LAdS}, since any path connecting the given endpoints in the complex $\lambda$ plane is allowed. These paths correspond to different representations of the same physical saddle contribution, as they can be continuously deformed into one another without crossing singularities or modifying the endpoint constraints, and therefore belong to the same homotopy class of admissible contours. The same reasoning applies to both timelike and spacelike  separations: the affine parameter $\lambda$ may also be complexified in the spacelike case, and the prescribed boundary data again select an admissible class of complex geodesics whose length in this case is real.

From eq.~\eqref{eq:LAdS}, we define the renormalized length $\delta {\mathcal L}\equiv{\mathcal L}+ 2 \log \eps$ by removing the divergent part of the geodesic length in pure AdS \cite{Balasubramanian:2011ur,Balasubramanian:2012tu}. Substituting 
into \eqref{eq:geodesicapprox}, this coincides with \eqref{eq:eq:2ptplaneTO} and directly leads to the non-divergent part of \eqref{eq:TSvac}. Alternatively, the regulated length $\mathcal{L}$ \eqref{eq:LAdS} can be directly matched to the CFT result with the trivial identification of the cutoffs $\delta=\eps$.

%%%%%%%%%%%%%%%%%%%%%%%%%%%%%%%%%%%%%%%%%%%%%
\subsection{Holographic timelike R\'enyi entropy}\label{sec:Renyi}

The holographic prescription for evaluating the R\'enyi entropy of a spacelike interval was worked out in \cite{Hung:2011nu,Lewkowycz:2013nqa,Dong:2016fnf,Dong:2018lsk}. Here we first revisit the case of a spacelike interval in a 2d CFT on a line in the vacuum state, with R\'enyi entropy given by \eqref{eq:SAnSpacelike}. 
We work out the result holographically along the lines of \cite{Dong:2016fnf,Dong:2018lsk}, where the refined entropy (a~derivative of the R\'enyi entropy with respect to the R\'enyi index $n$) is related to the area of a bulk codimension-two cosmic brane homologous to the entangling region. The brane has a specified $n$-dependent tension and backreacts on the ambient geometry by creating a conical deficit angle. To obtain the bulk geometry with a cosmic brane, one can solve for the classical solution to the equations of motion resulting from a total (Euclidean) action $I_{\rm bulk} + I_{\rm brane}$ \cite{Dong:2016fnf}. 

In the special case of AdS$_3$/CFT$_2$, where a replica twist construction is available in the CFT and three-dimensional gravity has no propagating degrees of freedom, the backreacted bulk geometry can also be constructed directly from the boundary stress tensor.  Here we use this approach  following \cite{Banados:1998gg, Fitzpatrick:2014vua,Asplund:2014coa}. In this way, the area of the cosmic brane that holographically evaluates the R\'enyi entropy is directly related to the regularized length of a geodesic on a three-dimensional geometry with a defect.

An alternative is to evaluate on-shell the full gravitational action. This was carried out explicitly in \cite{Hung:2011nu,Lewkowycz:2013nqa,Colin-Ellerin:2021jev}, both in Euclidean and Lorentzian signature. We instead compute the cosmic brane area as it directly extends the holographic TEE computation of the previous section via complex geodesics. Indeed, for timelike separations, the brane, the geodesic trajectory and its length all become complex; see eqs.~\eqref{eq:uTlambda} and \eqref{eq:RlengthFinal}. 

%%%%%%%%%%%%%%%%%%%%%%%%%%%%%%%%%%%%%%%%%%%%%
\subsubsection{Ba\~{n}ados geometries from replica twists} 

Consider the Euclidean asymptotically AdS$_3$ geometry \cite{Banados:1998gg}
\be \label{eq:Banados}
ds^2 = \frac{L}{2} \,dz^2 + \frac{\bar L}{2} \,d \bar z^2 + \left( \frac{1}{y^2} + \frac{y^2}{4} L \bar L \right) dz\,d\bar z  +  \frac{dy^2}{y^2}\,,
\ee
with 
\be 
L = L(z)\,, \qquad \bar L = \bar L (\bar z)\,,
\ee
where $z$ and $\bar z$ are complex coordinates on the boundary plane and $y$ is the bulk radial coordinate. This is a solution of the vacuum Einstein equations for arbitrary $L, \bar L$  (away from possible singularities). In such a state the holographic dictionary relates $L$ with the CFT stress tensor $T = \langle T_{zz} (z) \rangle$ as
\be 
T = - \frac{c_n}{12} L \,.
\ee
The relevant semiclassical stress tensor in the presence of two heavy insertions, here replica twists, at $z = z_1, z_2$, both with weights $(h_n, \bar h_n)$ is 
\be \label{eq:Tdefect}
T(z) = h_n \, \frac{(z_1 - z_2)^2} {(z - z_1)^2 (z - z_2)^2}\,, \qquad \bar T(\bar z) = \bar h_n \, \frac{(\bar z_1 - \bar z_2)^2} {(\bar z - \bar z_1)^2 (\bar z - \bar z_2)^2}\,. 
\ee
To be explicit, we have denoted the central charge of the replica theory where the twist operators are defined as $c_n = n c$, with $c$ being the central charge of the seed CFT.

\begin{figure}[t]
\centering
\includegraphics[width=0.43\textwidth]{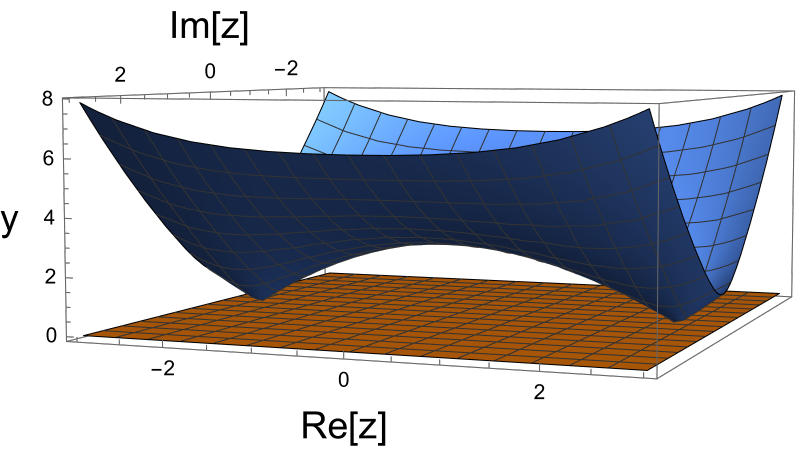}  \hfill \includegraphics[width=0.43\textwidth]{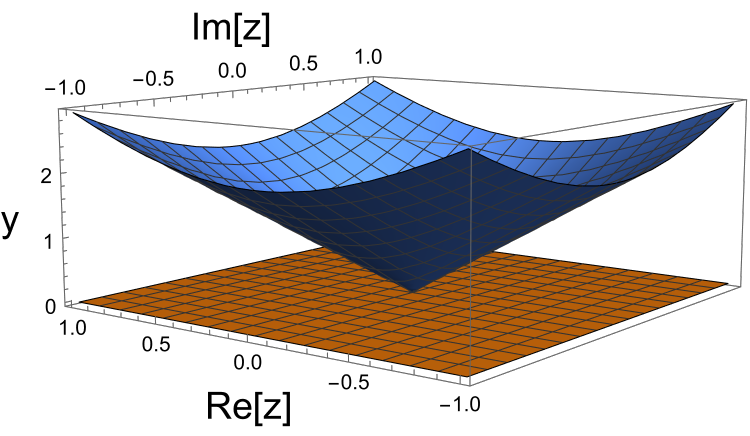}
 \caption{(Left) In blue, the wall surface of the Ba\~{n}ados geometry where $\det g =0$ for $z_1 - z_2 = \ell = 5$ and R\'enyi index $n=4$. (Right) For canonical insertions $z_1 =0, z_2 = \infty$, the wall is a cone emanating from the origin.
 } \label{fig:banana}
\end{figure}

The metric determinant of \eqref{eq:Banados} vanishes at the hypersurface
\be \label{eq:detg0}
y^4 L(z) \bar L(\bar z) =4 \,,
\ee
depicted in figure \ref{fig:banana}. 
This is the same surface that was dubbed \textit{wall} in the closely related \textit{spacetime banana} geometries of \cite{Abajian:2023jye,Abajian:2023bqv}. It signals a breakdown of the Ba\~{n}ados coordinates in parametrizing this  backreacted geometry as we move inward from the AdS boundary. 

To continue the geometry beyond the wall, we exploit the fact that all Euclidean solutions of three-dimensional gravity with negative cosmological constant are locally AdS$_3$ and therefore can be written in the form \eqref{eq:Banados} \cite{Banados:1998gg}. 
To do this, we look for a boundary conformal map that transforms the stress tensor \eqref{eq:Tdefect} to the vacuum stress tensor and then extend it to the corresponding bulk diffeomorphism.
The Euclidean geometry \eqref{eq:Banados} is dual to the stress tensor \eqref{eq:Tdefect}, while pure AdS$_3$ in Poincar\'e coordinates has $T = \bar T = 0$,
\be \label{eq:EAdS3metric}
ds^2 = \frac{dw\, d\bar w+ du^2}{u^2}\,.
\ee
Recall the conformal transformation law 
\be \label{eq:TConfMap}
T(z) = \(\frac{dw}{dz}\)^2 T(w(z)) + \frac{c_n}{12} \{w,z\}\,,
\ee
where
\be
\{w,z\} = \frac{w'''(z) w'(z) - \frac 3 2 w''(z)^2}{w'(z)^2} \,. 
\ee
Solving \eqref{eq:TConfMap} with $T(w)=0$ gives the three-parameter family
\be \label{eq:map}
w(z) = a_1 \frac{(z -z_2)^{\alpha_n}}{(z-z_1)^{\alpha_n} (z_1-z_2)^{\alpha_n} + a_2 (z-z_2)^{\alpha_n}} + a_3\,,
\ee
where 
\be
\alpha_n \equiv \sqrt{1- \frac{24 h_n}{c_n}} = \frac 1 n\,.
\ee
Choosing $a_1 =1, a_2 = a_3 =0$, the map takes the simple form
\be \label{eq:mapsimpleMain}
w(z) = \frac{(z-z_2)^{\alpha_n}}{(z-z_1)^{\alpha_n} (z_1-z_2)^{\alpha_n}}\, .
\ee
This is the uniformization map from the $n$-sheeted manifold to the quotient one. Its action is particularly simple to visualize for canonical insertions $z_1 =0, z_2 = \infty$, for which it reduces to $w(z) = z^{-1/n}$. It therefore maps the entire complex $z$-plane to a wedge of opening angle $2\pi/n$ in the complex $w$-plane with boundaries identified. 

\begin{figure}
\centering
\includegraphics[width=0.55\textwidth]{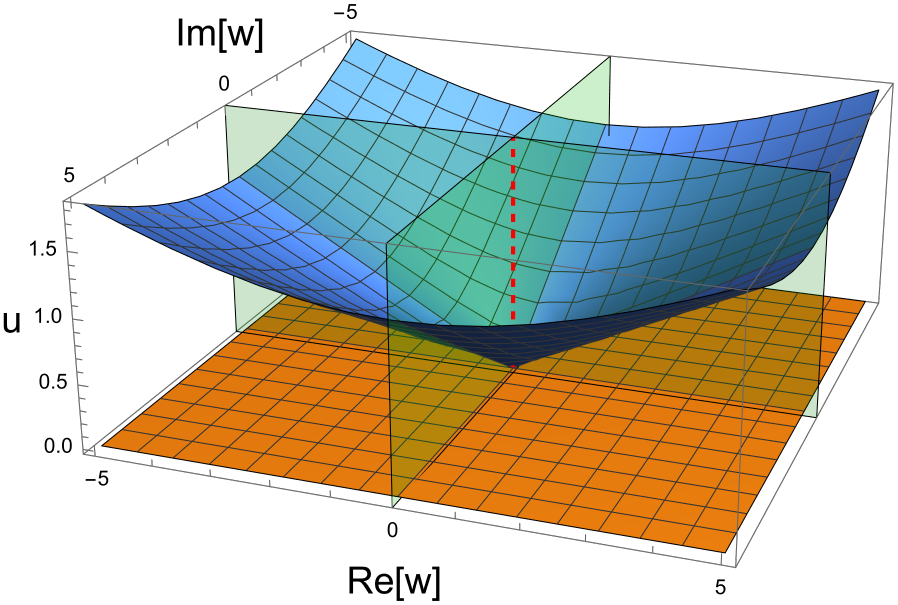}  \hfill \includegraphics[width=0.38\textwidth]{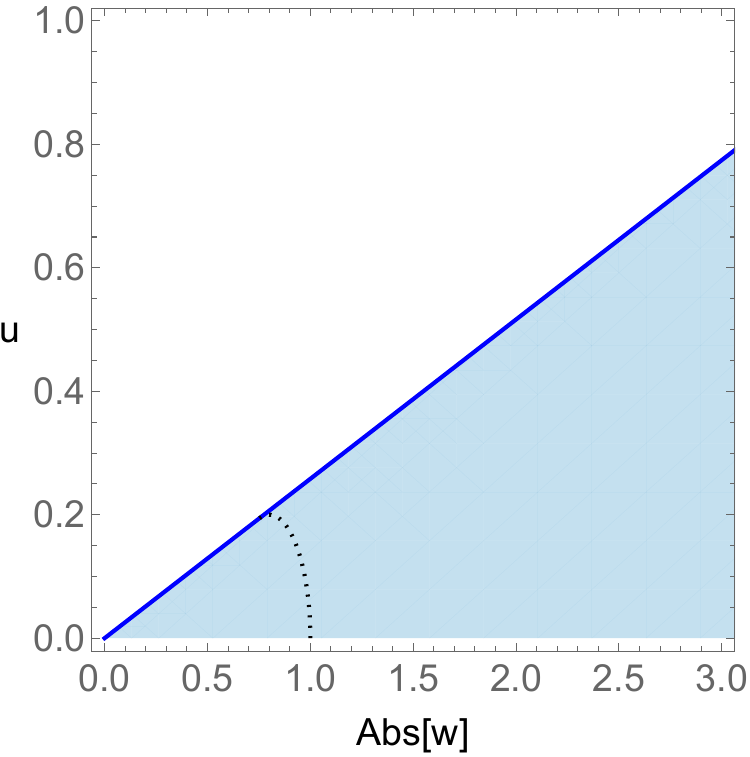} 
 \caption{(Left) The wall \eqref{eq:EAdSwall} depicted in EAdS$_3$ emanating from $w=0$ on the boundary. Here we set $n=4$ and used canonical insertions $z_1 =0, z_2 = \infty$. In red dashed, the AdS geodesic that connects $w=0$ and $w=\infty$ on the boundary. In light green, the periodic identifications. (Right) A section of the bulk wedge of opening angle $2\pi/n$. The Ba\~{n}ados coordinates only cover the region between the boundary and the wall of eq.~\eqref{eq:EAdSwall}, in blue solid. In black dashed, an illustration of the action of the diffeomorphism for $z \bar z =1$.
 } \label{fig:EAdSwall}
\end{figure}

The corresponding non-linear bulk diffeomorphism relating Poincar\'e AdS$_3$ to a generic geometry \eqref{eq:Banados} was derived by Roberts \cite{Roberts:2012aq} and reads
\be\label{eq:RobertsFull} 
\begin{aligned}
w &= w(z) - \frac{2 y^2 w'(z)^2 \bar w''(\bar z)}{4 w'(z) \bar w'(\bar z) + y^2 w''(z) \bar w''(\bar z)}  \\[.5em]
\bar w &= \bar w(\bar z) - \frac{2 y^2 \bar w'(\bar z)^2 w''(z)}{4 \bar w'(\bar z) w'(z) + y^2 \bar w''(\bar z) w''(z)}   \\[.5em]
u &= y \frac{4 (w'(z) \bar w'(\bar z))^{3/2}}{4 w'(z) \bar w'(\bar z) + y^2 w''(z) \bar w''(\bar z)}\,.  
\end{aligned}
\ee
Under this diffeomorphism, the boundary insertion points
\be\label{eq:Banadosdefect}
(z, \bar z, y) = (z_1, \bar z_1, 0)\,, \qquad (z,\bar z, y) = (z_2,\bar z_2, 0)\,,
\ee
map to boundary insertions in AdS$_3$
\be \label{eq:AdSdefect}
(w, \bar w, u) = (\infty, \infty, 0)\,, \qquad (w, \bar w, u) = (0,0, 0)\,.
\ee
Again, the geometric action of the diffeomorphism is easier to visualize for canonical insertions $z_1 =0, z_2 = \infty$. The boundary identifications of the wedge of opening angle $2\pi/n$ extend straight into the bulk for all $u$. The wall \eqref{eq:detg0} in the Ba\~{n}ados geometry \eqref{eq:Banados} maps to a cone emanating from the origin in Euclidean AdS$_3$ (EAdS$_3$)
 \be \label{eq:EAdSwall}
u = \frac{\sqrt{w \bar w}}{\sqrt{n^2-1}}\,,
 \ee
as depicted in figure \ref{fig:EAdSwall}~(left). The patch covered by the Ba\~{n}ados coordinates between the boundary and the wall maps to the region between the EAdS$_3$ boundary and the cone, see figure \ref{fig:EAdSwall} (right). 
The coordinate system $(w, \bar w, u)$ therefore provides a regular description of the geometry beyond the wall. In particular, they allow us to parametrize the surface where the defect in the original geometry lies, \ie the geodesic that connects $w=0$ to $w =\infty$ on the boundary, as represented in figure \ref{fig:EAdSwall} (left).

%%%%%%%%%%%%%%%%%%%%%%%%%%%%%%%%%%%%%%%%%%%%%
\subsubsection{Holographic R\'enyi entropy from codimension-two cosmic brane}
 
The Roberts diffeomorphism also provides a simple way to compute geodesic lengths in the defect geometry by mapping them to geodesics in Poincar\'e AdS$_3$.

A geodesic in Poincar\'e AdS$_3$, anchored near the boundary at
\be
(w, \bar w, u) = (w_1, \bar w_1, u_1)\,, \qquad (w, \bar w, u) = (w_2, \bar w_2, u_2)\,,
\ee
is given, to leading order in a near-boundary expansion, by (see appendix \ref{app:Banados})
\be  
\begin{aligned}
u(\lambda) &= \frac{\sqrt{w_1 - w_2} \sqrt{\bar w_1 -\bar w_2}}{2 \cosh \lambda }  \\
w(\lambda) &= \frac 1 2 \left\{ w_1 + w_2 +(w_1 - w_2) \tanh \lambda \right\}\\
\bar w(\lambda) &= \frac 1 2 \left\{ \bar w_1 + \bar w_2 +(\bar w_1 - \bar w_2) \tanh \lambda \right\}\,,
\end{aligned}
\ee
with geodesic length 
\be \label{eq:RobertsAdSshort}
{\mathcal L}  = \log \frac{(w_1 - w_2)(\bar w_1 -\bar w_2)}{u_1 u_2}\,.
\ee
Parameterizing the geodesic in terms of $u$ we can also write the solution as the two branches
\be \label{eq:wu}
w_\pm(u) =  \frac 1 2 \left\{ w_1 + w_2 \pm (w_1 - w_2) \sqrt{1- \frac{4 u^2}{(w_1 - w_2)(\bar w_1 -\bar w_2)}} \right\}\,,
\ee
and analogously for $\bar w(u)$. 

The length of a geodesic in the singular geometry \eqref{eq:Banados}--\eqref{eq:Tdefect}, anchored near the boundary at
\be
(z, \bar z, y) = (\tilde z_1, \bar{\tilde z}_1, \epsilon)\,, \qquad
(z, \bar z, y) = (\tilde z_2, \bar{\tilde z}_2, \epsilon)\,,
\ee
is straightforward to compute using the results of \cite{Roberts:2012aq}. The endpoints are first mapped to Poincar\'e AdS$_3$ through the diffeomorphism \eqref{eq:RobertsFull}, after which the geodesic length follows directly from \eqref{eq:RobertsAdSshort}. Since the endpoints lie close to the AdS boundary, only the asymptotic form of the diffeomorphism is needed,
\be\label{eq:Roberts1} 
\begin{aligned}
w_i &\approx w (\tilde z_i) \\
u_i &\approx \epsilon \sqrt{ w'(\tilde z_i) \bar w'(\bar{\tilde{z}}_i ) } \,.  
\end{aligned}
\ee
Our goal is to evaluate the length of the curve along which the defect itself sits, which has  endpoints \eqref{eq:Banadosdefect}. Under the Roberts map this corresponds to the geodesic joining the boundary points $w=0$ and $w=\infty$. Formally, this is obtained as the limiting case of the boundary-to-boundary geodesic \eqref{eq:wu}, whose two branches reduce to $w(u)=0$ and $w(u)=\infty$.

This length diverges both at the AdS boundary and at the defect endpoints. We regulate these divergences by displacing the boundary insertions to
\be \label{eq:regularizedinsertions}
(z, \bar z, y) = (z_1 + \delta_1,\bar z_1 + \bar \delta_1,  \epsilon )\,, \qquad (z, \bar z, y) = (z_2 +  \delta_2,\bar z_2 + \bar \delta_2, \epsilon )\,, 
\ee
with infinitesimal $\epsilon$ and $\delta_i$, $i=1,2$.  
% %
Using the regularized insertions in \eqref{eq:Roberts1} and substituting into \eqref{eq:RobertsAdSshort},  gives the regularized length in the defect geometry 
\be \label{eq:defectreg}
{\mathcal L} = \log \frac{(z_2 - z_1)^{\alpha_n} (\bar z_2 - \bar z_1)^{\alpha_n} }{\alpha_n^2 \epsilon^2 (\delta_1 \bar \delta_1)^{\frac{\alpha_n-1}{2}} (\delta_2 \bar \delta_2)^{\frac{\alpha_n-1}{2}}}\,. 
\ee
Notice that the bulk regulator $\epsilon$ and the twist regulators $\delta_i$ are independent, although their allowed ranges are not. 
Indeed, as one moves into the bulk from the boundary near $z=z_1,z_2$ upon reaching the wall \eqref{eq:detg0} emanating from the insertion points, the Ba\~nados coordinates cease to give a valid parametrization of the geometry. 
Consequently, the regularized endpoints \eqref{eq:regularizedinsertions} must be chosen to lie outside the wall, so that the coordinates $(z,\bar z,y)$ remain well defined along the computation.\footnote{In the limiting case in which the insertions approach the wall, this would set for instance
\be
\delta_1 \bar \delta_1= \delta_2 \bar \delta_2 =  \(1 - \alpha_n^2\) \, \frac{\epsilon^2}{4} \,.  
\ee}
Following \cite{Abajian:2023jye,Abajian:2023bqv}, which performed an equivalent derivation evaluating the total on-shell action of these solutions, we define instead a renormalized geodesic length $\delta\mathcal L$ by discarding the regulator-dependent contributions and retaining only the finite term that depends on the separation of the insertions\footnote{Notice that keeping these terms would only affect the results for the R\'enyi entropy by an additive term which does not depend on the insertions.}
\be \label{eq:LrenCosmic}
\delta {\mathcal L} = \frac 1 n \log (z_2 - z_1) (\bar z_2 - \bar z_1)\,.
\ee
In the language of \cite{Dong:2016fnf,Dong:2018lsk}, this is the (renormalized) area of the cosmic brane which, upon dividing by $4 G_N = 6/c$, evaluates the refined entropy
\be
\tilde S_{A,n} =  \frac{c}{3 n} \log \frac{\ell}{\delta} \,,
\ee
where we set explicitly $z_1 = - z_2 = \ell/2$ and reinstated the twist regulator $\delta$. 
To obtain the R\'enyi entropy we then perform the integration \cite{Dong:2016fnf,Dong:2018lsk}
\be \label{eq:refinedS}
S_{A,n} =\frac{n}{n-1} \int_1^n \frac{\tilde S_{n'}}{n'^2} dn'
\ee
and reproduce the R\'enyi expression \eqref{eq:SAnSpacelike}. 

%%%%%%%%%%%%%%%%%%%%%%%%%%%%%%%%%%%%%%%%%%%%%
\subsubsection{Timelike extension}\label{sec:timelike_Renyi}

Since we kept the insertions generic throughout, it is immediate to extend these results to the timelike case. For a purely timelike separation  $z_1 = - z_2 = i \Delta \tau/2$ we have for instance
\be \label{eq:Ltau}
\delta {\mathcal L} = \frac{1}{n} {\log \Delta \tau^2}\,,
\ee
and continuing this to Lorentzian with the time-ordered prescription and using \eqref{eq:refinedS}, we reproduce the expected CFT result \eqref{eq:TRSvac}.

To link this result to complex geodesics, we work out explicitly the cosmic brane profile. Again, this profile can only be obtained in coordinates that extend beyond the wall, namely the Poincar\'e AdS$_3$ coordinates. We first continue the geometries to Lorentzian signature by $z \to x - t, \bar z \to x+ t$. 
Under this continuation, the boundary coordinates defined by the conformal map \eqref{eq:mapsimpleMain} are in general complex, so we naturally map to a complex Poincar\'e AdS$_3$ metric. As before, we regularize the geodesic endpoints (keeping the notation introduced in Euclidean signature and, for simplicity, choosing equal, real twist regulators, $\delta_1=\delta_2\equiv\delta \ll \Delta t$)
\be
(z,\bar z, y) = \(- \frac{\Delta t}{2} + \delta ,\frac{\Delta t}{2} + \delta, \epsilon \)\,, \qquad
(z,\bar z, y) = \( \frac{\Delta t}{2} + \delta , - \frac{\Delta t}{2} + \delta ,\epsilon \) \,,
\ee
which correspond via \eqref{eq:Roberts1} to the complex AdS$_3$ endpoints 
\be
\begin{aligned} \label{eq:extrema}
(w_i, \bar w_i, u_i) &= \( \frac{1}{\delta^{\alpha_n}},  \frac{1}{\delta^{\alpha_n}} , \frac{1}{\delta^{\alpha_n}} \frac{\alpha_n \epsilon}{\delta }\) \\[.5em]
(w_f, \bar w_f, u_f) &= e^{ - i \pi \alpha_n} \( \frac{ \delta^{\alpha_n}}{\Delta t^{2\alpha_n}}, \frac{\delta^{\alpha_n}}{\Delta t^{2\alpha_n}}, \frac{ \delta^{\alpha_n}}{\Delta t^{2\alpha_n}} \frac{\alpha_n \epsilon}{\delta } \) \,.
\end{aligned}
\ee
The Poincar\'e AdS$_3$ solution is reviewed in appendix \ref{app:Banados} and has the complex profile
\be
\begin{aligned} \label{eq:uTlambda}
w(\lambda) &= \frac{1}{2}\[  w_f + w_i + \frac{\sqrt{(\Delta w^2 + (u_f - u_i)^2 )(\Delta w^2 + (u_f + u_i)^2)}}{\bar w_f -\bar w_i} \tanh \lambda + \frac{u_f^2 - u_i^2}{\bar w_f - \bar w_i} \] \\[.5em]
\bar w(\lambda) &= \frac{1}{2}\[  \bar w_f + \bar w_i + \frac{\sqrt{(\Delta w^2 + (u_f - u_i)^2 )(\Delta w^2 + (u_f + u_i)^2)}}{w_f - w_i} \tanh \lambda  + \frac{u_f^2 - u_i^2}{ w_f - w_i}\] \\[.5em]
u(\lambda) &= \frac{\sqrt{(\Delta w^2 + (u_f - u_i)^2 )(\Delta w^2 + (u_f + u_i)^2)}}{2 \Delta w \cosh \lambda}\ ,
\end{aligned}
\ee
where 
\be
\Delta w  \equiv \sqrt{w_f - w_i} \sqrt{\bar w_f -\bar w_i}\,.
\ee
The corresponding regularized geodesic length in the near-boundary limit \eqref{eq:RobertsAdSshort} here reduces to 
\begin{equation} \label{eq:RlengthFinal}
    \mathcal{L}=\log\frac{\Delta t^{2\alpha_n}}{\alpha_n^2\eps^2\delta^{2\alpha_n-2}} + i \pi \alpha_n \,.
\end{equation}
This is precisely the continuation of \eqref{eq:defectreg} to timelike-separated insertions, upon identifying $z_2-z_1=-\Delta t$ and $\bar z_2-\bar z_1=\Delta t$.

%%%%%%%%%%%%%%%%%%%%%%%%%%%%%%%%%%%%%%%%%%%%%
%%%%%%%%%%%%%%%%%%%%%%%%%%%%%%%%%%%%%%%%%%%%%
\section{Imaginary part of TEE}
\label{sec:Im_TEE}

As manifest in the example of the previous section and pointed out in \cite{Doi:2022iyj}, the TEE is in general complex. From the 2d CFT point of view, the imaginary part of the TEE has a clear origin: it arises from the overall phase in the time-ordered Lorentzian correlator of replica twists. While this takes a constant value in the vacuum state of a CFT on a line, it can in general vary as a function of time and space. In this section, we illustrate the information encoded in the imaginary part by making reference to the simplest non-trivial example, the vacuum state of a CFT on a circle. 

%%%%%%%%%%%%%%%%%%%%%%%%%%%%%%%%%%%%%%%%%%%%%
\subsection{TEE in the vacuum state of a CFT on a circle}
  
We consider the vacuum state of a 2d CFT on a spatial circle of size $2\pi$. In Euclidean signature, the result for the replica twist two-point function is obtained via a conformal map from the plane, $z = e^w=e^{i\phi+\tau}$:
\be 
\langle   \sigma_n(w_1,\bar w_1)   \tilde \sigma_n(w_2 ,\bw_2)     \rangle =   \left[ 2\( - \cosh\(\frac{w_{12} -\bw_{12}}{2}\)  +  \cosh\( \frac{w_{12} +\bw_{12}}{2}\) \) \right]^{-2h_n}\,.
\ee
Taking symmetric insertions $w_1 = i \Delta \phi /2 +\Delta \t /2 $ and $w_2 =  - i \Delta \phi /2 - \Delta \t/2 $, and continuing to Lorentzian time for spacelike separations, $-\Delta t^2 + \Delta \phi^2 >0 $, this becomes
\begin{equation}
\langle  \sigma_n( \Delta t/2, \Delta \phi/2)   \tilde \sigma_n(- \Delta t/2, - \Delta \phi/2)   \rangle  =  \left[ 2( - \cos \Delta \phi +  \cos  \Delta t ) \right]^{-2h_n}\,.  
\end{equation}
At equal times, it yields the familiar spacelike result for the entanglement entropy
\be \label{eq:TS}
S_A= \frac{c}{3}\log\left[  \frac{2}{\delta } \sin \frac{\Delta \phi}{2} \right] \, ,
\ee 
where we reinstated the UV regulator $\delta$ associated with the twist insertions. 

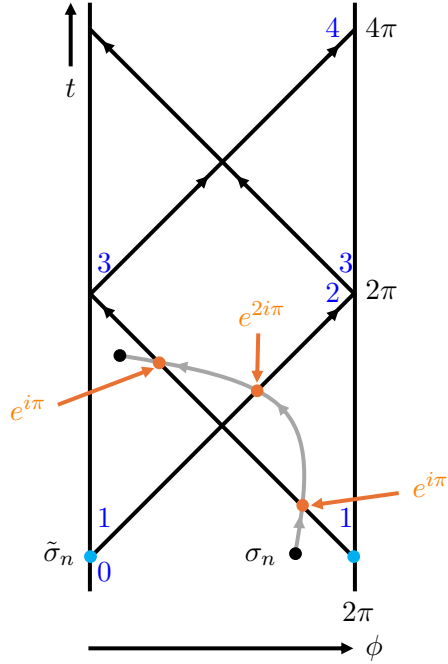
\begin{figure}[t]
\centering
\begin{overpic}[width=.3\linewidth]{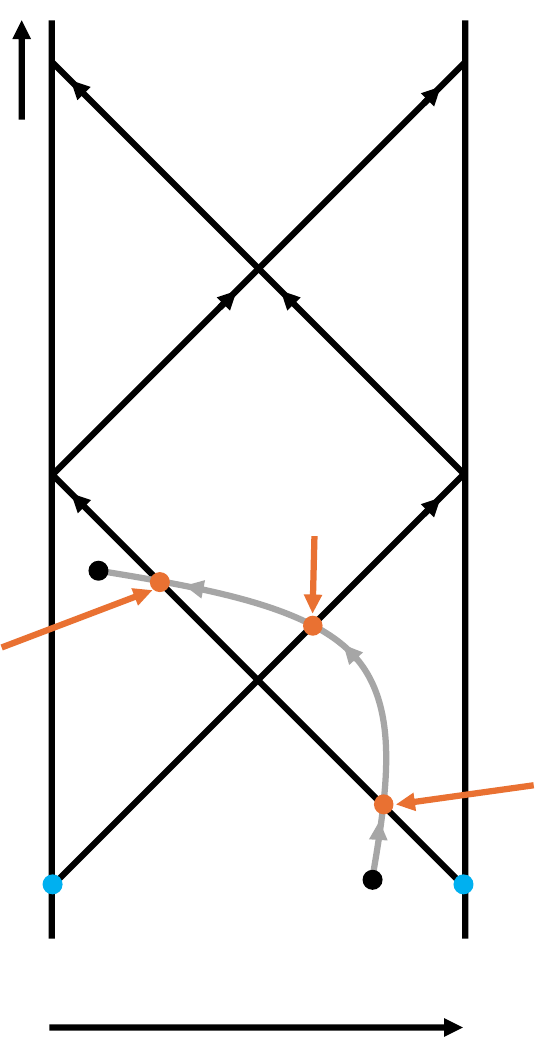}
    \put(-2,15){$\tilde \sigma_n$} 
    \put(1.3,84){$t$} 
    \put(46,1){$\phi$}
    \put(42.5,6){$2\pi$}
    \put(46,54){$2\pi$}
    \put(46,93){$4\pi$}
    \put(53,24.5){\color{orange}{$e^{i\pi}$}}
    \put(27,50){\color{orange}{$e^{2i\pi}$}}
    \put(-7,36){\color{orange}{$e^{i\pi}$}}
    \put(28,15){$\sigma_n$}
    \put(6,12){\color{blue}{$0$}}
    \put(6,20){\color{blue}{$1$}}
    \put(42,20){\color{blue}{$1$}}
    \put(40,53.5){\color{blue}{$2$}}
    \put(42,58){\color{blue}{$3$}}
    \put(6,58){\color{blue}{$3$}}
    \put(40,93){\color{blue}{$4$}}
\end{overpic}\caption{Analytic continuation of the two-point function of twist operators.
Each light-cone crossing changes the phase of the Lorentzian invariant entering the twist correlator  \eqref{eq:torderedcyl} by  $e^{\pm i\pi}$. The sign depends on the direction of the crossing, as represented in the figure for the time-ordered case.} \label{fig:ieps2}
\end{figure}

Continuing to timelike separations in the Lorentzian regime to evaluate $S_A^{(T)}$ as in \eqref{eq:TEEdef} requires time ordering the operators on the cylinder. As for the plane, this can be achieved by performing the analytic continuation from the spacelike to the timelike regime $-\Delta t^2 + \Delta \phi^2 <0 $ with the $i\varepsilon$ prescription  \cite{Kundu:2025jsm}
\be \label{eq:circlecorreps}
\langle  \sigma_n( \Delta t/2, \Delta \phi/2)   \tilde \sigma_n(- \Delta t/2, - \Delta \phi/2) \rangle =     \lim_{\varepsilon \to 0 } \left[2(  - \cos \Delta \phi  + \cos\( \Delta t - i \varepsilon \) ) \right]^{-2h_n}  \, .
\ee
The main difference from the planar case is that the cylinder contains an infinite sequence of causal diamonds, each corresponding to an additional branch cut of \eqref{eq:circlecorreps}. Accordingly, as one separates the twist operators in time, the phase accumulates. For a generic time separation and (anti-)time-ordered operators it is therefore not simply $\pm i\pi$, but $\pm i\pi k$ with $k\in\mathbb{Z}$. This integer labels the number of causal diamonds separating the insertions (see figure \ref{fig:ieps2}), and can be expressed as
\be\label{eq:kglobal}
k = \left\lfloor \frac{\Delta t - \Delta \phi}{2\pi} \right\rfloor + \left\lfloor \frac{\Delta t + \Delta \phi}{2\pi} \right\rfloor + 1 \, ,
\ee
where $\lfloor a \rfloor$ indicates the greatest integer less than or equal to $a$.

Therefore, for $\Delta t>0$ and when the insertions $\sigma_n,\tilde \sigma_n$ are separated by $k$ causal-diamond crossings,
\begin{equation} \label{eq:torderedcyl}
\langle  \sigma_n( \Delta t/2, \Delta \phi/2)   \tilde \sigma_n(- \Delta t/2, - \Delta \phi/2)   \rangle   =  \left[2e^{ i \pi k  } \left|  -\cos \Delta \phi  + \cos \Delta t \right| \right]^{-2h_n}   \,,
\end{equation}
which gives
\be \label{eq:TEEcirclegen}
S_A^{\rm (T)}= \frac{c}{6}\log\left[  \frac{2}{\delta^2 } \left|  -\cos \Delta \phi  + \cos \Delta t \right| \right]  + i \frac{c}{6} \pi k \,. 
\ee 
In particular, when $A$ is purely timelike ($\Delta \phi = 0 $) we have
\be \label{eq:TEEcircle}
S_A^{\rm (T)}= \frac{c}{6}\log\left[\frac{4}{\delta^2 } \sin^2 \frac{\Delta t}{2} \right]  + i \frac{c}{6} \pi k \,. 
\ee 
For $k = 1$ this directly coincides with the result obtained in \cite{Doi:2023zaf} by continuing the expression for the entanglement entropy. In the procedure outlined here, one performs the continuation to the Lorentzian regime of the replica twist operators and then evaluates the TEE. This provides a precise 2d CFT dual to the rotation procedure of bulk extremal surfaces described in \cite{Heller:2025kvp} and extends it. In particular, it allows one to overcome the limitation in \cite{Doi:2023zaf,Heller:2025kvp}, imposed by the fact that the spatial size of an interval is limited by the size of the system, which confines the analysis to time separations contained within a single causal diamond. In fact, here there is no obstruction to considering arbitrary separations, and the ambiguity in identifying the imaginary part of the TEE is resolved.

%%%%%%%%%%%%%%%%%%%%%%%%%%%%%%%%%%%%%%%%%%%%% 
\subsection{Holographic TEE in global AdS$_3$}

We now show that the field theory result \eqref{eq:TEEcirclegen} is reproduced naturally by a bulk complex geodesic computation. The geometry dual to the vacuum state of the 2d CFT on the Lorentzian cylinder is empty AdS$_3$ in global coordinates
\be\label{eq:globalAdSmetric}
ds^2 = -\(1+r^2 \)dt^2 + \frac{dr^2}{1+r^2} + r^2 d\phi^2\, ,
\ee
with the asymptotic boundary at $r\to\infty$.

Consider the near-boundary regularized points corresponding to the twist operator insertions in the dual CFT,
\be
(t_1,\phi_1,r_1) = \left( \frac{\Delta t}{2}, \frac{\Delta \phi}{2}, \frac{1}{\epsilon}\right)\,,\qquad
(t_2,\phi_2,r_2) = \left( -\frac{\Delta t}{2}, -\frac{\Delta \phi}{2}, \frac{1}{\epsilon}\right).
\ee
In the $\eps\ll1$ limit, the general profile of the spacelike geodesic connecting these two points can be written as (see appendix \ref{app:AAdS} for details)
\begin{equation}
\begin{aligned}
\tan t(\lambda) &= \tan\!\left(\frac{\Delta t}{2}\right)\tanh\lambda\\[.5em]
\tan \phi(\lambda) &= \tan\!\left(\frac{\Delta \phi}{2}\right)\tanh\lambda\\[.5em]
r(\lambda) &= \sqrt{\frac{\cos \Delta \phi+\cosh2\lambda}
{\cos \Delta t-\cos \Delta \phi}}\, ,
\end{aligned}
\end{equation}
with the parameter $\lambda$ varying between the extrema $\pm\lambda_*$, where
\be
\lambda_*=\frac{1}{2}\,\log\!\left[ 2\frac{ \cos\Delta t-\cos\Delta\phi}{\epsilon^2}\right]\,.
\ee
For spacelike separations, this straightforwardly reproduces the usual geodesic length and, via the Ryu-Takayanagi formula, the entanglement entropy \eqref{eq:TS}. In the timelike regime, the appropriate branch of the logarithm must be selected. We therefore apply the same regularization procedure as for the dual correlator \eqref{eq:circlecorreps}, namely $\Delta t\rightarrow\Delta t-i\varepsilon$.

Demanding that one recovers the spacelike boundary interval result as $\Delta t\rightarrow0$, and following the evolution of the argument of the logarithm in the complex plane as $\Delta t$ increases, yields precisely the same phase structure as for the dual correlator,
\be \label{lstargl}
\lambda_*= \frac{1}{2}\,\log\!\left[ 2\frac{ | \cos\Delta t-\cos\Delta\phi |}{\epsilon^2}\right]+i\frac{\pi}{2}k\,,
\ee
where $k$ is again given by \eqref{eq:kglobal}.

As in the planar case, timelike separations on the boundary of global AdS require the geodesic parametrization to become complex, so that geodesics with real boundary endpoints propagate through a complexified geometry. Compared with the planar case, the novelty here is that large time separations require a parametrization that extends by integer multiples of $\pi$ in the imaginary direction.

Using \eqref{lstargl} for the holographic evaluation of the TEE, together with the straightforward identification of the bulk and CFT cutoffs $\eps=\delta$,
\be
S_A^{(T)}=\frac{\mathcal L}{4G_N}=\frac{\lambda_*}{2G_N}\,,
\ee
one immediately reproduces the field theory prediction \eqref{eq:TEEcirclegen}. On both sides, the imaginary part is determined through the integer $k$, which encodes the relative causal separation of the endpoints. A detailed geometrical picture, illustrating the complex geodesics and comparing them with the piecewise curves of \cite{Doi:2022iyj,Doi:2023zaf}, is provided in appendix \ref{app:AAdS}.

%%%%%%%%%%%%%%%%%%%%%%%%%%%%%%%%%%%%%%%%%%%%%
%%%%%%%%%%%%%%%%%%%%%%%%%%%%%%%%%%%%%%%%%%%%%
\section{Locally excited states} \label{sec.locallyexcited}

In the cases discussed so far, the preparation of the CFT state in \eqref{eq:replicacorr} can be represented in terms of a path integral without any additional operator insertion. Here we are going to consider a broader class of states that can be prepared via the insertion of a heavy local primary operator, following \cite{Asplund:2014coa,Fitzpatrick:2014vua}. These are states that, upon conformal maps to the complex plane, can be expressed as 
\be
| \psi \rangle = \psi(z, \bar z)|0 \rangle\, , 
\ee
where $\psi$ is a heavy primary operator of the 2d CFT in the complex plane. 
Its dimensions are such that $h_\psi/c$ and $\bar h_\psi/c$ remain finite in the large $c$ limit.

For this class of states, the evaluation of the entanglement and R\'enyi entropies for a single interval can be conformally related to the evaluation of the Euclidean four-point function of two twist operators and two primary operators $\psi$ with canonical insertions \cite{Asplund:2014coa}
\be \label{eq:4pt=ID}
 G_n(z, \bar z) =  \langle \psi (\infty)  \sigma_n(z,\bz) \tilde \sigma_n(1)  \psi (0)\rangle \,. 
 \ee
Notice that, as mentioned around \eqref{eq:replicacorr}, the replica computation is actually defined in the cyclic orbifold CFT$^n/\mathbb Z_n$ and the state $|\Psi\rangle$ in  \eqref{eq:replicacorr} is the state in the orbifold theory obtained by inserting  $\psi$ in all $n$ copies.  That is, in \eqref{eq:4pt=ID} one should actually have  $\Psi = \psi_1\psi_2\dots\psi_n$ where the subscripts indicate different copies of the CFT. However, in the following we will confine our analysis to the limit where the correlator evaluates the (timelike) entanglement entropy. With a slight abuse of notation, we thus consider from the beginning the action of a single $\psi$ to define the state, as this does not change the final result (see, \eg \cite{Asplund:2014coa,Bernamonti:2024fgx} for a more detailed explanation).

In holographic CFTs, the sparseness of the spectrum of low conformal dimension operators and the exponentiation of Virasoro conformal blocks at large $c$ indicate that $ G_n(z, \bar z) $ is very well approximated by the identity Virasoro block (see, \eg \cite{Hartman:2013mia,Fitzpatrick:2014vua,Asplund:2014coa,Caputa:2014eta,Fitzpatrick:2015zha,Anous:2016kss}). Working under this assumption, we will therefore evaluate the correlator as 
\be  \label{eq:canon}
 G_n(z, \bar z) =  \langle \psi (\infty)  \sigma_n(z,\bz) \tilde \sigma_n(1)  \psi (0)\rangle  \approx   \Fcal_0(z)  {\bar\Fcal}_0(\bz)\,, 
\ee
where $\Fcal_0(z)  {\bar\Fcal}_0(\bz)$ indicate the holomorphic and antiholomorphic identity Virasoro blocks. 
In the limit where the correlator computes the entanglement entropy, $G_n(z, \bar z)$ is a heavy-heavy-light-light (HHLL) four-point function and the expression for $\Fcal_0(z)  {\bar\Fcal}_0(\bz)$ is known in closed form  and reads  \cite{Fitzpatrick:2014vua,Fitzpatrick:2015zha}
\be \label{eq:idb}
\log G_n(z, \bar z)     = \frac c 6 (1-n)\log \left[  \frac{ z^{\frac{1}{2}} \bz^{\frac{1}{2}}   ( z^{\alpha/2}-z^{-\alpha/2})( \bz^{\bar\alpha/2}-\bz^{-\bar\alpha/2}) }{\alpha \bar \alpha}  \right] + O\!\left[(n-1)^2\right], 
\ee
where 
\be  \label{alphadef}
\alpha \equiv \sqrt{1- \frac{24 h_\psi}{c} }\,, 
\ee
and $\bar \a$ is defined similarly in terms of  $\bar h_\psi$. Henceforth we restrict to scalar $\psi$ and set $\bar h_\psi=h_\psi$.

The four-point function evaluated in this approximation is not single-valued in the Euclidean plane. It exhibits non-trivial monodromies when one takes the twist operators around the heavy insertions $\psi$ preparing the state. 
For a given $z,\bar z$ in the complex plane, each $G_n(  e^{-2\pi i m } z,  e^{2\pi i m } \bar z)\approx \Fcal_0( e^{-2\pi i m } z )  {\bar\Fcal}_0( e^{2\pi i m }\bar z)$ for $m \in \mathbb{Z}$ defines a different Virasoro identity channel. The general prescription for evaluating the correlator in the identity-block approximation is then to maximize over all inequivalent identity channels \cite{Hartman:2013mia,Asplund:2014coa,Anous:2016kss}.

It is important to distinguish this choice of identity channel from the phases arising in the continuation of the four-point function to timelike separations in Lorentzian signature. While both may be viewed as choices of sheet of a multivalued function, the former is made entirely in the Euclidean regime and entails a correlated choice of monodromy for $z$ and $\bar z$. The latter is instead dictated by the operator ordering and the crossing of lightcones, and affects the holomorphic and antiholomorphic sectors independently, as $z$ and $\bar z$ evolve onto different sheets of the complex plane during the Lorentzian continuation.

%%%%%%%%%%%%%%%%%%%%%%%%%%%%%%%%%%%% 
\subsection{High-energy eigenstates}
\label{sec:high_energy_eigenstates}

The simplest set of states within this family are time-independent high-energy eigenstates of the Hamiltonian of the 2d CFT on a spatial circle, which admit a holographic dual description in terms of conical defect and BTZ geometries in AdS$_3$. 

These states are prepared on the Euclidean cylinder by radial quantization, through the insertion of a heavy primary $\psi$ at the origin of the complex plane  $z= e^w$
\be \label{eq:stateprep}
| \psi \rangle = \psi(0)|0 \rangle\,,    \qquad \quad \langle \psi | = \lim_{z \to \infty } z^{2h_\psi}\bar z^{2 h_\psi}\langle 0 | \psi(z, \bar z) \, .
\ee

The entanglement entropy of an interval $A$ in the normalized state on the spatial cylinder is computed as the two-point function of twist operators  
\be
\Tr \rho_A^n  
= \frac{\langle \psi(w_1, \bw_1 )  \sigma_n( w_2, \bw_2 ) \tilde \sigma_n (w_3,\bw_3) \psi(w_4, \bw_4 ) \rangle }{\langle \psi(w_1, \bw_1 ) \psi(w_4, \bw_4 ) \rangle}
= \(z \bz \)^{ h_n}  G_n(z, \bar z)\,,
\label{eq:4ptHHLL}
\ee
with conformal ratios
\be
z\equiv\frac{z_{12}z_{34}}{z_{13}z_{24}}= e^{ i \Delta \phi + \Delta \tau}\,, \qquad
\bz \equiv\frac{\bar z_{12}\bar z_{34}}{\bar z_{13}\bar z_{24}} =  e^{  -i \Delta \phi + \Delta \tau}\,,
\ee
for spatially symmetric Euclidean insertions $w_2= - w_3= \frac{1}{2} \( i \Delta \phi + \Delta \tau\)$.

By using \eqref{eq:idb} and considering all possible monodromies $G_n( e^{-2\pi i m } z, e^{2\pi i m }  \bar z )$, equation \eqref{eq:4ptHHLL}  yields for the entanglement entropy 
\be
\begin{aligned}\label{eq:SlocalqSL}
S_A &=\min_{m \in \mathbb{Z}}\frac {c}{6} \log \left[ \frac{4}{\alpha^2 \delta^2}    \sinh\(\frac{\alpha}{2} (\log z - 2 \pi i m) \)\sinh\(\frac{\alpha}{2} ( \log \bz +  2 \pi i m ) \)  \right] .
\end{aligned}
\ee
In writing this expression, we reinstated the UV regulator $\delta$ associated with the twist operator insertions.

To proceed further, it is useful to distinguish two ranges of conformal dimensions. States created by heavy operators with $0 < \frac{h_\psi}{c} < \frac{1}{24}$ have real $0<\alpha <1$. These states are dual to the conical defect AdS$_3$ geometries with global AdS$_3$ as the limiting case $h_\psi=0$.

For spacelike-separated twist operators,  we then continue \eqref{eq:SlocalqSL} to Lorentzian time $\Delta \tau \to i \Delta t$ in the spacelike regime, and \eqref{eq:SlocalqSL} gives
\be
S_A=\min_{m \in \mathbb{Z}}\frac {c}{6} \log \left[ \frac{2}{\alpha^2 \delta^2}   \big(\cos\( \alpha 
\Delta t \) - \cos\( \alpha (\Delta \phi - 2 \pi m) \) \big) \right]\, . \label{eq:EEam}
\ee
States prepared with heavier operators $\frac{h_\psi}{c} > \frac{1}{24}$ have instead purely imaginary $\alpha$, and it is more convenient to define
\be \label{eq:betadef}
\beta \equiv \frac{2\pi}{\sqrt{24 h_\psi/c-1 }}\,. 
\ee
Here $\beta>0$ plays the role of a dimensionless inverse temperature. Holographically, these states can indeed be thought of as BTZ microstates with energy fixed by the inverse effective temperature $\beta$ \cite{Asplund:2014coa}. The entanglement entropy in this case reads
\be
\begin{aligned}
S_A  =\min_{m \in \mathbb{Z}} \frac {c}{6} \log \left\{  \frac{\beta^2}{ 2\pi^2 \delta^2 }  \left[  \cosh\( \frac{2\pi}{\beta} (\Delta \phi - 2 \pi m) \)  - \cosh\( \frac{2\pi}{\beta}  \ \Delta t \)  \right] \right\}\,,  \label{eq:EEbm}
\end{aligned} 
\ee
where each channel has the thermal line form at effective inverse temperature $\beta$ \cite{Holzhey:1994we,Calabrese:2004eu}. 

To obtain the TEE defined in \eqref{eq:TEEdef}, we analytically continue the four-point correlator \eqref{eq:4ptHHLL} to the time-ordered two-point function of replica twists in the state $|\psi \rangle$. Namely, assuming without loss of generality $\Delta t >0$,
\be
 \langle T\{ \sigma_n (t)   \tilde \sigma_n(0) \} \rangle_\rho  
= \langle \psi(w_1, \bw_1 )  \sigma_n( \Delta t/2, \Delta \phi/2) \tilde \sigma_n ( -\Delta t/2,-\Delta \phi/2)\psi(w_4, \bw_4 ) \rangle\,.
\ee
The prescription in this case simply amounts to imposing the correct ordering between the twist operators with the regularization $\Delta t\to \Delta t - i\varepsilon$.

For CFT states with  $0 < \frac{h_\psi}{c} < \frac{1}{24}$ the picture that emerges is analogous to the one for the vacuum state on the cylinder, which is recovered for $\alpha =1$.
The difference is that the causal diamond structure on the cylinder is now described in terms of $\alpha$-rescaled coordinates.
The general expression for the TEE can be written as
\be \label{eq:TEECD}
S_A^{\rm (T)}= \min_{m \in \mathbb{Z}} \frac{c}{6} \log \left[    \frac{2}{\alpha^2 \delta^2}  \Bigg|\cos\( \alpha 
\Delta t \) -  \cos\( \alpha  \( \Delta \phi  - 2\pi m\)\)  \Bigg| \right] + i \frac{c}{6} \pi k_{\alpha,m} \,  
\ee 
where 
\be
k_{\alpha,m} =
\left\lfloor \frac{\alpha \(\Delta t-  \Delta \phi  + 2\pi m\) }{2\pi} \right\rfloor + \left\lfloor \frac{\alpha(\Delta t+  \Delta \phi  - 2\pi m )}{2\pi} \right\rfloor + 1 \,  . \label{eq:kalpham}
\ee
The minimization over the competing $m$-channel contributions is understood to act on the real part of the entropy, thereby selecting the corresponding imaginary part.

For states with  $ \frac{h_\psi}{c} > \frac{1}{24}$, on the other hand,
\begin{align}
S_A^{\rm (T)} &=\min_{m \in \mathbb{Z}} \frac {c}{6} \log \left[  \frac{\beta^2}{ 2\pi^2 \delta^2 }  \Big|  \cosh\( \frac{2\pi}{\beta} (\Delta \phi - 2 \pi m) \)  - \cosh\( \frac{2\pi}{\beta}  \Delta t \)  \Big| \right] \nonumber \\[.5em]
&\hspace{.5cm}+ i \frac{c}{6} \pi \,\Theta\!\left[ \Delta t^2- (\Delta \phi- 2\pi m)^2 \right]  \, . 
\label{eq:TEEbeta}
\end{align}
Again, the minimization is understood to be performed on the real part.
Unlike the previous parameter regime, and consistently with the observation that the ordinary entanglement entropy matches the expression for a finite-temperature 2d CFT on a line, the complex phases arising in the Lorentzian regime are compatible with the causal structure of Minkowski space.

In both \eqref{eq:TEECD} and \eqref{eq:TEEbeta}, if we  take purely timelike separations between the endpoints of the interval ($\Delta \phi =0$), we notice that for each pair $\pm |m|$ the entropy has the same real part. One may therefore wonder whether these contributions are to be summed, as advocated in a different boundary CFT setup in \cite{Kanda:2026jyk}.  
While all channels are expected to contribute to the full correlator, within the identity-block approximation they should instead be regarded as competing saddle contributions, each dominating in a different region of kinematic parameter space. 
This is precisely what happens in the Euclidean regime, where channels with the same $|m|$ are inequivalent. For purely timelike separations in Lorentzian signature, the pair of channels $\pm |m|$ becomes degenerate. However, continuing from a small but non-vanishing spatial separation, we again expect a unique dominant channel.

%%%%%%%%%%%%%%%%%%%%%%%%%%%%%%%%%%%%%%%%%%%%%%%%%%%%%%
\subsection{Conical defects and BTZ microstates}

The gravity dual is described by the asymptotically AdS$_3$ geometry
\begin{equation} \label{eq:alphametric}
    ds^2= -\( \alpha^2 + r^2 \)\,dt^2+\frac{dr^2}{\alpha^2 + r^2}+r^2d\phi^2\,.
\end{equation}
For $\alpha=1$, this reduces to global AdS$_3$ \eqref{eq:globalAdSmetric}. For $0<\alpha<1$, it describes a conical defect dual to the heavy primary state in the boundary theory. When $\alpha$ is purely imaginary, the geometry becomes the BTZ black hole, dual to a high-energy microstate with effective inverse temperature $\beta=2\pi/|\alpha|$.

We consider spacelike geodesics connecting the regulated boundary points
\begin{equation}
    (t_1,\phi_1,r_1)=\left(\frac{\Delta t}{2},\frac{\Delta \phi}{2},\frac{1}{\epsilon}\right),\qquad
    (t_2,\phi_2,r_2)=\left(-\frac{\Delta t}{2},-\frac{\Delta \phi}{2},\frac{1}{\epsilon}\right).
\end{equation}
Because the angular direction is periodically identified, there exists an infinite family of geodesics connecting the same boundary endpoints, distinguished by their winding number $m$. This is implemented by considering boundary angular separations of the form $\Delta\phi-2\pi m$, with $m\in\mathbb Z$. The spacelike geodesic for a given winding number $m$ is then (see appendix \ref{app:AAdS} for details)
\be  \label{eq:geodCDBTZ}
\begin{aligned}
t(\lambda) &=\frac{1}{\alpha }\arctan\[ \tan\!\left(\frac{\alpha\Delta t}{2 }\right)\tanh\lambda\]\\[.5em]
\phi(\lambda) &=\frac{1}{\alpha } \arctan\[ \tan\!\left(\frac{\alpha\,(\Delta\phi-2\pi m)}{2}\right)\tanh\lambda \] \\[.5em]
r(\lambda) &=\alpha  \sqrt{\frac{\cos (\alpha\,(\Delta\phi-2\pi m)) + \cosh (2\lambda)}
{\cos (\alpha\Delta  t)  - \cos (\alpha\,(\Delta\phi-2\pi m)) }}\,,
\end{aligned}
\ee
with length
\begin{equation}
\mathcal{L}=2\lambda_* 
=\log\left[\frac{2}{\alpha^2\epsilon^2}\left(\cos(\alpha\Delta t)-\cos(\alpha(\Delta\phi-2\pi m))\right)\right],
\end{equation}
where $\pm\lambda_*$ denote the values of the affine parameter at the two regulated endpoints.

The holographic entanglement entropy is therefore
\begin{equation}
S_A
=\min_{m\in\mathbb Z}
\frac{c}{6}\log\left[\frac{2}{\alpha^2\epsilon^2}\left(\cos(\alpha\Delta t)-\cos(\alpha(\Delta\phi-2\pi m))
\right)\right].
\end{equation}
For $0<\alpha<1$, this reproduces the standard result for conical defects and agrees with the CFT expression \eqref{eq:EEam} upon identifying $\epsilon=\delta$. For imaginary $\alpha$, it likewise reproduces the entanglement entropy of the BTZ microstate satisfying \eqref{eq:betadef}, namely \eqref{eq:EEbm}. Since we are describing a pure microstate rather than a thermal ensemble, we do not impose the usual homology condition. Accordingly, the disconnected Ryu-Takayanagi configuration involving a geodesic wrapping the horizon does not contribute~\cite{Asplund:2014coa,Asplund:2016koz}.

To access the timelike regime, we implement the standard $i\varepsilon$ prescription dictated by the time ordering of the boundary correlator. This does not modify the bulk geodesic profile \eqref{eq:geodCDBTZ}, for which $\varepsilon$ may be set to zero from the outset, but it affects the endpoint value of the  parameter 
\be\label{eq:ancontaads}
\lambda_*=\lim_{\vareps\to 0}\frac{1}{2}\log \left[\frac{2}{\alpha^2\epsilon^2}\left(\cos(\alpha(\Delta t-i\varepsilon) ) -\cos(\alpha(\Delta\phi-2\pi m))\right)\right].
\ee

We first consider the conical defect geometry, $0<\alpha<1$. After analytic continuation, the affine parameter becomes generically complex,
\be
\lambda_*=\frac{1}{2}
\log\!\left[\frac{2}{\alpha^2\epsilon^2}\left|\cos(\alpha\Delta t)-\cos(\alpha(\Delta\phi-2\pi m))\right|\right]+i\frac{\pi}{2}k_{\alpha,m},
\ee
where $k_{\alpha,m}$ is given by the same expression \eqref{eq:kalpham} above. The holographic TEE therefore becomes
\be
S_A^{\mathrm{(T)}} = \min_{m\in\mathbb Z} \frac{c}{6} \log\!\left[ \frac{2}{\alpha^2\epsilon^2} \left| \cos(\alpha\Delta t) - \cos(\alpha(\Delta\phi-2\pi m)) \right| \right] + i\frac{c}{6}\pi k_{\alpha,m}\,.
\ee

For BTZ microstates, by contrast, the analytic continuation acts on hyperbolic rather than periodic functions. As a consequence, the imaginary part depends only on whether the separation is timelike, namely $\Delta t^2-(\Delta\phi-2\pi m)^2>0$:
\begin{align}
S_A^{\mathrm{(T)}} &= \min_{m\in\mathbb Z}
\frac{c}{6} \log\left[ \frac{\beta^2}{2\pi^2\epsilon^2}
\Big| \cosh\!\left(\frac{2\pi}{\beta}(\Delta\phi-2\pi m)\right) - \cosh\!\left(\frac{2\pi}{\beta}\Delta t\right) \Big| \right] \nonumber\\[.5em]
&\hspace{.5cm} +i\frac{c}{6}\pi\, \Theta\!\left[\Delta t^2-(\Delta\phi-2\pi m)^2\right].
\end{align}
These expressions exactly match the CFT results \eqref{eq:TEECD} and \eqref{eq:TEEbeta} after the standard identification of the bulk and boundary UV cutoffs. The CFT channels labeled by $m$ are naturally identified with bulk geodesics of different winding number. As in the CFT analysis, for purely timelike separations the pair of windings $\pm|m|$ becomes degenerate.

%%%%%%%%%%%%%%%%%%%%%%%%%%%%%%%%%%%%%%%%%%%%%
\subsection{Local operator quench}\label{sec:localquench}

We now turn to a time-dependent extension of the previous construction and consider a local operator quench, obtained by locally exciting the ground state of the CFT through the insertion of a primary operator at the origin of the real line. The corresponding (spacelike) entanglement entropy was studied in \cite{Nozaki:2014hna,He:2014mwa,Nozaki:2014uaa} for rational CFTs and in \cite{Caputa:2014vaa,Asplund:2014coa} for holographic large-$c$ CFTs. The analysis was recently extended to TEE for rational CFTs in \cite{Guo:2025ase}.

The post-quench state on the real line is modeled by the insertion of a primary local operator $\psi$ with weight $h_\psi < c/24$ at the origin of the complex plane. To regularize the insertions and produce a normalizable state, it is necessary to introduce a small offset in the imaginary time direction.  
The density matrix describing the state at a time $t$ after the quench can then be written as 
\be
\rho(t) = {\cal{N}} e^{-iHt}\psi(z_1, \bz_1) |0 \rangle \langle 0 | \psi(z_4, \bz_4)e^{iHt}\,,
\ee
where the normalization factor $\cal{N}$ is fixed by $\tr \rho(t) = 1$, and
\be
\begin{aligned}
z_1 &=  i\mu \ , \quad && \bz_1=  -i\mu\ , \\ 
z_4 &=  -i\mu  \ , \quad && \bz_4 =  i\mu \, .  \\
\end{aligned}
\ee
The quench is local in the limit $\mu \to 0$, and we will work in the scaling limit where all length and time scales are much larger than $\mu$, unless explicitly stated otherwise. 

The entanglement entropy for a single spacelike interval $A$ in this time-dependent state at a time $t$ is evaluated via the two-point correlator of twist operators \eqref{eq:replicacorr}, which for the normalized state here amounts to the four-point function
\be \label{eq:4pt}
\begin{aligned}
\Tr \rho_A^n(t)  &=  \frac{\langle \psi(z_1, \bz_1) \sigma_n(z_2, \bz_2) \tsigma_{n} (z_3, \bz_3) \psi(z_4, \bz_4) \rangle}{\langle \psi(z_1, \bz_1)  \psi(z_4, \bz_4)  \rangle}\\
\end{aligned}
\ee
with
\be
\begin{aligned}
z_2 &= x_2 + i \tau_2 - t\,,  \quad &&\bz_2 =  x_2 - i \tau_2 + t \,,\\
z_3 &=  x_3 + i \tau_3 - t\,, \quad &&\bz_3 = x_3 - i \tau_3 + t \,,
\end{aligned}
\ee
for generic Euclidean endpoints $(x_i,\tau_i)$ of the interval $A$. 

Notice that the correlator \eqref{eq:4pt} is already Lorentzian. To evaluate the spacelike entanglement entropy one needs to make sure it gives the equal time two-point function of twist operators in the state prepared by $\psi$. The Euclidean preparation of the excited state actually automatically produces the desired ordering throughout the evolution (see appendix \ref{app:TO}). 
The computation of the spacelike entanglement entropy was carried out in \cite{Asplund:2014coa} (see also \cite{Hartman:2015lfa}), and we review it here.

Since \eqref{eq:4pt} is conformally related to the canonical four-point function  $G_{n}(z, \bz)$ as 
\be \label{eq:rhotq}
\Tr \rho_A^n(t)  = (z_{23} \bz_{23})^{-2 h_n}[(1-z)(1-\bz)]^{2 h_n}G_n(z, \bz) \, ,
\ee
where the cross ratios read
\be\label{eq:cr}
z =  \frac{z_{12} z_{34}}{z_{13} z_{24}}\,, 
\qquad \bz =  \frac{\bz_{12} \bz_{34}}{\bz_{13} \bz_{24}} \,,
 \ee
and using \eqref{eq:rhotq} with \eqref{eq:idb}, we can write
\be\label{smaster}
S_A(t) =\lim_{n\to 1}\frac{\log \Tr \rho_A^n(t)}{1-n} =  \frac{c}{6}\log\left[ \frac{z_{23} \bar z_{23} (z \bz)^{ \frac{1}{2}(1-\alpha)}(1-z^{\alpha})(1-\bz^{\alpha})}{\delta^2 \alpha^2(1-z)(1-\bz)} \right],
\ee
where we have reinserted the regulator $\delta$ associated to the twist operators. 

For an equal time interval extending from $x_0$ to $x_0 + \Delta x$,
\be
\begin{aligned}
z_2 &= x_0  - t\,,  \quad &&\bz_2 =  x_0   + t \ ,\\
z_3 &= x_0 + \Delta x   - t \,, \quad &&\bz_3 = x_0   + \Delta x  + t \,,
\end{aligned}
\ee
the cross ratios read
\be
\begin{aligned} \label{eq:zcquench}
z &=  \frac{(t - x_0 + i\mu)(t - x_0 - i\mu - \Delta x)}{(t - x_0 - i\mu)(t - x_0 + i\mu - \Delta x)}
  \simeq 1 - \frac{2 i \,\Delta x \,\mu}{(t - x_0)(t - x_0 -\Delta x)} + O(\mu^2)\,, \\[.5em]
 \bz &=    \frac{(t + x_0 + i\mu)(t + x_0 - i\mu + \Delta x)}{(t + x_0 - i\mu)(t + x_0 + i\mu + \Delta x)}   
 \simeq 1 + \frac{2 i \,\Delta x \,\mu}{(t + x_0)(t + x_0 + \Delta x)} + O(\mu^2) \, .\\
\end{aligned}
 \ee
Both cross ratios lie on the unit circle, $|z|=|\bar z|=1$, and $z \neq \bz$ since they are Lorentzian. 

To determine the time evolution of the entanglement entropy one must follow the trajectories of $z$ and $\bar z$ in the complex plane and keep track of the branch cuts crossed by the Lorentzian correlator $G_n(z,\bar z)$ \cite{Asplund:2014coa,Hartman:2015lfa}. These crossings are completely determined by the causal structure of the four-point function together with the operator ordering.
In fact,  $\psi(z_1, \bz_1)$ and  $ \psi(z_4, \bz_4)$ are respectively time- and anti-time-ordered with respect to both $\sigma_{n} (z_2, \bz_2)$ and $\tsigma_{n} (z_3, \bz_3)$. When $\sigma_{n} (z_2, \bz_2)$ crosses the right-moving lightcone of $ \psi(z_1, \bz_1)$ for $z_{12}=0$,  there is a phase $e^{- \pi i}$ associated with the $z_{12}$ factor in $z$. Similarly, when it crosses the right-moving lightcone of $ \psi(z_4, \bz_4)$ for $z_{24}=0$, there is a phase $e^{ \pi i }$ associated with  $z_{24}$. 
Combining these two contributions, the holomorphic cross ratio changes sheet according to $z \to e^{-2\pi i} z$ when $\sigma_{n}$ crosses the quench lightcone of $\psi \psi$. The same reasoning applies to $\tilde\sigma_n$, except that the phases enter with the opposite sign in the definition of $z$: $z_{13}$ and $z_{34}$ pick the same phases as $z_{12}$ and $z_{24}$, but contribute in the opposite way for $z$, as can be seen explicitly from \eqref{eq:zcquench}.\footnote{The same happens for $\bz$ when crossing the left-moving lightcones giving $\bz \to e^{-2\pi i} \bz$ for $\sigma_{n} (z_2, \bz_2)$ and $\bz \to e^{2\pi i} \bz$ for  $\tsigma_{n} (z_3, \bz_3)$.}

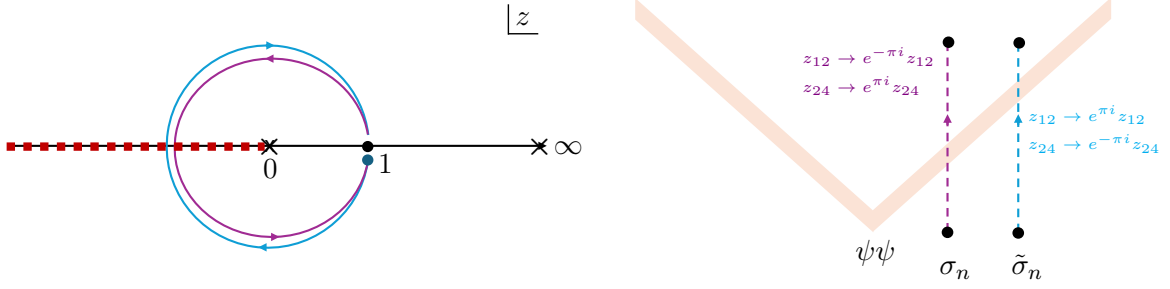
\begin{figure}[t] 
\centering
\begin{overpic}[width=.48\textwidth]{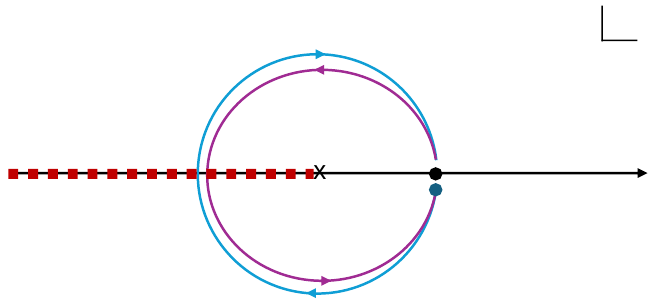}
    \put(48,14){0}
     \put(46.8,18.5){$\boldsymbol{\times}$}
    \put(69,15){1}
    \put(94,42){$z$}
    \put(101,18.5){$\infty$}
    \put(96,18.5){$\boldsymbol{\times}$} 
\end{overpic}
\hspace{.05\textwidth}
\begin{overpic}[width=.42\textwidth]{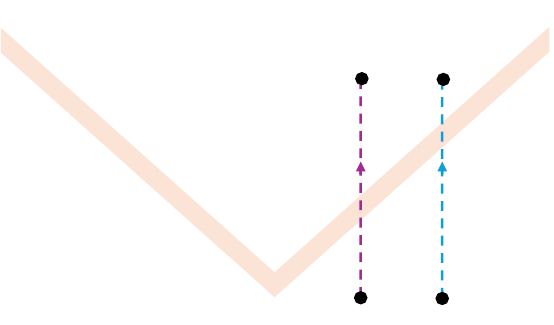}
    \put(46,-1){$\psi \psi$}
    \put(63.5,-4){$\sigma_n$}
    \put(78.5,-4){$\tilde\sigma_n$}
    \put(35,40){\textcolor{violet}{\tiny  $z_{12}\to e^{-\pi i}z_{12}$}}
    \put(35,34){\textcolor{violet}{\tiny  $z_{24}\to e^{\pi i}z_{24}$}}
    \put(82,28){\textcolor{cyan}{\tiny  $z_{12}\to e^{\pi i}z_{12}$}}
    \put(82,22){\textcolor{cyan}{\tiny  $z_{24}\to e^{-\pi i}z_{24}$}}
\end{overpic}
\vspace{.5cm}
  \caption{Schematic evolution of the holomorphic cross ratio $z$ in the complex plane for the case $x_0>0$. At $t\approx x_0$, $\sigma_n$ crosses the lightcones of the two $\psi$ insertions, causing $z$ to wind clockwise around the origin and move onto the adjacent sheet according to the monodromy $z\to e^{-2\pi i}z$. At $t\approx x_0+\Delta x$, $\tilde\sigma_n$ crosses the same lightcones, and $z$ retraces its path back to the principal sheet. The right panel illustrates the phases acquired by the coordinate differences entering the definition of the cross ratio.} 
 \label{fig:zbar}
\end{figure}

Let us first consider the case $x_0>0$, so that the operator insertion lies outside the interval $A$.
The antiholomorphic ratio $\bar z$ approaches $1$ at all times, while  $z$  has a non-trivial evolution. At early times, in the regime $0\leq t< x_0$, $z$ approaches 1 from below in the complex plane. Evaluating \eqref{smaster} on the principal branch of both $z$ and $\bz$ then yields
\be\label{eq:soutt0}
S_A =  \frac{c}{3}\log\left[ \frac{\Delta x }{\delta}\right] \,.
\ee
At $t\approx x_0$, $\sigma_n$ crosses the lightcones of the two $\psi$ insertions. Correspondingly, the holomorphic cross ratio $z$ winds clockwise around the origin, crosses the relevant branch cut of $G_n(z,\bar z)$, and moves onto the adjacent sheet according to the monodromy $z\to e^{-2\pi i}z$. It remains on this sheet throughout the interval $x_0<t<x_0+\Delta x$. 
Evaluating \eqref{smaster}  with $z \to z e^{-2\pi i}$ yields
\be\label{eq:branch}
S_A = \frac{c}{6}\log\left[\frac{ \Delta x (t-x_0)(x_0+\Delta x-t)}{\delta^2\mu }\frac{\sin (\pi \alpha)}{\alpha} \right] \, .
\ee
At $t \approx x_0 + \Delta x$, $\tilde \sigma$ crosses the lightcones of the two $\psi$ insertions, and $z$ retraces its path, crossing the branch cut of $G_n(z,\bz)$ back to the principal sheet. Then for large times $t> x_0 +\Delta x$, $z$ is back on the principal sheet and the answer is again given by the vacuum result \eqref{eq:soutt0}.\footnote{An equivalent way of describing this is to look at the small $\mu$ expansion. There $z\sim 1$ throughout, but it passes through infinity as the sign of the imaginary part changes at the time where the lightcones are crossed. Given the set of insertions here and the relative branch cut, this is topologically equivalent to the finite-$\mu$ path in figure \ref{fig:zbar}.} Figure \ref{fig:zbar} gives a pictorial illustration of this.
Throughout this discussion we have considered only a single identity-block channel, namely, only the Virasoro identity-block that has zero monodromy and that reproduces the expected OPE limit for $z \to 1$ and $\bz \to1$. In fact, one can check explicitly that in all time regimes  the other channels, corresponding to the combined Euclidean monodromies  $G( e^{-2\pi i m } z, e^{2\pi i m } \bz )$, give subleading contributions at all times. 

The pattern of dominant identity channels changes when the operator $\psi$ is inserted inside the interval. Using reflection symmetry, we take $x_0 < 0$ and $|x_0| \leq \Delta x/2$ without loss of generality. In this case both $z$ and $\bz$ move non-trivially in the complex plane. The pattern is determined by the crossing of $\psi \psi$'s lightcones by the twist operators. At $t\approx  |x_0|$, $\sigma$ crosses the left-moving lightcone of $\psi\psi$ and $\bz$ changes sheet $\bz \to e^{- 2\pi i} \bz$, while when $\tilde \sigma$ crosses the lightcones, at $t\approx x_0 + \Delta x$, $z \to e^{2\pi i} z$. After $z$ and $\bz$ have changed sheet and crossed the corresponding branch cut in $G(z,\bz)$, they  remain on that sheet for the rest of the evolution. 

Implementing these sheet changes and comparing the competing identity channels yields the following picture.
At early times $0<t< |x_0|$, both $z$ and $\bz$ approach 1 on the principal sheet. In this window all operators remain spacelike separated and \eqref{smaster} is evaluated on the principal branch, \ie for the identity Virasoro channel with $m=0$. This gives eq.~\eqref{eq:soutt0},
\be
S_A = \frac{c}{3}\log\left[\frac{\Delta x }{\delta}\right]  \ .  
\ee
At late times $t>x_0 + \Delta x$, the answer is again \eqref{eq:soutt0}, but now $z$ and $\bz$ are both on the next sheet, and the vacuum result is obtained starting with \eqref{smaster} for $G(e^{-2\pi i m} z,e^{2\pi im} \bz)$. Equivalently, the dominant Virasoro identity-block at late time is the one corresponding to $m=1$. At intermediate times  $z$ and $\bz$ are on different sheets. Correspondingly, one finds that in the window  $|x_0| < t < x_0 + \Delta x$ the principal sheet $m=0$ channel dominates initially, giving
\be
S_A = \frac{c}{6}\log\left[\frac{ \Delta x(t +x_0)(t +x_0 + \Delta x)}{\delta^2\mu }\frac{\sin (\pi \alpha)}{\alpha} \right] \, .  
\ee
At  $t> \sqrt{|x_0|(x_0+\Delta x )}$, the dominant saddle switches from the $m=0$ channel to the $m=1$ channel, giving 
\be
S_A = \frac{c}{6}\log\left[\frac{ \Delta x(t- x_0)(x_0 + \Delta x-t)}{\delta^2\mu }\frac{\sin (\pi \alpha)}{\alpha} \right] \, .  
\ee
We now turn to the TEE. We take the twist operators $\sigma(z_2,\bar z_2)$ and $\tilde\sigma(z_3,\bar z_3)$ to be purely timelike separated, while leaving the remaining operator insertions unchanged. This amounts to analytically continuing the Euclidean correlator to the Lorentzian configuration
\be
\begin{aligned}
z_2 &= x_0  - t \,, \quad &&\bz_2 =  x_0   + t \,,\\
z_3 &= x_0  - t - \Delta t  \,, \quad &&\bz_3 = x_0 + t + \Delta t \ .
\end{aligned}
\ee
To compute the TEE, as defined in section \ref{sec:SITEE}, the twist operators must also be correctly time-ordered with respect to each other. This is implemented through the prescription
 $\Delta t \to \Delta t - i \varepsilon$ while taking the limits $\mu,\varepsilon\to0$ with $\mu>\varepsilon$ (see appendix \ref{app:TO}). Here this is particularly simple, since the cross ratios become
\be
\begin{aligned} \label{eq:zcquenchT}
z &=  \frac{(t - x_0 + i\mu)(t - x_0 - i\mu +  \Delta t - i \varepsilon)}{(t - x_0 - i\mu)(t - x_0 + i\mu +  \Delta t - i \varepsilon)}
  \simeq
  1 + \frac{2 i \,( \Delta t - i \varepsilon) \,\mu}{(t - x_0)(t - x_0 + \Delta t - i \varepsilon)} + O(\mu^2)\,, \\[.5em]
\bz &=  \frac{(t + x_0 + i\mu)(t + x_0 - i\mu +  \Delta t - i \varepsilon)}{(t + x_0 - i\mu)(t + x_0 + i\mu +  \Delta t - i \varepsilon)}
  \simeq
  1 + \frac{2 i \,( \Delta t - i \varepsilon) \,\mu}{(t + x_0)(t + x_0 + \Delta t - i \varepsilon)} + O(\mu^2) \,.\\
\end{aligned}
 \ee
Since $\mu > \varepsilon$, the $i\varepsilon$ prescription does not alter the sheet trajectories of $z,\bar{z}$ in \eqref{eq:zcquenchT}. Its role here is to fix the phase of the conformal prefactor $z_{23}\bz_{23}= - (\Delta t - i \vareps)^2$  in  \eqref{smaster}.
Implementing the prescription therefore yields
\be \label{smasterT}
S^{\rm (T)}_A = \frac{c}{6}\log\left[  \frac{  \Delta t^2\,  (z \bz)^{ \frac{1}{2}(1-\alpha)}(1-z^{\alpha})(1-\bz^{\alpha})}{\delta^2 \alpha^2(1-z)(1-\bz)} \right]  + i \frac{c}{6}\pi  \,.
\ee
From this point onward one may safely set $\varepsilon=0$ in \eqref{eq:zcquenchT}.

The remainder of the analysis closely parallels the spacelike case. The evolution is again determined by the crossings of the lightcones of the heavy operator pair $\psi\psi$ by the twist operators, together with the competition among the different identity-block channels labeled by the Euclidean monodromy index $m$.

Analytically continuing from the spacelike configuration,  when $\tilde\sigma$ crosses the lightcone of $\psi\psi$ at $t= x_0 - \Delta t$,  the holomorphic cross ratio changes sheet according to $z\to e^{2 \pi i}z$. Later, when $\sigma$ crosses the same lightcone at $t= x_0$ the cross ratio $z$ moves in the opposite direction $z\to e^{-2 \pi i}z$, bringing the correlator back to its original sheet.

To give a concrete example, we take $x_0>0$ and fix the separation between the twist operators to $0<\Delta t<x_0$. One then sees the following. At early times $0<t< x_0- \Delta t$ and again at late times $t>x_0$, the TEE coincides with its vacuum value
\be \label{eq:vacT}
S_A^{\rm (T)}= \frac{c}{3}\log \[ \frac{ \Delta t }{\delta}\] +  i  \frac{ c }{6} \pi\, .
\ee 
In both time regimes, this follows from evaluating \eqref{smasterT} on the principal sheet for both $z$ and $\bar z$, since the two twist operators remain on the same side of the lightcone generated by $\psi\psi$.
At intermediate times $x_0- \Delta t < t < x_0 $,  only $\tsigma$ has crossed into the $\psi \psi$ lightcone,  while $\sigma$ has not. One therefore has 
\be\label{eq:nonvacT}
\begin{aligned}
S^{\rm (T)}_A & = \frac{c}{6}\log\left[   \frac{\Delta t \, (x_0-t)(t + \Delta t  -x_0 )}{\delta^2\mu } \frac{\sin(\alpha \pi )}{\alpha} \right]  + i \frac{c}{6}\pi \, .  \\
\end{aligned}
\ee
This result is obtained by starting from the principal identity-block ($m=0$) and implementing the sheet transition  $z\to e^{2 \pi i}z$.

In summary, the TEE displays the same qualitative causal structure as the spacelike entanglement entropy. The nontrivial contribution is confined to the interval during which only one of the twist operators has crossed the lightcone of the local excitation, and the vacuum result is recovered once both twist operators lie on the same side of the lightcone. The resulting time dependence of the TEE is therefore compatible with the familiar quasiparticle interpretation of local operator quenches~\cite{Nozaki:2014hna}. 

%%%%%%%%%%%%%%%%%%%%%%%%%%%%%%%%%%%%%%%%%%%%%
\subsection{Boosted conical defect in Poincar\'e AdS$_3$}

The holographic dual of a local operator quench in a 2d CFT is described by a locally AdS$_3$ geometry containing a boosted conical defect that represents the bulk trajectory of the heavy operator insertion. 
This geometry can be obtained as the Bañados geometry determined by the expectation value of the CFT stress tensor, as discussed in section \ref{sec:Renyi} and in \cite{Asplund:2013zba,Asplund:2014coa}. 
Here we adopt instead the equivalent construction of \cite{Horowitz:1999gf,Nozaki:2013wia,Asplund:2013zba}, which is better suited for describing the complex geodesics relevant for TEE.

One starts from the static conical defect geometry in global AdS$_3$ coordinates\footnote{This is just \eqref{eq:alphametric}, where we renamed the time coordinate to avoid notational conflicts.}
\begin{equation}  \label{eq:CDgeom}
ds^2 = - (\alpha^2+ r^2)d \ft^2 + \frac{dr^2}{ (\alpha^2+ r^2)} + r^2 d\phi^2,     
\end{equation}
with $0<\alpha<1$ and applies the map  \cite{Horowitz:1999gf,Nozaki:2013wia} 
\be   \label{eq:CDquenchemap}
\begin{aligned}
r &= \frac{1}{\mu y} \sqrt{\mu^2 x^2 + \frac{1}{4}(-\mu^2+y^2-t^2+x^2)^2} \\[.5em]
\tan \ft &= \frac{2\mu t}{\mu^2+y^2-t^2+x^2}   
\\[.5em]
\tan \phi &= \frac{2\mu x}{\mu^2 - y^2+t^2-x^2},    
\end{aligned}   
\ee
where $y$ is the radial coordinate in the  Poincar\'e AdS$_3$ metric with boundary at $y=0$. 
For $\alpha=1$, this map reduces to a transformation between global and Poincar\'e AdS$_3$. For $0<\alpha<1$, instead, it generates the boosted conical defect geometry dual to the local operator quench. In particular, it describes the backreaction of Poincar\'e AdS$_3$
\be 
ds^2 = \frac{1}{y^2} \( -dt^2  + dy^2 + dx^2\)
\ee
to a massive particle oscillating in time in the radial direction along $x=0$ and approaching the AdS boundary $y=0$ up to a distance $\mu$. 

The holographic entanglement entropy in this geometry can then be computed by exploiting this equivalence. Boundary endpoints in the physical boosted geometry are first mapped back to the static conical defect geometry through \eqref{eq:CDquenchemap}. One then computes the corresponding geodesic there and finally evaluates its length in the original geometry. The general details of this construction are summarized in appendix \ref{app:boosted_defect}.

In the  entanglement entropy case, this construction was analyzed in detail in \cite{Nozaki:2013wia,Asplund:2013zba,Asplund:2014coa}. Spacelike-separated boundary points in the boosted geometry are mapped to spacelike-separated points in the static conical defect geometry, while winding geodesics around the conical defect become geodesics winding around the boosted particle trajectory in Poincar\'e coordinates.
 The resulting holographic computation exactly matches the dual CFT computation described above \cite{Asplund:2014coa}.

%%%%%%%%%%%%%%%%%%%%%%%%%
For the TEE we instead consider timelike-separated boundary endpoints and search for geodesics connecting them through the complexified bulk geometry.  
In order to compare with the local quench results of our CFT analysis, we choose the segment joining $(t, x_0, \eps)$ and $(t+\Delta t, x_0, \eps)$ as our timelike interval and work in the instantaneous quench limit $\mu \to 0$.
As in the CFT computation, the continuation to timelike separations is fixed through the prescription $\Delta t \to \Delta t - i \varepsilon$, which removes any ambiguity in the analytic continuation. 
The details of this computation are described in appendix \ref{app:boosted_defect}. Here we summarize the main results. For the geodesic connecting these boundary points with  no winding around the conical defect, which is the only one relevant in the case at hand, one finds that the geodesic length can be written in the compact form
\be \label{eq:geoeqblock}
\mathcal{L}   =\log\!\left[ \frac{4 \Delta t^2 \sqrt{z \bz}\, \sinh\!\left(\frac{\alpha}{2}\log z \right)\sinh\!\left(\frac{\alpha}{2}\log \bz\right)}{ \alpha^2\epsilon^2 (1- z) (1- \bz) }\right],
\ee
where $z$ and $\bar z$ are precisely the CFT cross ratios introduced in \eqref{eq:zcquenchT}.  
This makes the agreement with the CFT result essentially manifest already at the level of the geodesic length. Indeed, including the appropriate factor, $S_A^{(T)} = \mathcal L/ 4 G_N$,  one immediately recovers the previous expression \eqref{smasterT}.

Evaluating explicitly the geodesic length in the instantaneous quench limit, as described in appendix \ref{app:boosted_defect}, gives then 
\begin{equation}
S_A^{\rm (T)} =  \frac{c}{3}\log\frac{\Delta t}{\epsilon} + i\frac{c}{6}\pi\,,
\end{equation}
for $t < x_0-\Delta t $ or $t > x_0$ and 
\begin{equation}
S_A^{\rm (T)} = \frac{c}{6}\log\left[ \frac{\Delta t(t+\Delta t-x_0)(x_0-t)\sin \pi \alpha}{\mu \alpha \epsilon^2} \right]  + i \frac{c}{6}\pi
\end{equation}
if $x_0 - \Delta t < t < x_0$, in exact agreement with the CFT expressions \eqref{eq:vacT} and \eqref{eq:nonvacT} upon identifying the bulk and boundary UV cutoffs.

To illustrate the structure of the complex geodesics, consider the simple case $x_0=0$.
Working with $y_i=y_f=\eps \ll 1$, which in turn implies $r_i, r_f \gg1$, in the conical defect coordinates the geodesic profile reads
\be
\begin{aligned}
r(\lambda) &= \frac{\alpha \left( e^{\lambda} r_f + e^{-\lambda}r_i  \right)}
{i \sqrt{2 r_f r_i \left(1-\cos\!\left(\alpha \Delta \ft \right)\right)}} \\[6pt]
\ft(\lambda) &= \frac{1}{\alpha} \arctan \[ \frac{e^{2\lambda} r_f \sin\!\left(\alpha \ft_f\right) + r_i \sin\!\left(\alpha \ft_i\right)}
{e^{2\lambda} r_f \cos\!\left(\alpha \ft_f\right) + r_i \cos\!\left(\alpha \ft_i\right)}\]\\
\phi(\lambda) &= 0\,,
\end{aligned}
\ee
with the parameter $\lambda$ varying between the complex extrema $-\lambda_*$ and $\lambda_*$, with 
\be
 \lambda_*  = \frac{1}{2} \log \[\frac{2 r_i r_f \left(1-\cos\!\left(\alpha \Delta \ft\right) \right)}{\alpha^2} \] + i \frac{\pi}{2} \, .
\ee
This makes explicit that the holographic dual of TEE is naturally described by complex geodesics whose parametrization, profile and length all become complex after the Lorentzian continuation.

%%%%%%%%%%%%%%%%%%%%%%%%%%%%%%%%%%%%%%%%%%%%%
%%%%%%%%%%%%%%%%%%%%%%%%%%%%%%%%%%%%%%%%%%%%%
\section{Globally excited states}
\label{sec.excited}

%%%%%%%%%%%%%%%%%%%%%%%%%%%%%%%%%%%%%%%%%%%%%
In this section, we consider a class of globally excited states as a non-trivial cross-check, and apply the prescription introduced in section \ref{sec:SITEE} to states dual to time-dependent geometries. We first revisit the case of a thermal state on a line, dual to the planar AdS$_3$ black hole (see   \cite{Balasubramanian:2012tu,Doi:2023zaf,Heller:2024whi}). This provides an additional simple example of the analysis developed in the two previous sections and serves as a preparation for states quenched via a shell of local operator insertions, dual to planar AdS$_3$-Vaidya spacetime, to be discussed afterward. 

%%%%%%%%%%%%%%%%%%%%%%%%%%%%%%%%%%%%%%%%%%%%%
\subsection{Thermal state on a line}

Consider a thermal 2d CFT on a line. The thermal state is prepared by the Euclidean path integral on a cylinder $w=x+i\tau$, with $\tau\sim\tau+\beta$ and temperature $T=1/\beta$.
Consider an interval $A$ with endpoints $w_1 = \Delta x/2 + i \Delta \tau /2$ and $w_2 = -\Delta x/2 - i \Delta \tau /2$. Using the conformal map $z=e^{\frac{2\pi}{\beta}w}$ from the cylinder to the plane, one obtains the Euclidean two-point function
\be 
\langle   \sigma_n(w_1,\bar w_1)   \tilde \sigma_n(w_2 ,\bw_2)     \rangle = \left(  \frac{2 \pi^2}{\beta^2}\right)^{2 h_n} \left[  -\cos  \(\frac{2\pi  }{\beta} \Delta \tau \) +  \cosh\( \frac{2 \pi }{\beta} \Delta x \)  \right]^{-2h_n}.
\ee
The entanglement entropy of a spatial interval of size $\Delta x$ with the UV cutoff reinstated reads then
\be
S_A 
 = \frac{c}{3}\log\left[  \frac{\beta}{ \pi \delta } \sinh  \( \frac{\pi  }{\beta} \Delta x \) \right] \, .
\ee 
To obtain the TEE, we analytically continue the correlator to Lorentzian signature.  For time-ordered operators, this is implemented again via the regularization $\Delta t\to \Delta t-i\vareps$ \cite{Hartman:2015lfa,Kundu:2025jsm}. Assuming without loss of generality $\Delta t\geq0$, we have
\begin{equation}
\begin{aligned}
\langle  \sigma_n( \Delta t/2, \Delta x/2)   &\tilde \sigma_n(- \Delta t/2, - \Delta x/2)   \rangle\\[.5em]
&=\lim_{\varepsilon \to 0 }   \left(  \frac{2 \pi^2}{\beta^2}\right)^{2 h_n} \left[ - \cosh \frac{2\pi  }{\beta} (\Delta t - i \varepsilon )  + \cosh  \frac{2 \pi }{\beta} \Delta x   \right]^{-2h_n}  \, \\[.5em]
 &=     \left(  \frac{2 \pi^2}{\beta^2}\right)^{2 h_n} \left[  e^{i \pi} \( \cosh \( \frac{2\pi  }{\beta}\Delta t   \)  - \cosh \( \frac{2 \pi }{\beta} \Delta x \) \)   \right]^{-2h_n}\, .
 \end{aligned}
\end{equation}
The correlator structure is the same as on the plane, and so is the resulting imaginary part of the TEE, see \eqref{eq:TSvac},
\be \label{eq:TEEThermal}
S_A^{\rm (T)}= \frac{c}{6}\log\left[  \frac{\beta^2}{2 \pi^2 \delta^2 } \left( \cosh \( \frac{2\pi  }{\beta} \Delta t \) - \cosh \( \frac{2 \pi }{\beta} \Delta x   \) \right)\right]  + i \frac{c}{6} \pi   \,.
\ee 
When $A$ is purely timelike ($\Delta x = 0 $) this reduces to
\be \label{eq:Tline}
S_A^{\rm (T)}= \frac{c}{3}\log\left[  \frac{\beta}{\pi  \delta } \sinh \( \frac{\pi  }{\beta} \Delta t\)\right]  + i \frac{c}{6} \pi \, ,
\ee 
which coincides with the analytic continuation of the entanglement entropy expression to timelike intervals found in  \cite{Doi:2023zaf}. Notice the imaginary part is completely independent of the system temperature $\beta$.

%%%%%%%%%%%%%%%%%%%%%%%%%%%%%%%%%%%%%%%%%%%%%
\subsection{Planar AdS$_3$ black hole}

The holographic dual of this state is the planar AdS$_3$ black hole
\begin{equation} \label{eq:planarBH}
    ds^2=\frac{1}{y^2}\left[ -\( 1-\frac{y^2}{y^2_H} \) \,dt^2+\frac{dy^2}{ 1-\frac{y^2}{y^2_H}}+dx^2\right]\,.
\end{equation}
For endpoints
\be
(t_1,x_1,y_1)=\(\frac{\Delta t}{2},\frac{\Delta x}{2},\epsilon\)\,, \qquad (t_2,x_2,y_2)=\(- \frac{\Delta t}{2},- \frac{\Delta x}{2},\epsilon\)
\ee
on the regularized boundary, the geodesic solution in the $\epsilon\to 0$ limit reads 
\be
\begin{aligned} \label{eq:sol_thermal}
t(\lambda) &=y_H\operatorname{arcsinh}\left[\frac{\sqrt{2}\sinh{\frac{\Delta t}{2y_H}}\sinh{\lambda}}{\sqrt{\cosh{\frac{\Delta t}{y_H}}+\cosh{2\lambda}}}\right] \\[.5em]
x(\lambda) &= y_H\operatorname{arcsinh}\left[\frac{\sqrt{2}\sinh{\frac{\Delta x}{2y_H}}\sinh{\lambda}}{\sqrt{\cosh{\frac{\Delta x}{y_H}}+\cosh{2\lambda}}}\right]\\[.5em]
y(\lambda) &= y_H\sqrt{\frac{\cosh{\frac{\Delta x}{y_H}}-\cosh{\frac{\Delta t}{y_H}}}{\cosh{\frac{\Delta x}{y_H}}+\cosh{2\lambda}}}\,.
\end{aligned}
\ee
The endpoints are reached at $\lambda=\pm \lambda_*$, with
\begin{equation}
    \lambda_*=\frac 1 2 \log\left[\frac{4 y_H^2}{\epsilon^2} \sinh{\left(\frac{\Delta x+\Delta t}{2y_H}\right)}\sinh{\left(\frac{\Delta x-\Delta t}{2y_H}\right)}\right]\, . 
\end{equation}
For a timelike interval on the boundary, assuming the same time-ordering as in the CFT, we obtain the analytic continuation uniquely fixed by the stated $i\varepsilon$ prescription and continuity from the spacelike region
\be
\begin{aligned}
    \lambda_* &=\lim_{\varepsilon\to 0} \frac 1 2 \log{\left[\frac{4 y_H^2}{\epsilon^2} \sinh{\left(\frac{\Delta x+\Delta t-i\varepsilon}{2y_H}\right)}\sinh{\left(\frac{\Delta x-\Delta t+i\varepsilon}{2y_H}\right)}\right]}\\[.5em]
    &=\frac 1 2 \log{\left[\frac{4 y_H^2}{\epsilon^2} \sinh{\left(\frac{\Delta x+\Delta t}{2y_H}\right)}\sinh{\left(\frac{\Delta t-\Delta x}{2y_H}\right)} \right]}+i\frac{\pi}{2}\,,
\end{aligned}
\ee
and  complex length
\begin{equation}
    \mathcal{L}= \log{\left[\frac{4 y_H^2}{\epsilon^2} \sinh{\left(\frac{\Delta x+\Delta t}{2y_H}\right)}\sinh{\left(\frac{\Delta t-\Delta x}{2y_H}\right)} \right]}+i\pi \,. 
\end{equation}
This exactly reproduces the CFT result \eqref{eq:TEEThermal} upon identifying $y_H=\beta/2\pi$ and matching the bulk and boundary UV cutoffs.
These results reduce to the ones of \cite{Heller:2024whi} for purely timelike separations ($\Delta x=0$).  We refer to appendix \ref{app:AdSPoincare} for more details regarding the geodesic computation.

%%%%%%%%%%%%%%%%%%%%%%%%%%%%%%%%%%%%%%%%%%%%%
\subsection{Global operator quench}\label{sec:global_quench}

We now consider the TEE for a global quench in 2d CFT dual to the collapse of a shell of matter in AdS$_3$, following the framework developed in \cite{Anous:2016kss}. There, the global quench is realized in the 2d CFT through a homogeneous shell of local operator insertions. In this protocol, the state before the quench, for $t<0$, is the vacuum state of the 2d CFT. At $t=0$ the system is quenched with a shell of local insertions which injects a large amount of energy into the system. 
The post-quench system is a time-dependent excited state, which thermalizes as the injected matter equilibrates.

Here we exploit the fact that the correlator defining the TEE of the interval $A=[t_1,t_2]$ can be identified with the Lorentzian equal-space autocorrelation function of \cite{Anous:2016kss},
\be \label{eq:auto}
G(t_1,t_2)  = \langle T \left\{ {\mathcal O_Q (t_1)} {\mathcal O_Q (t_2)} \right\} \rangle\,.
\ee
Rather than reproducing the analysis of \cite{Anous:2016kss}, we briefly summarize the ingredients relevant for our discussion.
This is technically more involved, but directly related to the locally excited states of section \ref{sec.locallyexcited} and \cite{Fitzpatrick:2014vua,Asplund:2014coa}. 
The Euclidean state is prepared by multiple heavy operator insertions, which in the large-$c$ limit generate a classical stress tensor background. 
The heavy operators have conformal dimensions satisfying $h_H/c=O(1)$, while the probe operators entering the autocorrelation function $G(t_1,t_2)$ are perturbatively light, with conformal dimension $h_Q/c \ll 1$. In this regime, the evaluation of the probe correlator reduces in a nontrivial way to evaluating heavy–heavy–light–light correlators, which in the semiclassical limit are dominated by the Virasoro identity-block.
In particular, the resulting Lorentzian correlator is obtained by analytic continuation of the Euclidean configuration with appropriate contour choices around branch cuts, as discussed in \cite{Anous:2016kss}. 
The autocorrelator $G(t_1,t_2)$ reduces to the vacuum result when both insertions lie before the quench, and to the thermal equilibrium result when they both lie after it.
When the two operators are instead across the quench, \ie for $t_1<0$ and $t_2 > 0$,  the result of \cite{Anous:2016kss} reads
\begin{equation} \label{eq:utoshell}
    G(t_1,t_2)=e^{- 2 \pi i  h_Q}\left[\frac{\beta}{\pi}\sinh\( \frac{\pi t_2}{\beta}\) -t_1\cosh\(\frac{\pi t_2}{\beta}\)\right]^{-4 h_Q}\,.
\end{equation}
The corresponding TEE is obtained by identifying $h_Q$ with $h_n=\frac{c}{24}\left(n-\frac{1}{n}\right)$, as $n \to 1$. Reinserting the UV regulator for replica twists, this yields  
\be\label{eq:S_global_quench}
S_A^{\rm (T)}=\lim_{n\to 1}\frac{1}{1-n}\log G(t_1,t_2) =\frac{c}{3}\log\left[\frac{1}{\delta}\left(\frac{\beta}{\pi}\sinh\( \frac{\pi t_2}{\beta}\) -t_1\cosh\(\frac{\pi t_2}{\beta}\)\right)\right]+i\frac{c}{6}\pi\,.
\ee
This is the result for $t_1<0$ and $t_2>0$, while for  $t_1, t_2<0$ one has the vacuum result \eqref{eq:TSvac}, and for $t_1,t_2>0$ the thermal result \eqref{eq:Tline}, both depending only on the time difference  $\Delta t = t_2 - t_1$. 
Note that, in all cases, the imaginary part of the entropy equals that of the corresponding time-independent states defined on a line. It is therefore independent of both the size of the interval and the quench time. As for the local operator quench, the time dependence of the single interval TEE is compatible with a quasiparticle picture. In the next section we reproduce these results holographically through a computation of complex geodesics in planar AdS$_3$-Vaidya spacetime.

As a final remark, the original analysis of \cite{Anous:2016kss} included the case where the CFT lives on a spatial circle. For a circle of size $2\pi R$, the autocorrelation function is given by \cite{Anous:2016kss}
\begin{equation}
    G(t_1,t_2)=e^{- 2 \pi i  h_Q} \left[\frac{\beta}{\pi} \cos\( \frac{t_1}{2 R}\)\,\sinh\(\frac{\pi t_2}{\beta}\)- 2R \sin\(\frac{t_1}{2 R}\)\cosh 
    \(\frac{\pi t_2}{\beta}\)\right]^{-4 h_Q},
\end{equation}
which returns \eqref{eq:utoshell} upon taking the $R \to \infty$ limit. Even without performing a detailed analysis, we notice that the periodic factors associated with the insertion at pre-quench times $t_1<0$ suggest that the phase acquires a non-trivial dependence on $t_1$. We expect the imaginary part for the resulting TEE to increase with $|t_1|$ by multiple integers corresponding to the number of causal diamonds traversed, in the same way as discussed for the vacuum state on the circle in section \ref{sec:Im_TEE}. By contrast, this is not the case for $t_2$, the insertion probing the post-quench part of the geometry.

%%%%%%%%%%%%%%%%%%%%%%%%%%%%%%%%%%%%%%%%%%%%%
\subsection{Planar AdS$_3$-Vaidya}\label{sec:ads3_vaidya}

The dual spacetime to the global operator quench is planar AdS$_3$-Vaidya, whose metric can be written in advanced Eddington-Finkelstein coordinates as
\begin{equation}\label{eq:Vaidyametric}
    ds^2=\frac{1}{y^2} \left[-(1-m(v) y^2)\,dv^2- 2\,dv\,dy+ dx^2 \right] \,, 
\end{equation}
see, \eg \cite{Balasubramanian:2010ce,Balasubramanian:2011ur,Balasubramanian:2011at,Balasubramanian:2012tu}. With the mass profile $m(v)=\Theta(v)/ y_H^2$, the metric \eqref{eq:Vaidyametric} describes the collapse of an infinitesimally thin shell of null dust along $v=0$. Before the shell, for $v<0$, the geometry reduces to Poincar\'e AdS$_3$ in null ingoing coordinates, from which we recover \eqref{eq:PAdSmetric} through
\be
v = t - y\,,
\ee
while after, for $v>0$, via
\be
v = t + \frac{y_H}{2} \log \frac{y_H- y}{y_H + y} \,,
\ee
it reduces to the planar AdS$_3$ black hole metric \eqref{eq:planarBH} with horizon radius $y_H =\beta / 2\pi $. 

The interesting configuration for the TEE is that of an interval $A$ across the quench, in which case it is dual to the length of a complex geodesic connecting two timelike-separated boundary points at the same $x$ coordinate, $(t_1,y_1)=(t_1<0,\epsilon)$ in the empty AdS$_3$ region and $(t_2,y_2)=(t_2>0,\epsilon)$ in the black hole region. The two branches connect on the shell at $(v,y) = (v_s, y_s)$, whose location is fixed by extremizing the total length as detailed below. This provides a refraction condition at the interface of the two media. Such geodesics were first worked out in \cite{Balasubramanian:2012tu} to study the spectral function thermalization in this holographic strongly coupled setting. This involved solving for the same type of correlators we are considering, and we here revisit the computation of \cite{Balasubramanian:2012tu} in a slightly different fashion, explicitly imposing real boundary conditions at the AdS boundary. 

For the AdS part, we have the solution
\be
 \begin{aligned}
    t_\mathrm{AdS}(\lambda) &= \frac{y_1(v_s+y_s)\sinh(\lambda - \lambda_1) - t_1 y_s \sinh(\lambda - \lambda_s)}{ y_1 \sinh(\lambda - \lambda_1) - y_s \sinh(\lambda - \lambda_s)} \\[.5em]
    y_\mathrm{AdS}(\lambda) &= \frac{ y_1 y_s \sinh(\lambda_s-\lambda_1) } { y_1 \sinh(\lambda-\lambda_1) - y_s \sinh(\lambda-\lambda_s) }\,,
\end{aligned}
\ee
and thus in terms of the ingoing coordinate 
\be\label{AdS_matching}
 v_\mathrm{AdS}(\lambda) = \frac{y_s [y_1 \sinh (\lambda_1-\lambda_s)- t_1 \sinh (\lambda -\lambda_s)]+ y_1 (v_s+y_s) \sinh (\lambda -\lambda_1)}{y_1 \sinh (\lambda -\lambda_1)-y_s \sinh (\lambda - \lambda_s)}\,,
\ee
with 
\be\label{AdS_distance}
\begin{aligned}
\cosh (\lambda_s -\lambda_1) &= \frac{(v_s-t_1) (t_1-v_s-2 y_s) + y_1^2}{2 y_1 y_s  }
\approx \frac{(v_s-t_1) (t_1-v_s-2 y_s)}{2 y_s \epsilon }\ ,
\end{aligned}
\ee
where we used the near-boundary expansion in writing the last approximation. The length of the portion of geodesic contained in empty AdS is then given by
\be\label{eq:VlenAdS}
\lambda_s - \lambda_1 = \log \[ \frac{(v_s-t_1) (t_1-v_s-2 y_s)}{y_s \epsilon }\] \,.
\ee
For the black hole part, instead, we have 
 \be\label{BTZ_matching}
 \begin{aligned} 
   \tanh \(\frac{t_\mathrm{BH}(\lambda)}{y_H}\) &= \frac{y_2\left[y_H \sinh \left(\frac{v_s}{y_H}\right)+y_s \cosh \left(\frac{v_s}{y_H}\right)\right]\sinh (\lambda -\lambda_2)- y_s\sqrt{y_H^2 -y_2^2}\sinh \left(\frac{t_2 }{y_H}\right) \sinh (\lambda -\lambda_s)}{y_2\left[y_s \sinh \left(\frac{v_s }{y_H}\right)+y_H \cosh \left(\frac{v_s}{y_H}\right)\right]\sinh (\lambda -\lambda_2)-y_s\sqrt{y_H^2 -y_2^2} \cosh \left(\frac{t_2}{y_H}\right) \sinh (\lambda -\lambda_s)} \\[.5em]
    y_\mathrm{BH}(\lambda)  &= \frac{ y_2 y_s \sinh(\lambda_2-\lambda_s) }{ y_2 \sinh(\lambda-\lambda_2)-y_s \sinh(\lambda-\lambda_s) }\\[.5em]
    v_\mathrm{BH}(\lambda) &=t_\mathrm{BH}(\lambda) + \frac{y_H}{2} \log \[ \frac{y_H- y_\mathrm{BH}(\lambda)}{y_H + y_\mathrm{BH}(\lambda)} \]\ ,
\end{aligned}
\ee
with
\begin{equation}\label{BTZ_distance}
\begin{aligned}
\cosh (\lambda_2 -\lambda_s) 
&=\frac{y_H^2 - \sqrt{y_H^2 - y_2^2}\left[y_H \cosh\left(\frac{t_2-v_s}{y_H}\right) - y_s \sinh\left(\frac{t_2-v_s}{y_H}\right)\right]}{y_2 \ y_s}\\[.5em]
&\approx \frac{y_H^2}{y_s \eps}\left[ 1 - \cosh\left(\frac{t_2-v_s}{y_H}\right) + \frac{y_s}{y_H} \sinh\left(\frac{t_2-v_s}{y_H}\right)\right]\,,
\end{aligned}
\end{equation}
where we used again the near-boundary expansion in writing the last expression. The length of this portion is 
\be\label{eq:VlenBTZ}
\lambda_2 - \lambda_s = \log \[\frac{2 y_H^2}{y_s \eps}\left\{ 1 -  \cosh\left(\frac{t_2-v_s}{y_H}\right) + \frac{y_s}{y_H} \sinh\left(\frac{t_2-v_s}{y_H}\right)\right\}\].
\ee
The total geodesic length is then given by the sum of \eqref{eq:VlenAdS} and \eqref{eq:VlenBTZ} as a function of the junction point parameters on the shell $(v_s,y_s)$. 

In \cite{Balasubramanian:2012tu,Balasubramanian:2010ce,Balasubramanian:2011ur} the joining condition between the two geodesic portions was imposed at the shell $v_s =0$, and extremizing the length with respect to $y_s$ fixes it to 
\be \label{eq:shellys}
y_s = \left[\frac{1}{t_1}+\frac{1}{2y_H}\coth\left(\frac{t_2}{2y_H}\right) \right]^{-1}\,,
\ee
and the total length to
\be \label{eq:lenVaidya}
\lambda_2 - \lambda_1 = %
  2\log \left[\frac{1}{\eps}\left( 2y_H\sinh \frac{t_2}{2y_H}-t_1\cosh \frac{t_2}{2y_H}\right)\right]+i\pi\,.
\ee
In particular, the AdS contribution reads
\be
\lambda_s - \lambda_1 = \log\[\frac{t_1}{2y_H\eps}\left(2y_H-t_1 \coth\!\left(\frac{t_2}{2y_H}\right)\right)\]\,,
\ee
and given that $t_1<0$, $t_2>0$, this always brings an $i \pi$ factor corresponding to a complex profile in the AdS part of the Vaidya geometry. For the black hole part instead
\be
\lambda_2 - \lambda_s =\log\[\frac{2 y_H}{t_1\epsilon}\sinh{\left(\frac{t_2}{2y_H}\right)}\left(t_1 \cosh{\left(\frac{t_2}{2y_H}\right)}-2y_H\sinh{\left(\frac{t_2}{2y_H}\right)}\right)\]\,,
\ee
where the argument of the logarithm is always positive. 

Since we are allowing for complex excursions of the geodesics, it is however also natural to verify that this is indeed the optimal geodesic among those that ``cross the shell'' along the imaginary $v_s$ axis, and not strictly at $v_s =0$. Indeed, if we extremize the total length with respect to $y_s$, while keeping $v_s$ free, this fixes it to 
\be
y_s = \left[\frac{1}{t_1-v_s}+\frac{1}{2y_H}\coth\left(\frac{t_2-v_s}{2y_H}\right) \right]^{-1} .
\ee
Substituting this in the total length to express it solely as a function of $v_s$ and minimizing the real part of the length along the purely imaginary $v_s$ direction, one finds that the real part is minimized at $\operatorname{Im} v_s=0$. We conclude therefore that, at least in the thin-shell limit considered here, the junction occurs at $v_s=0$, in agreement with \cite{Balasubramanian:2012tu}.

Also notice that, while $y_s$ in eq.~\eqref{eq:shellys} is always real, it changes sign at $t_1/(2 y_H) = - \tanh(t_2/2y_H)$, where it also diverges. Correspondingly, the joining point on the real shell at $v=0$, can lie either outside or inside the event horizon, but also within a real extension of the original geometry to a negative $y$-coordinate, as the intersection point crosses the Poincar\'e horizon along $v=0$. 

Since the argument of the logarithm in \eqref{eq:lenVaidya} is strictly positive for any $t_1<0$ and $t_2>0$, the imaginary part of the geodesic length is always equal to $i\pi$, with no dependence on the boundary times. The TEE that stems from the length \eqref{eq:lenVaidya} is then given by
\begin{equation}\label{eq:SVaidya}
    S_A^{\rm (T)}=\frac{c}{3}\log \left[\frac{1}{\eps}\left(2y_H\sinh \frac{t_2}{2y_H}-t_1\cosh \frac{t_2}{2y_H}\right)\right]+i\frac{c}{6}\pi\,, \quad {\rm for} \,\, t_1 < 0 < t_2\,,
\end{equation}
and precisely matches the one obtained from the CFT side \eqref{eq:S_global_quench} with the usual identification between the event horizon radius and inverse temperature, $y_H=\beta/2\pi$. Obviously, for times $t_1< t_2 <0$ we have instead the pure AdS result \eqref{eq:TSvac}, and for $0< t_1 < t_2 $ the planar black hole one \eqref{eq:Tline}.

Notice that the holographic TEE in this AdS$_3$-Vaidya setup was also studied in \cite{Katoch:2025bnh,Katoch:2026dzs}, using an extension of the piecewise curve holographic prescription advocated in \cite{Doi:2022iyj, Doi:2023zaf}. Our findings based on the analysis of complex geodesics differ from those results both in the profiles of the curves and in the final result for the TEE, but do reproduce correctly the known explicit CFT results of \cite{Anous:2016kss} for the correlator. 

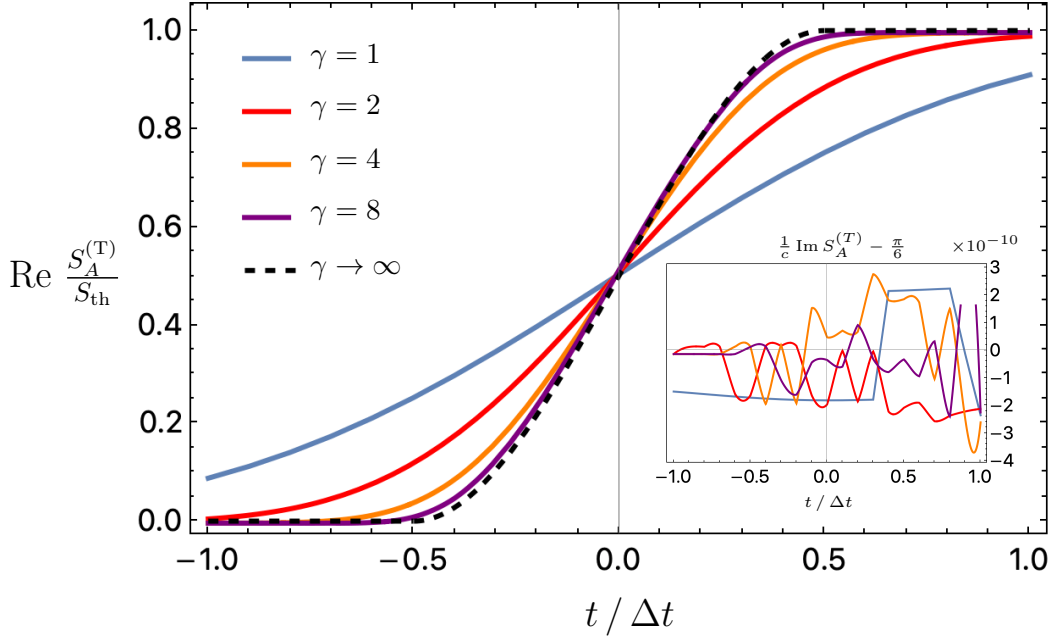
\begin{figure}[t]
    \centering
    \begin{overpic}[width=.8\linewidth]{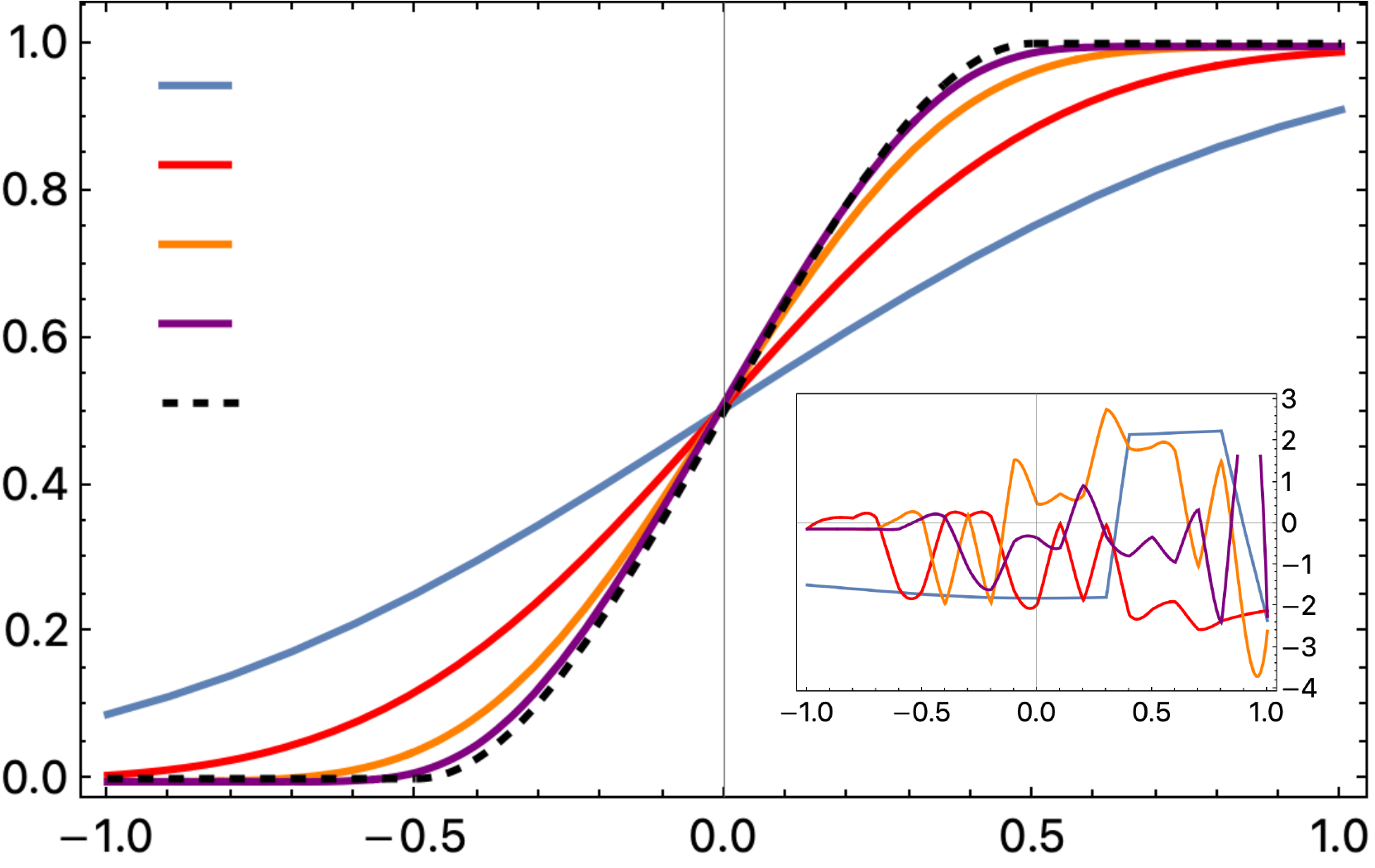}
    \put(19,56.5){$\gamma=1$}
    \put(19,51){$\gamma=2$}
    \put(19,45){$\gamma=4$}
    \put(19,39.5){$\gamma=8$}
    \put(19,33.5){$\gamma\to \infty$}
    \put(49,-5){\Large $t\,/\,\Delta t$}
    \put(-14,32){\Large Re $\frac{S_A^{\rm (T)}}{S_{\text{th}}}$}
    \put(70,36){\tiny $\frac{1}{c}\operatorname{Im} S_A^{(T)}-\frac{\pi}{6}$}
    \put(89,36){\tiny $\times 10^{-10}$}
    \put(73,8){\tiny $t\,/\,\Delta t$}
    \end{overpic}
    \vspace{.8cm}
    \caption{Real part of the TEE in the Vaidya-like quench with mass profile \eqref{eq:massprofile} for different values of $\gamma$, expressed as a function of the mean time $t=\frac{1}{2}(t_1+t_2)$ normalized to the interval size $\Delta t$. Here $m_0=1$ and $\Delta t = 1$. The values for the entropy are normalized to the thermal value $S_{\rm th}$ attained for $t/\Delta t \to \infty$. The analytically solvable thin-shell limit, denoted by the black dashed line, is recovered as $\gamma\to \infty$. In the inset, we plot the difference $\frac 1 c {\rm Im}S_A^{\rm (T)} - \frac \pi 6$ between the imaginary part and its expected value from the thin-shell computation, which is zero within numerical error tolerance for all values of $\gamma$, also away from the thin-shell limit.}
    \label{fig:VaidyaComplex}
\end{figure}

Finally, and as a further check, our result for the geodesic length and holographic TEE can also be extended to a thick-shell profile of the form
\begin{equation}\label{eq:massprofile}
    m(v)=\frac{m_0}{2}(1+\tanh{(\gamma v)})\,,
\end{equation}
in the geometry \eqref{eq:Vaidyametric}. In this case, no analytic solution is available, but, as detailed in appendix \ref{app:Vaidya}, the geodesic equations can be solved numerically and through our prescription this allows one to evaluate the TEE. We indeed find that the numerical computation reproduces the expected result \eqref{eq:SVaidya} in the thin-shell limit $\gamma\to \infty$, as illustrated in figure \ref{fig:VaidyaComplex}, and also extends beyond it.

%%%%%%%%%%%%%%%%%%%%%%%%%%%%%%%%%%%%%%%%%%%%%
%%%%%%%%%%%%%%%%%%%%%%%%%%%%%%%%%%%%%%%%%%%%%
\section{Discussion \label{sec:discussion}}
%%%%%%%%%%%%%%%%%%%%%%%%%%%%%%%%%%%%%%%%%%%%%
%%%%%%%%%%%%%%%%%%%%%%%%%%%%%%%%%%%%%%%%%%%%%

The derivation presented in this work not only establishes TEE as a boundary defined quantity with a controlled semiclassical dual in AdS$_3$/CFT$_2$, but also opens multiple new directions for future research. Below we elaborate on those we consider the most relevant.

\paragraph{Multiple intervals.}~One of the key aspects highlighted in this work is the role of operator ordering in the context of TEE. Since our analysis focused on the two-point function of replica twist operators, or equivalently on a single timelike interval, the only relevant choice was between time-ordered and anti-time-ordered correlators. Although choosing one of the two is necessary to define TEE consistently, these are simply related by complex conjugation.

As for ordinary entanglement entropy, however, it is natural to generalize this picture to subregions determined by multiple twist-antitwist insertions. The corresponding correlation function admits multiple inequivalent operator orderings, giving rise to distinct quantities that fall under the umbrella of temporal entanglement. One example is the \emph{entanglement in time} recently introduced in~\cite{Milekhin:2025ycm}, corresponding to two spatial intervals that are timelike separated. It would be very interesting to classify the resulting possibilities and understand their physical properties. We plan to investigate these aspects in the near future. 

While some tools from real-time quantum field theory can be borrowed, as we have done here, there is an important subtlety stemming from the fact that replica twist operators are local fields of the orbifold theory, but endpoints of replica permutation defects, which are nonlocal from the seed theory perspective. Although this subtlety did not play a role in our analysis for two insertions, it already becomes apparent for correlation functions of two twist-antitwist pairs. A familiar example is provided by the spacelike configuration, where, depending on the replica construction and ordering of twist fields, the same four-point function evaluates either the mutual information or the entanglement negativity~\cite{Calabrese:2012ew}.

\paragraph{Quasiparticle picture.}~The quasiparticle picture of entanglement production after a quench has been one of the central paradigms for understanding entanglement dynamics in integrable many-body systems~\cite{Calabrese:2005in,Alba:2017ekd}. In this picture, an instantaneous quench creates pairs of entangled quasiparticles that propagate ballistically, and the entanglement entropy  of a spatial interval is obtained  by counting the pairs shared between the interval and its complement at a given time.

In the present work, we found that the real part of the single-interval TEE has a time dependence compatible with the usual quasiparticle counting in all time-dependent states we studied, both for local quenches, discussed in section \ref{sec:localquench}, and for global quenches, discussed in section \ref{sec:global_quench}.
By contrast, the imaginary part of the TEE remains constant throughout the transient regime in both classes of quenches, suggesting that, at least in the examples considered here, it does not encode dynamical entanglement production.

A natural next step would be to extend this comparison to free quantum field theories.
Indeed, already for spacelike entanglement entropy, the picture survives in structure but the details of the propagation dynamics are non-universal and depend on the post-quench state (see, \eg~\cite{Calabrese:2007rg,Fagotti:2008mlc,Alba:2017ekd}). Along the lines of~\cite{Heller:2025kvp}, it would also be interesting to understand how the quasiparticle interpretation is modified by boosts of the entangling region.
Finally, one could consider generalizations of TEE to multiple intervals. This direction is especially relevant because, even in conventional spatial setups, the quasiparticle picture is known to have limitations for disjoint intervals~\cite{Asplund:2015eha}.

\paragraph{Imaginary part.}~A characteristic feature of TEE, as compared to the entanglement entropy of spacelike regions, is its complex-valued nature. While the real part appears to play a very similar role and encodes information about the entangling region and the state of the system, understanding what information is encoded in the non-Hermiticity, and in particular in the imaginary part, remains a central open question. In our analysis, the latter takes a very simple form, and is directly determined by the effective causal structure of the setting under consideration. It would be worthwhile to understand whether this separation between the real and imaginary parts of the TEE holds more generally.

On the other hand, in higher-dimensional holographic settings, the imaginary part of TEE exhibits a nontrivial dependence on the size of the boundary subregion \cite{Heller:2024whi,Heller:2025kvp}. The simple behavior found in the present work may therefore be a consequence of the constraining power of the two-dimensional conformal group. Indeed, our preliminary analysis of a three-dimensional bulk setup with explicitly broken conformal symmetry, together with global quenches in free quantum field theories~\cite{freequenches}, indicates that the imaginary part generally depends nontrivially on the size of the boundary subregion and other parameters of the problem. Systematizing our understanding of the imaginary part beyond the 2d CFT cases considered in this work is a natural avenue for future research, which also aligns with recent imaginary-part-based $c$-theorem considerations~\cite{Grieninger:2023knz,Roychowdhury:2025ukl,Giataganas:2025div,Roychowdhury:2025ebs,Giataganas:2025ize}.

\paragraph{Higher dimensions.} A natural generalization of holographic TEE to higher dimensions involves codimension-two extremal surfaces, whose physics differs substantially from that of geodesics in AdS$_3$. First, even in the spacelike regime, the standard holographic entanglement entropy exhibits a richer structure; for instance, in thermal states, nontrivial phase transitions can take place before the lightcone is reached~\cite{Heller:2025kvp}. Second, as discussed in~\cite{Heller:2024whi,Heller:2025kvp}, in the timelike regime there may exist branches of complex extremal surfaces that violate the UV/IR correspondence~\cite{Susskind:1998dq}. 

To make further progress in the understanding of higher-dimensional TEE, it would be very helpful to develop a degree of boundary control comparable to the one achieved in the present work. In particular, in AdS$_4$, a natural next step in this direction would be to understand the timelike generalization of Wilson loop expectation values, whose holographic duals are two-dimensional surfaces~\cite{Rey:1998ik,Maldacena:1998im}. These observables are not subject to a homology constraint, making them a particularly clean setting in which to investigate the role of complex saddles in higher-dimensional holography.

More generally, important progress in arbitrary spacetime dimension may come from an appropriate generalization of the entanglement first law and, more broadly, the physics of modular Hamiltonians~\cite{Chen:2021lnq}. To this end, the modular Hamiltonian for a spherical subregion in any CFT in its ground state is given by a weighted integral of the energy-momentum tensor operator~\cite{Casini:2011kv}. A natural starting point would be to understand if any of this structure holds when such a spherical subregion is analytically continued to become timelike, starting with 2d CFT.

The higher-dimensional analysis also motivates revisiting more general asymptotically AdS$_3$ spacetimes to determine whether they admit geodesic solutions, perhaps complex, that violate the UV/IR correspondence.

\paragraph{Homology constraint.}
One of the defining features of holographic entanglement entropy is the homology constraint~\cite{Fursaev:2006ih,Headrick:2007km,Haehl:2014zoa}, which requires the bulk extremal surface to be homologous to the boundary subregion. From the boundary perspective, this admits a natural interpretation in terms of replica twist operators, whose associated topological defect may belong to different homotopy classes~\cite{Haehl:2014zoa}.

In the cases analyzed in this work, involving a single interval on boundary geometries conformally equivalent to the plane, the situation is straightforward. A more instructive example is provided by the torus. There, different homotopy classes of the replica defect correspond to distinct replica channels, while the choice of bulk filling determines which of these channels admit a holographic realization. 

An interesting open question is whether holographic TEE admits a timelike analogue of the homology constraint and, if so, what its boundary interpretation and physical role should be. In the spacelike setting, purity of the underlying quantum state and entropy inequalities provided important motivation for this constraint, suggesting that a better understanding of analogous properties of holographic TEE may ultimately clarify this issue.

\paragraph{Kinematic-space organization.}
The conformal family of spherical spatial regions, together with observables
assigned to them, admits a natural geometric organization in kinematic
space~\cite{Czech:2016xec,deBoer:2016pqk}. In two dimensions, this relation is
especially suggestive for the present construction: a spatial interval and a
timelike-separated pair specify, respectively, the spatial corners and the
temporal tips of the same causal diamond. TEE should
therefore not simply be regarded as the ordinary kinematic-space entropy
evaluated on an unfamiliar type of region. Rather, it is natural to ask whether
the time-ordered twist correlator defines a different, complex-valued
continuation of the entanglement data associated with the same diamond.

Our results show that such an extension must retain information beyond the
local geometry of kinematic space. In particular, the imaginary part of the TEE is
fixed by operator ordering and changes by quantized amounts under successive effective
lightcone crossings. Since this information is locally constant, it would be
invisible to constructions based only on derivatives of the entropy, such as
the usual kinematic-space metric. This suggests that TEE should instead be
formulated on a covering or complexification of kinematic space, with the sheet
and monodromy data recording the causal continuation of the twist correlator.

\paragraph{Tensor-network realizations.}
At the boundary level, a natural next step is to connect our CFT construction
with temporal density matrices obtained by transverse contraction of tensor
networks describing critical spin chains. The universal CFT results derived
here provide quantitative benchmarks for such simulations, while comparison
with holographic CFTs may help distinguish generic properties of temporal
entanglement from those associated with semiclassical gravity.

At the bulk level, dynamical tensor networks, spacetime random tensor networks
and path-integral circuits reproduce aspects of covariant extremization and
bulk reconstruction~\cite{May:2016dgv,Qi:2018shh,Takayanagi:2018pml}. A tensor-network realization of TEE, however, cannot be an ordinary
minimal-cut prescription, because such a prescription produces a real,
nonnegative entropy and does not retain operator ordering. A more concrete
target would be a Lorentzian spacetime network in which timelike-separated
twist operators source a replica-permutation defect. Its semiclassical
description should involve a generally complex domain-wall saddle, with the
dominant contribution selected by the smallest real part of its action. In the
replica limit this saddle should reproduce the complex geodesic found in this
work, while at finite replica index it should provide a discrete counterpart
of our complex cosmic brane. Such a construction would also offer a controlled
setting in which to investigate a timelike analogue of the homology
constraint.

\paragraph{De Sitter holography.}
The relation to dS/CFT is not merely a parallel motivation: complex
entropies and their connection to temporal entanglement already played a
central role in the works that introduced TEE to holography
~\cite{Doi:2022iyj,Doi:2023zaf}, as well as in~\cite{Narayan:2022afv}; see also the broader dS/CFT literature
~\cite{Witten:2001kn,Strominger:2001pn,Strominger:2001gp,
Maldacena:2002vr,McFadden:2009fg,Anninos:2011ui,Thavanesan:2025ibm}.
Our results suggest a more stringent continuation than analytically
continuing only the final entropy or extremal-surface area. One should instead
continue the complete replica construction: the ordered twist correlator, its
replica-index dependence, the complex brane saddle and the rule selecting the
dominant saddle.

It would be particularly interesting to determine whether the operator
ordering that fixes the imaginary part in AdS/CFT becomes an
$i\varepsilon$ or integration-contour prescription in dS/CFT, and whether the
changes of sheet encountered at successive lightcone crossings become Stokes
transitions between complex de Sitter saddles. These questions could be
approached through analytic continuation of the AdS construction developed
here or directly within Kosmic Field Theories~\cite{Thavanesan:2025ibm}.

%%%%%%%%%%%%%%%%%%%%%%%%%%%%%%%%%%%%%%%%%%%%%%%%%%
%%%%%%%%%%%%%%%%%%%%%%%%%%%%%%%%%%%%%%%%%%%%%%%%%%
\section*{Acknowledgements}

It is a pleasure to thank Stefano Carignano, Xi Dong, Rob Myers, Ronak Soni, Luca Tagliacozzo, Tadashi Takayanagi and Erik Tonni for discussions on the topics of relevance to the present manuscript. We would like to thank Federica Colombo and Meng-Ting Wang for collaboration on related topics and comments on the draft. This project has received funding from the European Research Council (ERC) under the European Union’s Horizon 2020 research and innovation programme (grant number: 101089093 / project acronym: High-TheQ). Views and opinions expressed are however those of the authors only and do not necessarily reflect those of the European Union or the European Research Council. Neither the European Union nor the granting authority can be held responsible for them. This work was also partially supported  by the Priority Research Area Digiworld under the program Excellence Initiative  - Research University at the Jagiellonian University in Krakow.
AB and FG are grateful for the hospitality of the Galileo Galilei Institute for Theoretical Physics (GGI) where part of this work was carried out. 
FO is supported by the Research Foundation Flanders (FWO) doctoral fellowship 1182825N. FO thanks the Yukawa Institute for Theoretical Physics in Kyoto, where part of this work was carried out, for its hospitality and acknowledges financial support from the FWO grant for long stay abroad V439125N.

%%%%%%%%%%%%%%%%%%%%%%%%%%%%%%%%%%%%%%%%%%%%%
%%%%%%%%%%%%%%%%%%%%%%%%%%%%%%%%%%%%%%%%%%%%%
\appendix 

%%%%%%%%%%%%%%%%%%%%%%%%%%%%%%%%%%%%%%%%%%%%%
\section{Geodesic computations} \label{App:geo}

Here we collect and summarize some useful results for geodesic computations in asymptotically AdS$_3$ spacetimes.

We consider $\mathrm{AdS}_3$ embedded in a flat space $\mathbb{R}^{2,2}$ with coordinates
\be
X^A = (X^0, X^1, X^2, X^D)\,,
\ee
endowed with the metric
\be
\eta_{AB} = \mathrm{diag}(-1, +1, +1, -1)\,,
\qquad
A,B = 0,1,2,D\,.
\ee
AdS space is defined as the (universal covering of the) hyperboloid
\be
\eta_{AB} X^A X^B = -1 \,.
\ee
Geodesics in AdS can be obtained directly from the embedding. They are given by the intersections of the hyperboloid with two-dimensional planes through the origin. In the spacelike case these are the curves
\be
X^A(\lambda) = A^A \cosh\lambda + B^A \sinh\lambda\,,
\ee%
with  $A^A$ and $B^A$  such that 
\be
\eta_{AB} A^A A^B = -1\,, 
\qquad
\eta_{AB} B^A B^B = 1\,,
\qquad
\eta_{AB} A^A B^B = 0\,.
\ee
As one can check, they lie entirely on the AdS hyperboloid  $\eta_{AB} X^A(\lambda) X^B(\lambda) = -1$ and their tangent vector satisfies 
\be
\eta_{AB} \frac{dX^A}{d\lambda}\frac{dX^B}{d\lambda} = 1\,.
\ee 
Given two points $X_i^A = X^A(\lambda_i)$ and $X_f^A = X^A(\lambda_f)$ on the geodesic, their embedding-space contraction is
\be
\eta_{AB} X_i^A X_f^B = -\cosh(\lambda_f - \lambda_i)\,.
\ee
The geodesic distance $\mathcal{L} = \int^{\lambda_f}_{\lambda_i}d \lambda   $ between the two points therefore satisfies
\be
\cosh\!\left(\mathcal{L}\right) =  \cosh(\lambda_f - \lambda_i)=  \cosh(2 \lambda_*)=   -\eta_{AB} X_i^A X_f^B \,.
\ee
Above, we defined a symmetric parameterization through    $\lambda_*= \lambda_f = - \lambda_i$.  We will use this parametrization in most of what follows.

%%%%%%%%%%%%%%%%%%%%%%%%%%%%%%%%%%%%%%%%%%%%%
\subsection{Asymptotically planar AdS$_3$ spacetimes}
\label{app:AdSPoincare}

In this appendix we collect the derivation of the geodesic profiles in planar asymptotically AdS$_3$ spacetimes. Our results are consistent with those of \cite{Balasubramanian:2012tu}, where the same solutions were obtained using a different approach.

Let us start from the planar AdS$_3$ black hole with event horizon at $y=y_H$, \ie a BTZ black brane. The embedding in this case is given by
\begin{equation}\label{eq:emb_BTZplanar}
\begin{aligned}
    X^0 &=\frac{y_H}{y}\sqrt{1-\frac{y^2}{y_H^2}}\sinh{\frac{t}{y_H}} \\[.5em]
    X^1&=\frac{y_H}{y}\sqrt{1-\frac{y^2}{y_H^2}}\cosh{\frac{t}{y_H}} \\[.5em]
    X^2&=\frac{y_H}{y}\sinh{\frac{x}{y_H}}\\[.5em]
    X^D&=\frac{y_H}{y}\cosh{\frac{x}{y_H}}\, ,
\end{aligned}
\end{equation}
with induced metric
\begin{equation}
    ds^2=\frac{1}{y^2}\left[-\left(1-\frac{y^2}{y_H^2}\right)dt^2+\frac{dy^2}{1-\frac{y^2}{y_H^2}}+dx^2\right]\,.
\end{equation}
The inverse relations are
\be
\begin{aligned}
    t &=y_H \operatorname{arcsinh}\frac{X^0}{\sqrt{-\left(X^2\right)^2+\left(X^D\right)^2-1}}\\[.5em]
    x &= y_H \operatorname{arcsinh}\frac{X^2}{\sqrt{-\left(X^2\right)^2+\left(X^D\right)^2}} \\[.5em]
        y &=y_H \frac{1}{\sqrt{-\left(X^2\right)^2+\left(X^D\right)^2}} \,. 
\end{aligned}
\ee
The geodesic connecting two points $(t,x,y)=(t_i,x_i,\epsilon)$ and $(t,x,y)=(t_f,x_f,\epsilon)$ in the $\epsilon\to 0$ limit has the profile
\be
\begin{aligned}
t(\lambda) &=y_H\operatorname{arcsinh}\left[\frac{\sqrt{2}\sinh{\frac{\Delta t}{2y_H}}\sinh{\lambda}}{\sqrt{\cosh{\frac{\Delta t}{y_H}}+\cosh{2\lambda}}}\right]\\[.5em]
x(\lambda) &= y_H\operatorname{arcsinh}\left[\frac{\sqrt{2}\sinh{\frac{\Delta x}{2y_H}}\sinh{\lambda}}{\sqrt{\cosh{\frac{\Delta x}{y_H}}+\cosh{2\lambda}}}\right]\\[.5em]
y(\lambda) &= y_H\sqrt{\frac{\cosh{\frac{\Delta x}{y_H}}-\cosh{\frac{\Delta t}{y_H}}}{\cosh{\frac{\Delta x}{y_H}}+\cosh{2\lambda}}}\,,
\end{aligned}
\ee
and length
\begin{equation}
        \mathcal{L}=2 \lambda_*=2 \log{\left[\frac{2 y_H}{\epsilon}\sqrt{\sinh{\left(\frac{\Delta x+\Delta t}{2y_H}\right)}\sinh{\left(\frac{\Delta x-\Delta t}{2y_H}\right)}}\right]}\,.
\end{equation}
In writing these expressions we have defined $\Delta t=t_f-t_i$, $\Delta x=x_f-x_i$, chosen a symmetric parametrization, and worked in the small $\epsilon$ near-boundary limit. 

This shows that going to the timelike regime $\Delta x^2-\Delta t^2<0$ requires complexifying the affine parameter and yields a complex length for the geodesic. More specifically, implementing the regularization corresponding to the time ordering of boundary operators, $\Delta t \to \Delta t  - i \varepsilon$,
\be
\begin{aligned}
    \lambda_* &=\lim_{\varepsilon\to 0} \frac{1}{2}\log{\left[\frac{4 y_H^2}{\epsilon^2} \sinh{\left(\frac{\Delta x+\Delta t-i\varepsilon}{2y_H}\right)}\sinh{\left(\frac{\Delta x-\Delta t+i\varepsilon}{2y_H}\right)}\right]}\\[.5em]
    &=\frac{1}{2}\log{\left[\frac{4 y_H^2}{\epsilon^2} \sinh{\left(\frac{\Delta x+\Delta t}{2y_H}\right)}\sinh{\left(\frac{\Delta t-\Delta x}{2y_H}\right)} \right]}+i\frac{\pi}{2}\, .
\end{aligned}
\ee

\begin{figure}[t]
\centering
    \begin{overpic}[width=.3\textwidth]{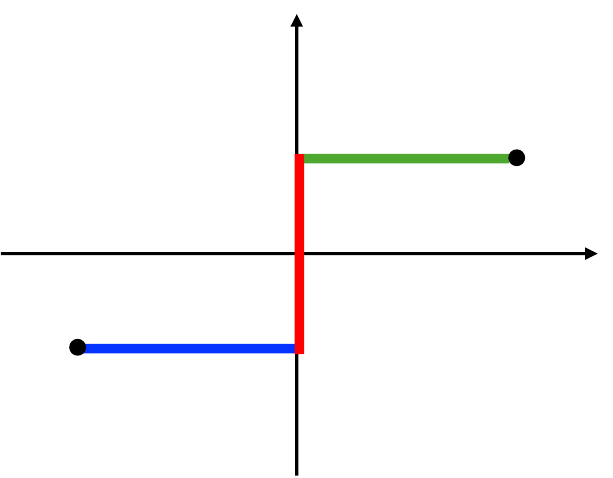}
    \put(4,11){$-\lambda_*$}
    \put(81,57){$\lambda_*$}
    \put(82,27){$\operatorname{Re}\lambda$}
    \put(41,80){$\operatorname{Im}\lambda$}
    \put(52,18)  {$-\frac{i\pi}{2}$} 
    \put(38,50) {$\frac{i\pi}{2}$}
    \end{overpic}
     \hspace{.02\textwidth}
    \begin{overpic}[width=0.3\textwidth]{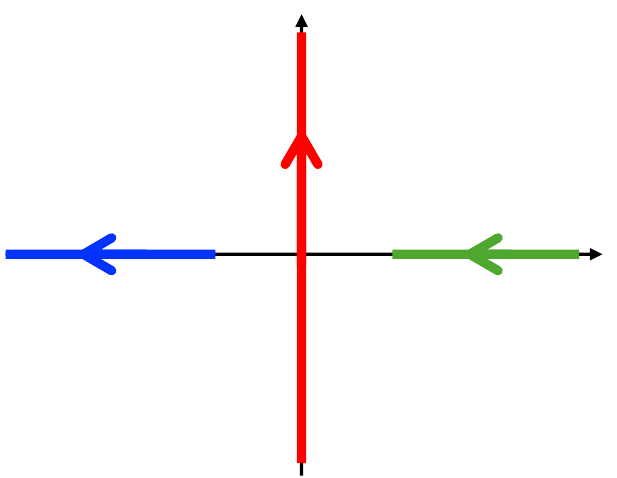}
    \put(83,25){$\operatorname{Re}t$}
    \put(41,80){$\operatorname{Im}t$}
    \put(25,25){$-\frac{\Delta t}{2}$}
    \put(60,25){$\frac{\Delta t}{2}$}
    \end{overpic}
    \hspace{.02\textwidth}
    \begin{overpic}[width=0.3\textwidth]{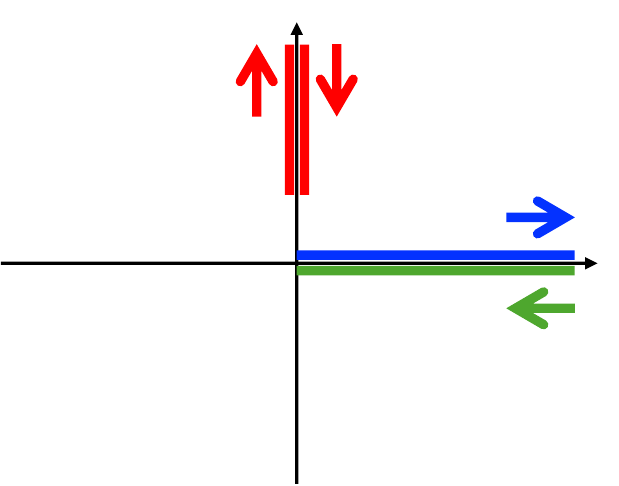}
    \put(82,18){$\operatorname{Re}y$}
    \put(41,80){$\operatorname{Im}y$}
    \put(31,45){$i\frac{\Delta t}{2}$}
    \end{overpic}
\caption{Profile of the complex geodesic in Poincar\'e AdS$_3$ for a purely timelike boundary interval of size $\Delta t$. (Left) One possible choice of  contour in the complex $\lambda$ plane connecting the regulated endpoints $\lambda=\pm\lambda_*$, separated by an imaginary shift of $i\pi$.  (Middle) The corresponding profile in the complex $t$ plane. Starting from one endpoint, $t$ evolves along the real axis to infinity, then follows the imaginary direction at vanishing real part before coming back along the real axis to the second endpoint. (Right) The corresponding profile for the holographic coordinate $y$. After departing from the real boundary at $y=0$ and reaching the deep bulk, $y\to\infty$, the geodesic continues along the imaginary axis, reaches a turning point at $y= i \Delta t/2$, and finally comes back crossing through infinity to the real boundary. For simplicity, we choose a piecewise contour. Picking instead a smooth $\lambda$ contour yields a smooth geodesic profile.}
\label{fig:Pgeod}
\end{figure}
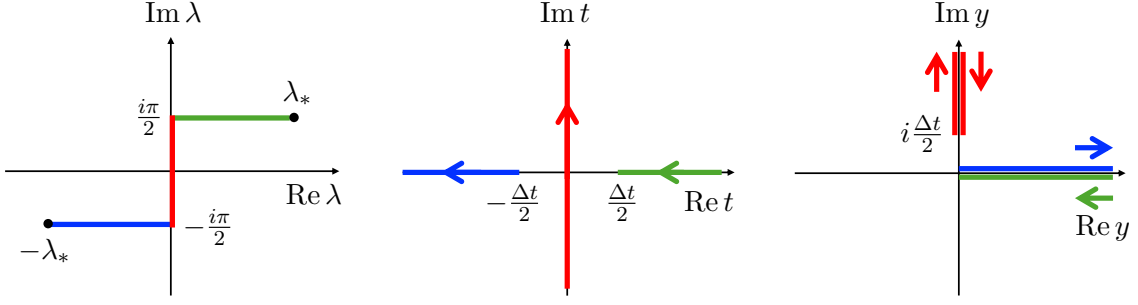

The case of empty AdS$_3$ in planar coordinates can be obtained from the previous one by sending $y_H\to \infty$, which immediately gives the geodesic profile
\begin{equation} \label{eq:spacelikegeo}
\begin{aligned}
t(\lambda) &= \frac{\Delta t}{2}\tanh\lambda \\[.5em]
x(\lambda) &= \frac{\Delta x}{2}\tanh\lambda \\[.5em]
y(\lambda) &= \frac{1}{2}\sqrt{(\Delta x)^2-(\Delta t)^2}\,\operatorname{sech}\,\lambda\,,
\end{aligned}
\end{equation}
with
\be
\lambda_* = \frac{1}{2} \log\!\left[ \frac{-\Delta t^2 + \Delta x^2}{\epsilon^2}\right]\,.
\ee 
Again, for timelike separations $\Delta x^2-\Delta t^2<0$ the affine parameter $\lambda$ needs to be complex, and so is the geodesic \cite{Balasubramanian:2012tu},
\be
\lambda_*  = \lim_{\vareps \to 0} \frac{1}{2}\log\!\left[ \frac{-(\Delta t- i \varepsilon)^2 + \Delta x^2}{\epsilon^2}\right] =   \frac{1}{2}\log\!\left[ \frac{\Delta t^2 -  \Delta x^2}{\epsilon^2}\right] + i \frac{\pi}{2} \,.
\ee
In both cases, $\lambda$ follows a contour in the complex plane connecting $-\lambda_*  $ to $\lambda_*$. These are $i \pi$ apart in the imaginary direction, as explained in \cite{Heller:2024whi}. The corresponding geodesic now connects real boundary points through a complex AdS$_3$ geometry. Although the geodesic is any smooth path in the complex plane between $-\lambda_*$ and $\lambda_*$, for illustration purposes in figure \ref{fig:Pgeod} we consider a purely timelike separation ($\Delta x=0$) and follow as a proxy a sample piecewise path, as was done in \cite{Heller:2024whi}. 
The general profile can be described as a smoothed out version of this particular case, where the geodesic emanates from the boundary points, reaches the past and future Poincar\'e horizons along two real branches and connects in a closed trajectory along a purely imaginary branch. 

\begin{figure}
\centering
\includegraphics[width=.7\textwidth]{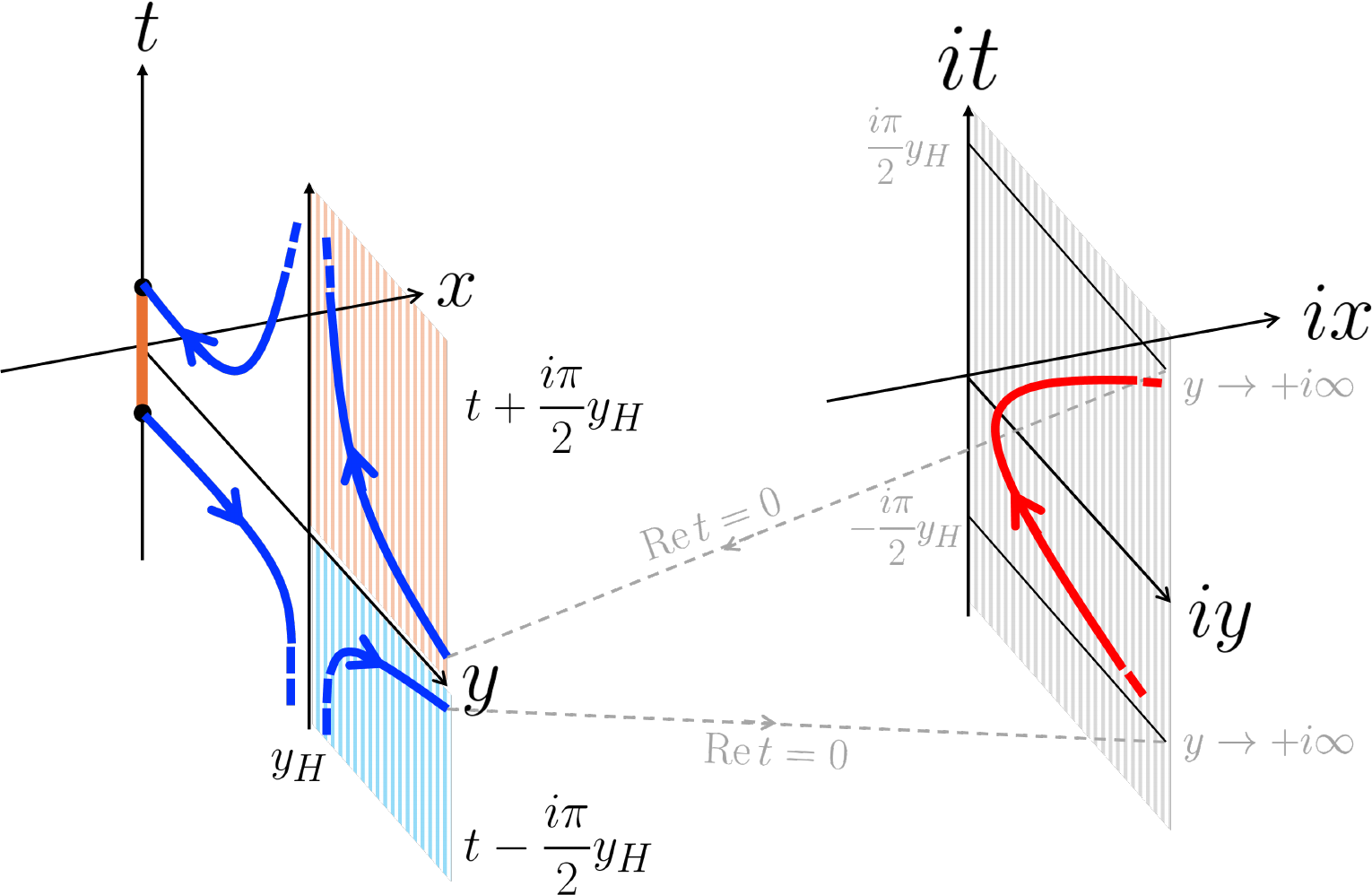}
\caption{Sketch of the timelike geodesic for the black brane in AdS$_3$ with the same choice of boundary region and path as in figure \ref{fig:Pgeod}. After reaching the deep bulk $y\to\infty$ along the real section of the spacetime, the geodesic moves in the imaginary-time and imaginary-radial directions, compare with figure \ref{fig:Pgeod}. Furthermore, note that time diverges at the event horizon $y_H$ and acquires an imaginary shift in the black hole interior $y>y_H$, as it usually happens when extending Poincaré coordinates to this region. Adapted from \cite{Heller:2024whi}.}
\label{fig:BBgeod}
\end{figure}

We can visualize the geodesic profile in the black brane case in an equivalent manner. Again for a purely timelike separation and choosing the same path as in figure \ref{fig:Pgeod}, this leads to the trajectory represented in figure \ref{fig:BBgeod}. The main difference from the vacuum case is a divergence of the time coordinate at the horizon and an additional imaginary shift in the interior, which is well-known and expected from the properties of static black hole metrics.

%%%%%%%%%%%%%%%%%%%%%%%%%%%%%%%%%%%%%%%%%%
%%%%%%%%%%%%%%%%%%%%%%%%%%%%%%%%%%%%%%%%%%
\subsubsection{Geodesics in Euclidean AdS$_3$ for generic insertions} \label{app:Banados}

We report here the case of general insertions in Euclidean AdS$_3$ Poincar\'e, which is employed in evaluating the timelike R\'enyi entropies in the auxiliary complex AdS$_3$ geometry of section \ref{sec:timelike_Renyi}. 
All the results that follow can be obtained from the embedding formalism as described before with a standard identification of coordinates leading to the metric
\be
ds^2 = \frac{dw d\bar w+ du^2}{u^2}\, .
\ee
The geodesic connecting  generic points $(w_i, \bw_i, u_i )$ and $(w_f, \bw_f , u_f )$
can be written as 
\be
\begin{aligned}
w(\lambda) &= \frac{1}{2}\[  w_f + w_i + \frac{\sqrt{(\Delta w^2 + (u_f - u_i)^2 )(\Delta w^2 + (u_f + u_i)^2)}}{\bar w_f -\bar w_i} \tanh \lambda + \frac{u_f^2 - u_i^2}{\bar w_f - \bar w_i}\] \\[.5em]
\bar w(\lambda) &= \frac{1}{2}\[  \bar w_f + \bar w_i + \frac{\sqrt{(\Delta w^2 + (u_f - u_i)^2 )(\Delta w^2 + (u_f + u_i)^2)}}{w_f - w_i} \tanh \lambda + \frac{u_f^2 - u_i^2}{ w_f - w_i}\]  \label{eq:wbfull}\\[.5em]
u(\lambda) &= \frac{\sqrt{(\Delta w^2 + (u_f - u_i)^2 )(\Delta w^2 + (u_f + u_i)^2)}}{2 \Delta w \cosh \lambda}\, ,
\end{aligned}
\ee
where
\be
\begin{aligned}
\lambda_f &= \operatorname{arcsinh} \(\frac{\Delta w}{2u_f} - \frac{u_f^2 - u_i^2}{2 u_f \Delta w}\) \\[.5em]
\lambda_i  &= - \operatorname{arcsinh} \(\frac{\Delta w}{2u_i} + \frac{u_f^2 - u_i^2}{2 u_i \Delta w}\)\,,
\end{aligned}
\ee
and we have defined 
\be
\Delta w  \equiv \sqrt{w_f - w_i} \sqrt{\bar w_f -\bar w_i}\,. 
\ee
The corresponding geodesic length is given by
\be
\begin{aligned} \label{eq:RobertsAdS}
{\mathcal L} = \log \left[ \frac{\Delta w^2 + \sqrt{(\Delta w^2 + (u_f - u_i)^2 )(\Delta w^2 + (u_f + u_i)^2)}}{2 u_f u_i}+ \frac{u_f}{2 u_i} + \frac{u_i}{2 u_f}  \right] \,.
\end{aligned}
\ee
%

%%%%%%%%%%%%%%%%%%%%%%%%%%%%%%%%%%%%%%%%%%%%%%%%%%%%%
%%%%%%%%%%%%%%%%%%%%%%%%%%%%%%%%%%%%%%%%%%%%%%%%%%%%%
\subsection{Asymptotically global AdS$_3$ spacetimes} \label{app:AAdS}
 
In this appendix, we review the computation of complex geodesics in the family of asymptotically AdS$_3$ geometries described by the metric
\begin{equation}
 ds^2=- \( \alpha^2 + r^2\) \,dt^2+\frac{dr^2}{\alpha^2 + r^2}+r^2d\phi^2\, .
\end{equation}
When $\alpha=1$, the metric reduces to global AdS$_3$. When $0<\alpha<1$, it represents a conical defect geometry with angular deficit $2\pi (1 - \alpha)$.
When $\alpha$ is purely imaginary, the geometry becomes the BTZ black hole, with inverse temperature $\beta=2\pi/|\alpha|$.

The metric is induced from the following embedding coordinates:
\be
\begin{aligned}
    X^0 &= \sqrt{1 + \frac{r^2}{\alpha^2}}\,\sin\!\left(\alpha t\right) \\[.5em]
    X^1 &= \frac r \alpha \,\sin(\alpha \phi) \\[.5em]
    X^2 &= \frac r \alpha \,\cos(\alpha \phi) \\[.5em]
    X^D &= \sqrt{1 + \frac{r^2}{\alpha^2}}\,\cos\!\left(\alpha t\right)\,.
\end{aligned}
\ee
The inverse coordinate relations are
\be
\begin{aligned}
    t&=\frac{1}{\alpha} \arctan \frac{X^0}{X^D} \\[.5em]
    \phi&=\frac{1}{\alpha} \arctan \frac{X^1}{X^2}\\[.5em]
    r&= \alpha\sqrt{(X^1)^2+(X^2)^2}\,.
\end{aligned}
\ee
A geodesic connecting two generic real regulated boundary points $(t_i,\phi_i,r_i)$ and $(t_f,\phi_f,r_f)$ admits the parametrization
\be
\begin{aligned}
    \tan\!\( \alpha \, t(\lambda) \)&= \frac{e^{2\lambda} \, r_f \sin\!\left(t_f \alpha\right) + r_i \sin\!\left(t_i \alpha\right)}{e^{2\lambda} \, r_f \cos\!\left(t_f \alpha\right) + r_i \cos\!\left(t_i \alpha\right) }\\[.5em]
   \tan\!\( \alpha \,   \phi(\lambda) \) &=  \frac{e^{2\lambda} \, r_f \sin(\alpha \phi_f) + r_i \sin(\alpha \phi_i)}{e^{2\lambda} \, r_f \cos(\alpha \phi_f) + r_i \cos(\alpha \phi_i)}\\[.5em]
     r(\lambda) &= \alpha
\sqrt{\frac{e^{2\lambda} r_f^2 + e^{-2\lambda} r_i^2 + 2 r_f r_i \cos(\alpha \Delta \phi) }{2 r_f r_i \left(\cos(\alpha \Delta t) - \cos(\alpha \Delta \phi) \right)}}\,.
\end{aligned}
\ee
Here we keep the radial positions of the endpoints generic, while assuming that they lie close to the boundary, $r_{i,f}\sim 1/\epsilon$, and define  $\Delta \phi = \phi_f-\phi_i$ and  $\Delta t = t_f-t_i$. Using the symmetric parametrization $\lambda_f=-\lambda_i=\lambda_*$, the boundary endpoints are reached at
\be
\lambda \to \pm \lambda_* = \pm \frac{1}{2}\log\!\left[\frac{2 r_i r_f}{\alpha^2}
\left(\cos(\alpha\Delta t)-\cos(\alpha\Delta\phi)\right)\right]\,.
\ee 
The corresponding regulated geodesic length is
\begin{equation}
    \mathcal{L} =
    \log\!\left[\frac{ 2 r_i r_f}{ \alpha^2}\left(\cos(\alpha \Delta t)-\cos(\alpha \Delta \phi)\right)\right] \, .
\end{equation}
For timelike-separated boundary points, the geodesics become complex, leading to a potential ambiguity in their imaginary part associated with the logarithm. This is resolved by adopting the $i\varepsilon$ prescription that matches the one used for the dual correlator, as described in the main text.

Moreover, as we discuss below, the conical defect and BTZ geometries admit families of geodesics with non-trivial winding around the angular direction. These winding geodesics are obtained by formally replacing the boundary angular separation by $\Delta\phi\rightarrow\Delta\phi-2\pi m$, with $m\in\mathbb{Z}$.
 
\begin{figure}[t]
    \begin{overpic}[width=.3\textwidth]{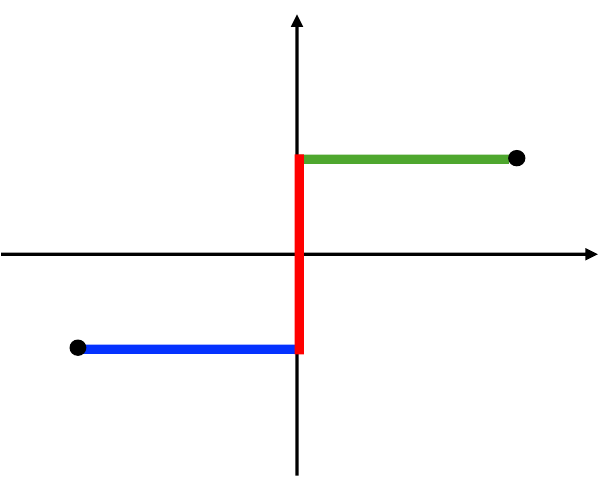}
    \put(4,11){$-\lambda_*$}
    \put(81,57){$\lambda_*$}
    \put(80,27){$\operatorname{Re}\lambda$}
    \put(41,80){$\operatorname{Im}\lambda$}
    \put(52,18){$-\frac{i\pi}{2}$}
    \put(38,50){$\frac{i\pi}{2}$}
    \end{overpic}
   \hspace{.02\textwidth}
    \begin{overpic}[width=0.3\textwidth]{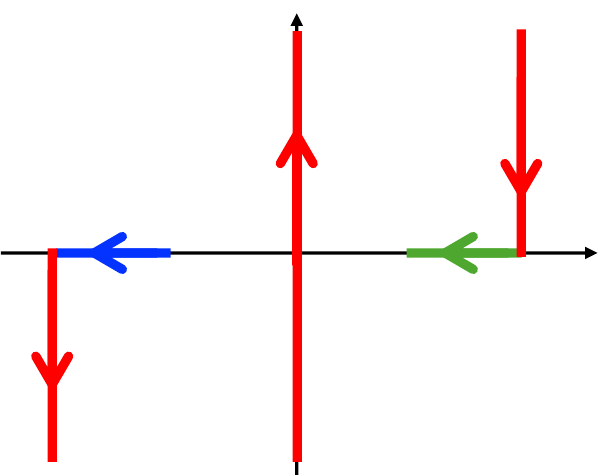}
    \put(80,25){$\operatorname{Re}t$}
    \put(41,80){$\operatorname{Im}t$}
     \put(0,40){ {$-\pi/2$}}
    \put(21,25){$-\frac{\Delta t}{2}$}
    \put(62,25){$\frac{\Delta t}{2}$}
     \put(90,40){{$\pi/2$}}
    \end{overpic}
    % %
        \hspace{.02\textwidth}
    \begin{overpic}[width=0.3\textwidth]{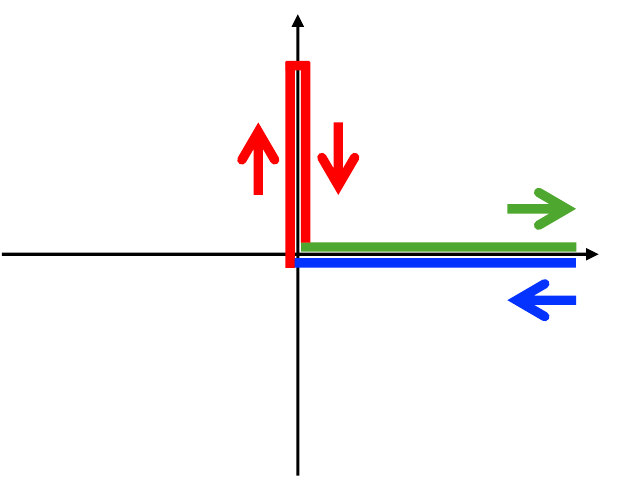}
    \put(80,20){$\operatorname{Re}r$}
    \put(41,80){$\operatorname{Im}r$}
    \put(55,65){\footnotesize $i\sqrt{1+\cot^2\!\frac{\Delta t}{2}}$}
    \end{overpic} 
 \caption{Profile of the complex geodesic dual to TEE in global AdS$_3$ for $\Delta t<2\pi$, corresponding to the piecewise contour in the complex $\lambda$ plane shown in the left panel. The profile is analogous to that in planar coordinates (figure \ref{fig:Pgeod}). The main difference is that, after departing from the interval endpoints, the time coordinate reaches the boundary of the causal diamond at $t=\pm\pi /2$ and then evolves along a purely imaginary direction. The piecewise behavior reflects the chosen contour in the complex $\lambda$ plane; a smooth contour would instead yield a smooth geodesic profile.}
 \label{fig:Ggeod}
\end{figure}
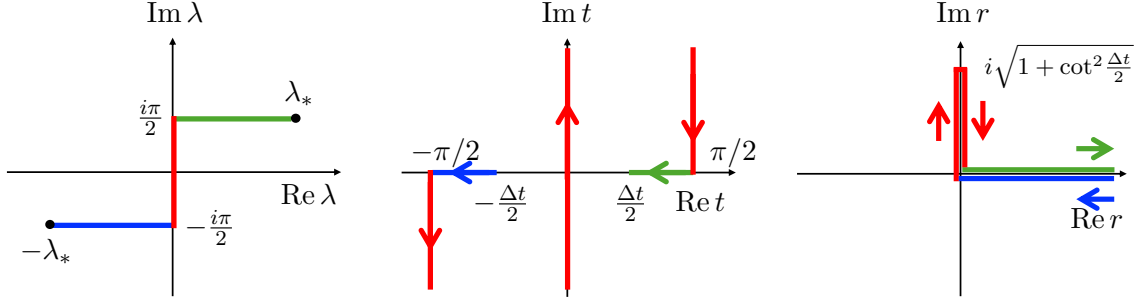

\begin{figure}
\centering
    \vspace{.5cm}
    \begin{overpic}[width=.25\textwidth]{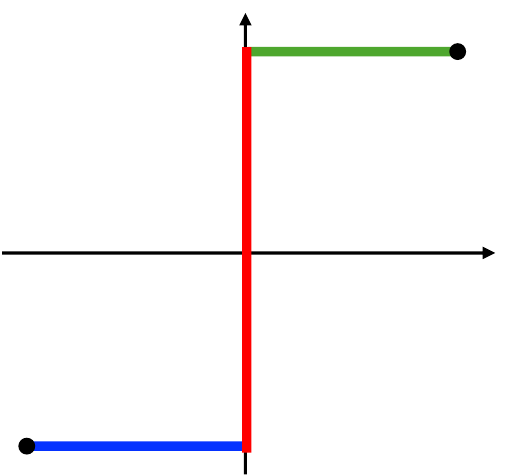}
    \put(0,11){$-\lambda_*$}
    \put(86,70){$\lambda_*$}
    \put(85,33){$\operatorname{Re}\lambda$}
    \put(41,95){$\operatorname{Im}\lambda$}
    \put(52,3){$-\frac{ik\pi}{2}$}
    \put(32,80){$\frac{ik\pi}{2}$}
    \end{overpic}
    \hspace{.03\textwidth}
    \begin{overpic}[width=0.33\textwidth]{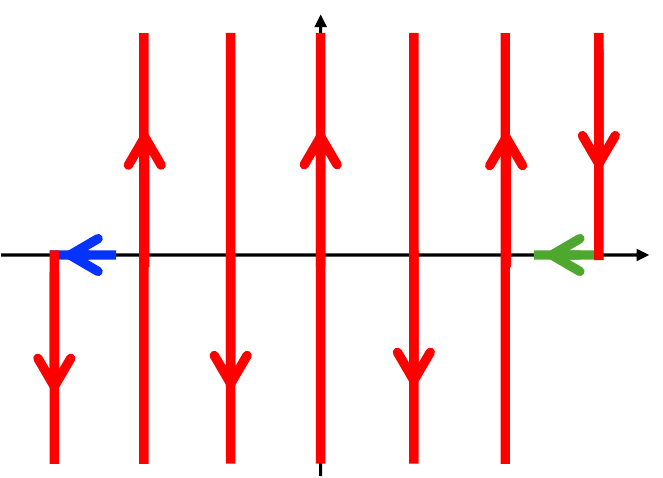}
    \put(95,36){$\operatorname{Re}t$}
    \put(43,73){$\operatorname{Im}t$}
    \put(-1,40){$-\frac{\Delta t}{2}$}
    \put(86,21){$\frac{\Delta t}{2}$}
    \end{overpic}
    \hspace{.03\textwidth}
    \begin{overpic}[width=0.27\textwidth]{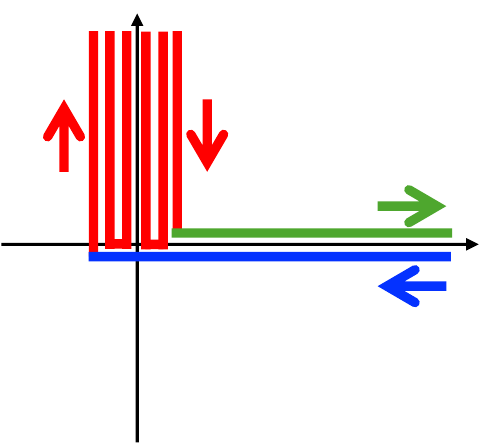}
    \put(95,44){$\operatorname{Re}r$}
    \put(21,91){$\operatorname{Im}r$}
    \end{overpic}
 \caption{Complex geodesic profile dual to TEE in global AdS$_3$ for a time separation $(k-1)\pi<\Delta t<(k+1)\pi$, with $k$ odd for $\Delta \phi=0$. The  piecewise contour of $\lambda$  connects the endpoints at $\lambda=\pm\lambda_*$, now separated by an imaginary shift of $ik\pi$. Compared to figure \ref{fig:Ggeod}, the geodesic repeatedly traverses successive imaginary branches before reaching the second endpoint. Consecutive branches in $t$ are separated by $\pi/2$ in real time and the radial coordinate traverses the positive imaginary axis. For  illustration purposes we set $k=3$.
 } \label{fig:Ggeod2}
\end{figure}
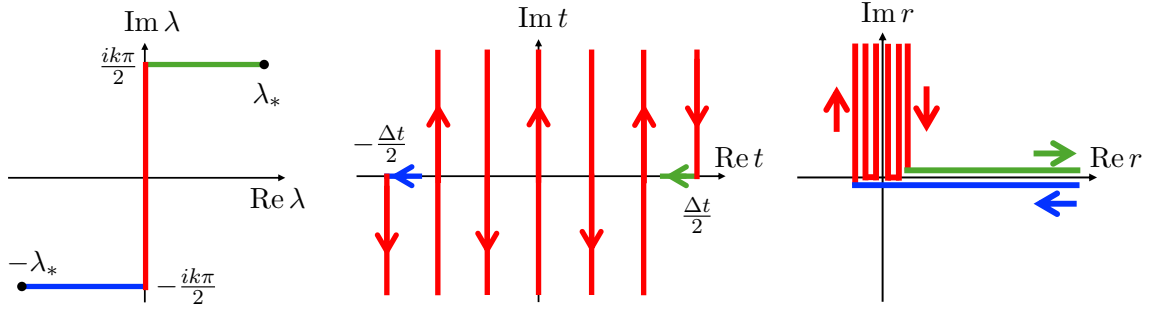

%%%%%%%%%%%%%%%%%%%%%%%%%%%%%%%%%%
\paragraph{Global AdS$_3$} Setting $\alpha=1$ reproduces global AdS$_3$. We consider symmetric boundary endpoints $t_f=-t_i=\Delta t/2$, $\phi_f=-\phi_i=\Delta\phi/2$, with radial regulator $r_{i,f}=1/\epsilon$. The geodesic profile then simplifies to
\begin{equation}  \label{eq:GADSgeo}
\begin{aligned}
\tan t(\lambda) &=\tan\!\left(\frac{\Delta t}{2}\right)\tanh\lambda\\[.5em]
\tan \phi(\lambda) 
&=\tan\!\left(\frac{\Delta \phi}{2}\right) \tanh\lambda \\[.5em]
r(\lambda) &=\sqrt{\frac{\cos \Delta \phi + \cosh 2\lambda}
{\cos \Delta t - \cos \Delta \phi}} \,,
\end{aligned}
\end{equation}
with the endpoints reached at
\be
\lambda \to \pm \lambda_* = \pm \frac{1}{2}\,\log\!\left[\frac{2}{\epsilon^2}\left(\cos\Delta t-\cos{\Delta \phi}\right)\right] \,.
\ee 
For spacelike-separated boundary points this reproduces the usual real geodesics. For timelike separations, instead, the affine parameter becomes complex. The corresponding branch of the logarithm is fixed by implementing the $i\varepsilon$ prescription, which gives
\be \label{lstarglobal}
\begin{aligned}
\lambda_* & = \lim_{\vareps \to 0}  \frac{1}{2}\,\log\!\left[\frac{2}{\epsilon^2}\left(\cos\Delta t-\cos\Delta\phi+ i\,\varepsilon\,\sin \Delta t\right)\right] \\[.5em]
&=    \frac{1}{2}\,\log\!\left[ \frac{2}{\epsilon^2}\left|\cos\Delta t-\cos{\Delta \phi}\right|\right] + i \frac{\pi}{2}k\,,
\end{aligned}
\ee
with %
\be\label{eq:app:kglob}
k =
\left\lfloor \frac{\Delta t-\Delta \phi}{2\pi} \right\rfloor + \left\lfloor \frac{\Delta t+\Delta \phi}{2\pi} \right\rfloor + 1 \, .
\ee
To understand the geometrical origin of the imaginary part, consider first the simplest example with $\Delta t<2\pi$ and $\Delta\phi=0$.
As for the Poincar\'e case, $\lambda$ follows a contour in the complex plane connecting $-\lambda_*  $ to $\lambda_*$ and these are $i \pi$ apart in the imaginary direction. Choosing the same contour as in the planar case leads to an analogous geometric picture, shown in figure \ref{fig:Ggeod}.
For $\Delta t>2\pi$ and $\Delta\phi=0$, the same contour gives $\lambda_*$ in \eqref{lstarglobal} an imaginary part  $i\pi k/2$, with odd $k$ for purely timelike separations. 
The corresponding picture is analogous to figure \ref{fig:Ggeod}, except that the contour in the complex $\lambda$ plane now spans a larger imaginary displacement. As a consequence, the geodesic repeatedly traverses successive complex branches before reaching the second boundary endpoint, as illustrated in figure \ref{fig:Ggeod2}. 
For $r$ this always happens at vanishing real part. 
For the time $t$, once $t \to - i \infty $ has been reached along $\text{Re}\,t= - \pi k /2$, the geodesic re-emerges from $t=- i \infty$ along $\text{Re}\,t = -\pi(k-1) /2$ and continues up to $t \to  i \infty $ and so forth, with $2k$ shifts of $\pi/2$ in total,  until the geodesic re-emerges from $t=+i\infty$ at $\text{Re}\,t = \pi k /2$.
Choosing a different branch of the logarithm would amount to selecting a different value of $k$, which is not continuously connected to the spacelike geodesic through the $i\varepsilon$ prescription. This is most easily visualized by adopting a smooth parametrization in the complex $\lambda$ plane.
%
 
%%%%%%%%%%%%%%%%%%%%%%%%%%%%%%
\paragraph{Conical defect} The conical defect geometry $0<\alpha<1$ parallels the global AdS$_3$ one with two important distinctions. The  periodicity in  \eqref{eq:GADSgeo} is now rescaled with $\alpha$. 
 Furthermore, there are now winding geodesics labeled by an integer $m$, the number of (oriented) windings of the geodesic around the conical defect. These are obtained geometrically by considering angular separations on the boundary of the form $\Delta \phi-2\pi m$, for any $m\in \mathbb{Z}$. Explicitly,
\be \label{eq:CDgeod}
\begin{aligned}
\tan ( \alpha\,  t(\lambda) ) &=\tan\!\left(\frac{\alpha\Delta t}{2}\right)\tanh\lambda\\[.5em]
\tan ( \alpha \, \phi(\lambda) ) &=\tan\!\left(\frac{\alpha\,(\Delta\phi-2\pi m)}{2}\right) \tanh\lambda \\[.5em]
r(\lambda) &=\alpha\sqrt{\frac{\cos(\alpha\,(\Delta\phi-2\pi m)) + \cosh 2\lambda}
{\cos \alpha\Delta  t - \cos(\alpha\,(\Delta\phi-2\pi m)) }} \,,
\end{aligned}
\ee
and 
\be
\lambda_*=\frac{1}{2}\log\left[\frac{2}{\alpha^2\epsilon^2}\left|-\cos\left(\alpha\,\Delta t\right)+\cos\left(\alpha\,(\Delta\phi-2\pi m)\right)\right|\right]+i\frac{\pi}{2}k_{\alpha,m}\,,
\ee
where $k_{\alpha,m}$ is the analogue of $k$ in \eqref{eq:app:kglob}, but rescaled by the parameter $\alpha$ and including the winding number $m$,
\be
k_{\alpha,m} =
\left\lfloor \frac{\alpha\,(\Delta t-\Delta \phi+2\pi m)}{2\pi} \right\rfloor + \left\lfloor \frac{\alpha\,(\Delta t +\Delta \phi-2\pi m)}{2\pi} \right\rfloor + 1 \,.
\ee
The dependence on the winding number thus enters both through the real part of the geodesic length and through the integer $k_{\alpha,m}$ fixing the appropriate logarithmic branch.
Notice also that when $\alpha$ is rational, only finitely many windings correspond to geometrically distinct geodesics, since larger windings are identified by the orbifold action.

%%%%%%%%%%%%%%%%%%%%%%%%%%%%%%
\paragraph{BTZ black hole}   In this case, $\alpha^2  =  - (2\pi / \beta)^2$ with $\beta>0$. The periodic dependence of the geodesic profile and of the corresponding length is lost, but there still exist infinitely many geodesics connecting the boundary points. This is seen directly from the explicit expressions for the geodesic profile
\be \label{eq:BTZgeod}
\begin{aligned}
\tanh \(\frac{2\pi}{\beta} t(\lambda) \)&=\tanh\!\left(\frac{\pi }{\beta}\Delta t\right)\tanh\lambda\\[.5em]
\tanh \( \frac{2\pi }{\beta} \phi(\lambda)\) &=\tanh\!\left( \frac{\pi }{\beta} (\Delta\phi-2\pi m)\right) \tanh\lambda \\[.5em]
r(\lambda) &= \frac{2\pi }{\beta}\sqrt{\frac{\cosh\( \frac{2\pi }{\beta} \,\(\Delta\phi-2\pi m\)\) + \cosh \( 2\lambda\)}
{\cosh\(\frac{2\pi }{\beta} \,\(\Delta\phi-2\pi m\)\) -\cosh \(\frac{2\pi }{\beta} \Delta  t\) }}\,,
\end{aligned}
\ee
and is reflected in the analytic structure of the endpoint value
\be
\begin{aligned}
\lambda_{*} =& \frac{1}{2}\log{\left[\frac{\beta^2}{2\pi^2\epsilon^2}\left|-\cosh{\frac{2\pi }{\beta}\,\Delta t}+\cosh{\frac{2\pi }{\beta}\,(\Delta\phi-2\pi m)}\right|\right]} \\[.5em]
&\hspace{6cm}+ i \frac{\pi }{2} \,\Theta\!\left[ \Delta t^2- (\Delta \phi- 2\pi m)^2 \right]\,,
\end{aligned}
\ee
whose imaginary part remains constant throughout the timelike regime, \ie as long as $\Delta t^2-(\Delta \phi-2\pi m)^2>0$.

\begin{figure}
    \centering
    \begin{overpic}[width=.3\textwidth]{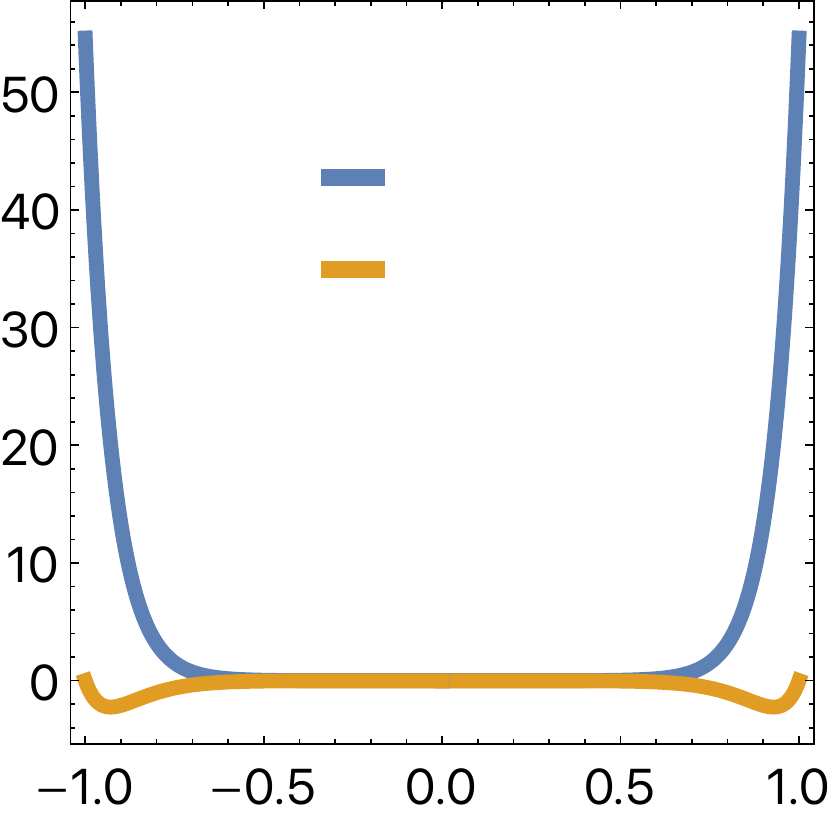}
    \put(50,-10){$\lambda'$}
    \put(50,75){\small $\operatorname{Re}r(\lambda')$}
    \put(50,64){\small $\operatorname{Im}r(\lambda')$}
    \end{overpic}
    \hspace{.015\linewidth}
    \begin{overpic}[width=.302\textwidth]{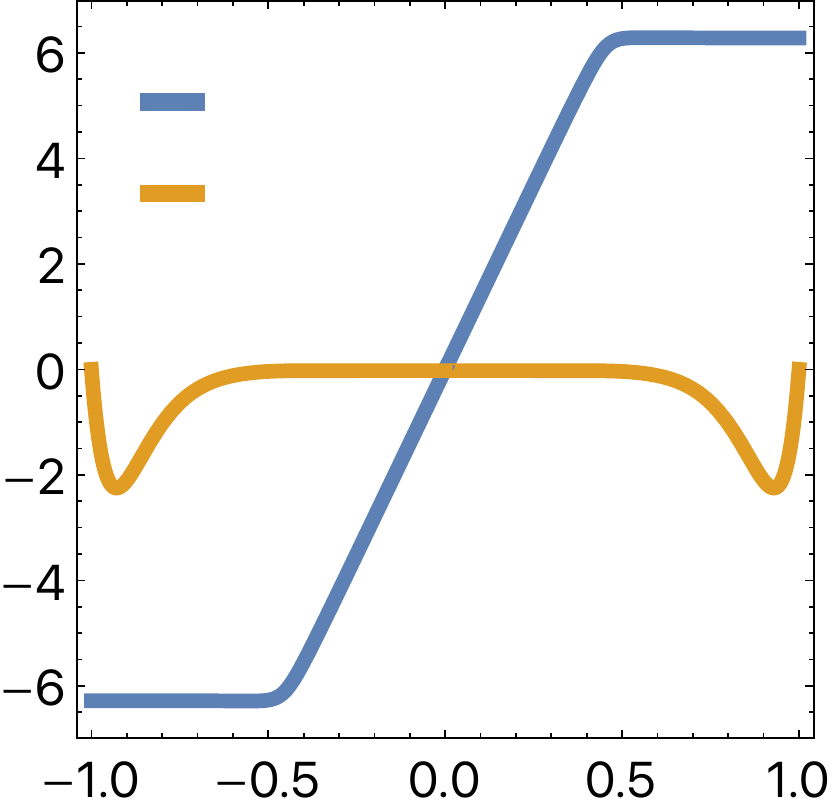}
    \put(50,-10){$\lambda'$}
    \put(28,83){\small $\operatorname{Re}\phi(\lambda')$}
    \put(28,72){\small $\operatorname{Im}\phi(\lambda')$}
    \end{overpic}
    \hspace{.015\linewidth}
    \begin{overpic}[width=.312\textwidth]{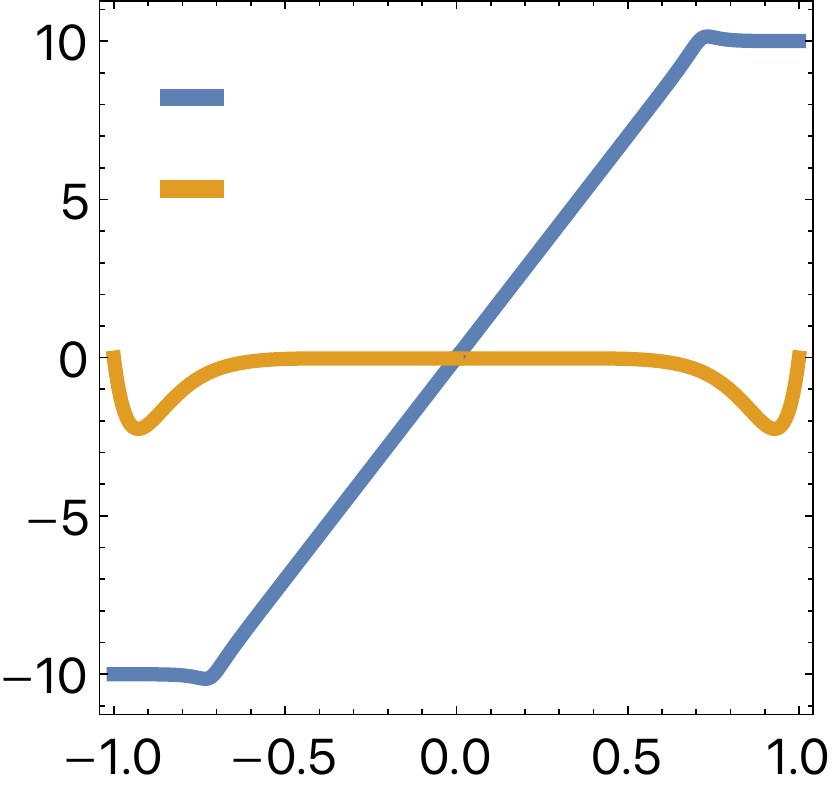}
    \put(51,-9){$\lambda'$}
    \put(30,81){\small $\operatorname{Re}t(\lambda')$}
    \put(30,70){\small $\operatorname{Im}t(\lambda')$}
    \end{overpic}
    \hspace{.015\linewidth}
    \vspace{.15cm}
    \caption{Complex geodesic in the BTZ black hole ($\beta =1$) connecting timelike-separated points on the boundary over an effective angular separation $4\pi$: $\Delta t=20$, $\Delta \phi=0$ and $m=2$. Here, $\lambda=\lambda_*\lambda'$ ($\lambda'$ moves along a straight line connecting $\pm \lambda_*$) and we plot in terms of the parameter $\lambda'\in(-1,1)$ to capture the entire geodesic. The boundary anchoring points at $\lambda'=\pm 1$ are real. At the turning point $\lambda'= 0$, $t$ and $\phi$ vanish, while the radial coordinate, despite being numerically small in the plot, is purely imaginary.}
    \label{fig:geoWind2}
\end{figure}
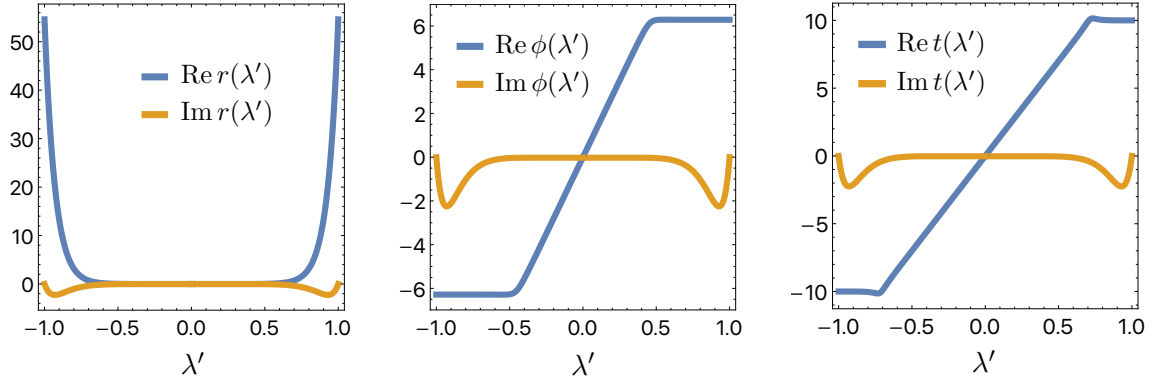

The role of winding geodesics in the timelike regime deserves a brief comment. We illustrate the discussion using the BTZ geometry, although the same considerations apply to the conical defect.
Suppose the boundary points are timelike separated in the $m=0$ sector, namely $\Delta t^2-\Delta\phi^2>0$. Since increasing $|m|$ increases the effective angular separation, there always exist sufficiently large values of $|m|$ for which $\Delta t^2-(\Delta\phi-2\pi m)^2<0$ and the corresponding geodesics are therefore real and locally spacelike. By contrast, for winding numbers satisfying $\Delta t^2-(\Delta\phi-2\pi m)^2>0$ the geodesics are complex, and the parameter connects the endpoints $\pm\lambda_*$ with an imaginary shift of $i\pi$.
In this case, although the real part of the angular coordinate still spans an angular interval of $2\pi m$, the usual geometric interpretation in terms of a real geodesic winding around the event horizon no longer applies, since the trajectory lies on a complexified section of the spacetime rather than on the real Lorentzian geometry. An example of such a complex winding geodesic with $m=2$ is shown in figure \ref{fig:geoWind2}.

%%%%%%%%%%%%%%%%%%%%%%%%%%%%%%%%%%%%%%%%%%%%%%%%%%%%
\subsection{Boosted conical defect in planar AdS$_3$}\label{app:boosted_defect}

To obtain the boosted conical defect geometry dual to the local operator quench, we start from the conical defect geometry in global AdS$_3$ \cite{Horowitz:1999gf,Nozaki:2013wia,Asplund:2013zba},
\begin{equation} 
ds^2 = - (r^2+\alpha^2)d\ft^2 + \frac{dr^2}{(r^2+\alpha^2)} + r^2 d\phi^2,     
\end{equation}
with $0 < \alpha <1$, and use the map  
\be  \label{mapquench}
\begin{aligned}
r &= \frac{1}{\mu y} \sqrt{\mu^2 x^2 + \frac{1}{4}(-\mu^2+y^2-t^2+x^2)^2}, \\[.5em]
\tan \ft &= \frac{2\mu t}{\mu^2+y^2-t^2+x^2}   \\[.5em]
\tan\phi &= \frac{2\mu x}{\mu^2 - y^2+t^2-x^2}\,.  
\end{aligned}   
\ee
For $\alpha=1$ this maps global AdS$_3$ to Poincar\'e AdS$_3$, while for $0<\alpha<1$ it produces the boosted conical defect geometry dual to the local operator quench.

Rather than solving the geodesic equations directly in the boosted geometry, we obtain the geodesics by mapping those of the global conical defect geometry through \eqref{mapquench}. For boundary endpoints $(t_i,x_i,\epsilon)$ and $(t_f,x_f,\epsilon)$, the corresponding geodesics are obtained from the generic conical defect solution. For large radius, near-boundary insertions, the geodesic profile can be written as
\be
\begin{aligned} \label{eq:CDgeneric}
\tan(\alpha\phi) =&\frac{e^{2\lambda}r_f \sin(\alpha \phi_f) +r_i \sin(\alpha \phi_i)}
{e^{2\lambda}r_f \cos(\alpha \phi_f) +r_i \cos(\alpha \phi_i)} \\[6pt]
\tan(\alpha\ft)=&\frac{e^{2\lambda}r_f \sin\!\left(\alpha \ft_f\right) +r_i \sin\!\left(\alpha \ft_i\right)}
{e^{2\lambda}r_f \cos\!\left(\alpha \ft_f\right) +r_i \cos\!\left(\alpha \ft_i\right)} \\[6pt]
r(\lambda)=& \alpha \sqrt{ \frac{e^{2\lambda}r_f^2+e^{-2\lambda}r_i^2+2r_i r_f\cos(\alpha\Delta\phi) }{2r_i r_f\left(\cos(\alpha\Delta \ft)-\cos(\alpha\Delta\phi)\right)}}\,,
\end{aligned}
\ee
with the  parameter  $\lambda$  ranging between
\be \label{eq:lsquench}
\pm \lambda_*  = \frac{1}{2} \log \[
\frac{2 r_ir_f \left(\cos\!\left(\alpha \Delta \ft\right) - \cos(\alpha \Delta \phi  )\right)}
{\alpha^2 \ } 
\]\, ,
\ee
where $ \Delta \ft = \ft_f(t_f, x_f,\eps)- \ft_i(t_i, x_i,\eps)$, $\Delta \phi = \phi_f(t_f, x_f,\eps) -\phi_i(t_i, x_i,\eps)$, and $(\ft,\phi)$ obtained from the map \eqref{mapquench}.  As before, winding geodesics are generated by introducing the shift $\Delta \phi \to \Delta \phi - 2\pi m$.

Using the map \eqref{mapquench} one can get the length and the profile of the geodesic in the geometry describing the boosted conical defect in AdS$_3$ Poincar\'e.

We now specialize to the geodesic connecting the regulated timelike interval  $(t_i, x_0, \epsilon)$ and $(t_f, x_0, \epsilon)$, for $t_f > t_i > 0$. We restrict to the $m=0$ sector, which is the dominant saddle relevant for the TEE considered in the main text.  
Substituting \eqref{eq:CDgeneric} into \eqref{eq:lsquench} one finds
\be
\begin{aligned}
\mathcal{L} = 2 \lambda_* 
&=\log\!\left[ \frac{4 (t_f- t_i)^2 \sqrt{ab}\, \sinh\!\left(\frac{\alpha}{2}\log a\right)\sinh\!\left(\frac{\alpha}{2}\log b\right)}{(1- a) (1-b)\alpha^2\epsilon^2}\right]\,,
\end{aligned}
\ee
where we introduced the shorthand notation
\be
\begin{aligned}
a &= \frac{(t_f - x_0 - i\mu)(-t_i + x_0 - i\mu)}{(t_f - x_0 + i\mu)(-t_i + x_0 + i\mu)}\, , \\[.5em]
b &= \frac{(t_f + x_0 - i\mu)(t_i + x_0 + i\mu)}{(t_i + x_0 - i\mu)(t_f + x_0 + i\mu)} \, .
\end{aligned}
\ee

Writing  $\mathcal{L} = 2 \lambda_* = \log \mathcal{A}$, the quantity $\mathcal A$ admits the expansion
\begin{equation}
\mathcal{A} = \frac{1}{\epsilon^2}\left(\frac{\mathcal{A}_{-2}}{\mu^2} +\frac{\mathcal{A}_{-1}}{\mu} + \mathcal{A}_0 + \ldots\right) +  \ldots    
\end{equation}
Although $a,b\to1$ as $\mu\to0$, they approach this limit along different trajectories in the complex plane depending on the causal relation between the interval endpoints and the quench. One can check that $\mathcal{A}_{-2}$ is identically zero in any time regime while
\be
\begin{aligned}
\mathcal{A}_{-1} = & \frac{(t_f-t_i)\,|t_f^2-x_0^2|\,|t_i^2-x_0^2|\,\sin(\pi q \alpha)}{(t_f+x_0)(t_i+x_0)\,\alpha} \\[.5em]
\mathcal{A}_{0} = &\frac{(t_f-t_i)^2\,|t_f^2-x_0^2|\,|t_i^2-x_0^2|\,\cos(\pi q\alpha)}{(t_f^2-x_0^2)(t_i^2-x_0^2)}
\end{aligned}
\ee
with $q=0$ or $1$, corresponding to the two branches selected by the $i\varepsilon$ prescription.

Assuming $x_0\geq0$, there are two regimes to consider:
\begin{itemize}
\item[(I)] Both endpoints of the timelike interval are in the future ($t_i>x_0$) or in the past ($t_f < x_0$) of the lightcone emanating from $x=0$. In this case $q=0$, $\mathcal{A}_{-1}$ vanishes, and at leading order in $\mu \to 0$, the geodesic length is only controlled by $\mathcal{A}_0$, which simplifies to 
\begin{equation}
\mathcal{A}_0 = - (t_f-t_i)^2\,.     
\end{equation}
\item[(II)] The upper endpoint of the timelike interval is located in the future of the lightcone emanating from $x=0$ ($t_f > x_0$), while the lower endpoint is located in its past ($t_i < x_0$). In this case $q=1$, $\mathcal{A}_{-1}$ does not vanish and provides the leading contribution, which simplifies to
\begin{equation}
\mathcal{A}_{-1} = -\frac{(t_f-t_i)(t_f-x_0)(x_0-t_i)\sin (\pi \alpha)}{\alpha}\, .    
\end{equation}
$\mathcal{A}_0$ is also nonzero, but it only gives a subleading contribution to the geodesic length in this case. 
\end{itemize}

The resulting geodesic length is therefore
\begin{equation}
\displaystyle \mathcal{L} = \left\{
\begin{array}{ll}
\displaystyle 2 \log\frac{t_f-t_i}{\epsilon} + i \pi  &\qquad \text{if}~t_f < x_0~\text{or}~t_i > x_0\,,\\[.5em]
\displaystyle \log \frac{(t_f-t_i)(t_f-x_0)(x_0-t_i)\sin (\pi \alpha)}{\mu \alpha \epsilon^2} + i\pi &\qquad \text{if}~t_f > x_0~\text{and}~t_i < x_0\,, 
\end{array}  \right. 
\end{equation}
where the branch of the imaginary part has again been fixed by the same $i\varepsilon$ prescription used in the CFT computation.

 %%%%%%%%%%%%%%%%%%%%%%%%%%%%%%%%%%%%%%
\section{Geodesics in the thick-shell Vaidya spacetime}
\label{app:Vaidya}

To compute the holographic TEE in the thick-shell Vaidya geometry considered in section \ref{sec:ads3_vaidya} we consider the metric
\eqref{eq:Vaidyametric},
\begin{equation} 
    ds^2=\frac{1}{y^2} \left[-(1-m(v) y^2)\,dv^2- 2\,dv\,dy+ dx^2 \right] \,, 
\end{equation}
where the mass aspect function $m(v)$ corresponds to the time-dependent areal radius squared of the apparent horizon. This geometry represents the formation of a black brane from the gravitational collapse of a shell of null dust.

Let us consider the purely temporal segment extending between $v=t_1<0$ and $v=t_2>0$ at constant $x$ as our boundary subregion. The Lagrangian to be extremized in order to find the complex geodesics associated with this boundary subregion is  
\begin{equation}\label{L_Vaidya}
\mathscr{L} = \frac{1}{y(\lambda)}\sqrt{- [ 1- m(v(\lambda)) y(\lambda)^2 ]v'(\lambda)^2 - 2 v'(\lambda)y'(\lambda)},       
\end{equation}
with $\lambda$ an arbitrary parameter. 

Let us assume that $\lambda$ is a complexified affine parameter, and that initial and final boundary conditions are respectively imposed at $-\lambda_*$ and $\lambda_*$
\begin{equation}
y(-\lambda_*) = y(\lambda_*) = y_c, \quad v(-\lambda_*) = t_1,~v(\lambda_*) = t_2,     
\end{equation}
where $y_c \ll 1$ is a UV cutoff. Any trajectory in the complex $\lambda$-plane joining $-\lambda_*$ and $\lambda_*$ provides a valid section of the complex geodesic.
Let this trajectory be parameterized as 
\begin{equation}
\lambda = 
\lambda_* \tilde\lambda(\xi),
\end{equation}
where $\xi \in [0,1]$ and $\tilde\lambda(\xi)$ is a curve in the complex $\lambda$-plane with $\tilde{\lambda}(0)=-1$ and  $\tilde{\lambda}(1)=1$. The length functional becomes 
\begin{equation}
\mathcal{L} = \int d\lambda = \int d\xi \lambda_* \tilde{\lambda}'(\xi) = \int \mathscr{L} d\xi \, ,       
\end{equation}
and  we can trade $\lambda$ in \eqref{L_Vaidya} for $\xi$, and consider the equations
\begin{equation}
\begin{aligned}\label{eom_Vaidya}
\frac{d}{d\xi}& \frac{\delta \mathscr{L}}{\delta v'(\xi)} - \frac{\delta \mathscr{L}}{\delta v(\xi)} = 0\, \\[5pt] \mathscr{L} &= \lambda_* \tilde{\lambda}'(\xi) \, .  
\end{aligned}
\end{equation}
To solve these equations of motion, we impose the boundary conditions at $\xi = 0$, 
\begin{equation}
y(0) = y_c, \quad v(0) = t_1,     
\end{equation}
and employ a Newton-Raphson method to find $v'(0)$ and $\lambda_*$ such that 
\begin{equation}
y(1) = y_c, \quad v(1) = t_2.     
\end{equation}
The external information this method needs consists of the mass aspect function $m(v)$ and the $\tilde{\lambda}(\xi)$ trajectory in the complex $\lambda$-plane. 

\begin{figure}
    \centering
    \includegraphics[width=0.5\linewidth]{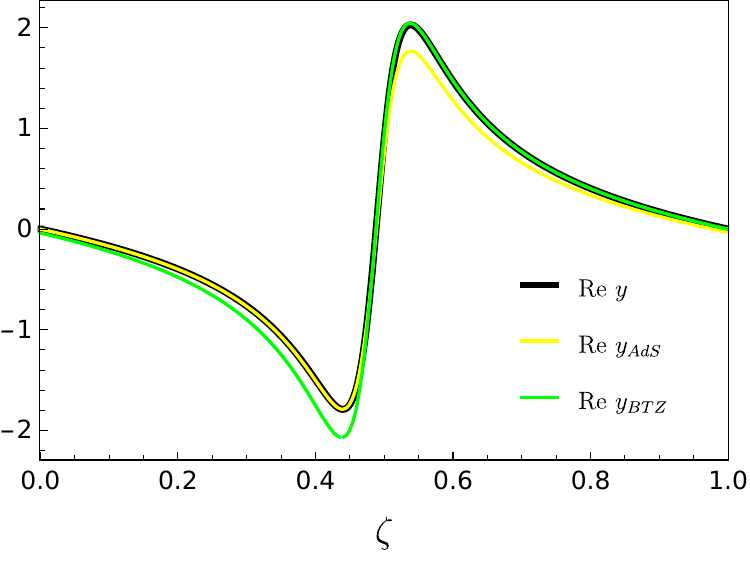}~    \includegraphics[width=0.5\linewidth]{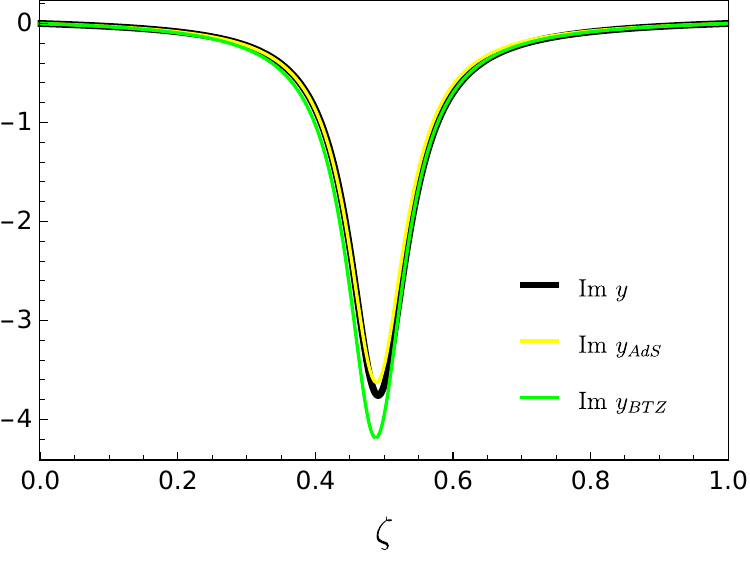}
    \caption{For a thick-shell spacetime with $m_0=1$, $\gamma=16$, boundary insertions $t_2=-t_1=0.75$, and path \eqref{path} in the complex $v$-plane with $\eta = 0.05$, we show the numerical solution of the geodesic equation $y(\zeta)$ (thick black lines), together with the prediction of the thin-shell matching computation described in section \ref{sec:ads3_vaidya}, both for the AdS piece $y_{\rm AdS}(\zeta)$ (yellow lines) and the BTZ piece $y_{\rm BTZ}(\zeta)$ (green lines). The left plot shows the real parts, while the right plot shows the imaginary ones.}
    \label{fig:comparison}
\end{figure}

For the computations presented in the main text, we chose as in \eqref{eq:massprofile} 
\begin{equation}
m(v)  = \frac{m_0}{2}\left(1+\tanh({\gamma v}) \right),  \end{equation}
such that the radius of the apparent horizon starts at zero at $v=-\infty$, ends at $m_0^\frac{1}{2}$ at $v=\infty$, and the global quench has a characteristic width $\gamma^{-1}$. As for the path $\tilde{\lambda}(\xi)$, we have set 
\begin{equation}
\tilde{\lambda}(\xi) = - \cos  (\pi\,  \xi) + i\,\eta\,\sin( \pi\,  \xi), \quad  \eta \in \mathbb R,~ \eta>0.      
\end{equation}
For the numerical computation, we fix $t_1 = t - \Delta t/2$, $t_2= t+\Delta t/2$, and keep the width $\Delta t$ of the boundary subregion fixed as $t$ ranges from large negative to large positive values. For $t \ll 0$, we employ the geodesic in vacuum AdS$_3$ as the initial guess; once the Newton-Raphson method finds a solution in the Vaidya spacetime, we slightly increase $t$ and take the known solution as initial guess for the next. For each choice of $t$ and $\Delta t$, the holographic TEE is simply given by 
\begin{equation}
S_{A}^{\rm (T)} = \frac{\lambda_*}{2G_N}. 
\end{equation}
We can compare the numerical computation for a thick-shell Vaidya spacetime with $\gamma \gg 1$ with the prediction of the thin-shell matching computation described in section \ref{sec:ads3_vaidya}. To do so, let us focus on the complex function $y(v)$ evaluated along a particular path $v=v(\zeta)$ in the complex $v$-plane. We choose 
\beq\label{path}
v(\zeta) = t_1 + g(\zeta)(t_2-t_1), \quad g(\zeta) = \zeta + i \eta \sin\left(\pi\zeta\right),~\eta > 0,~\eta \ll 1, 
\eeq
such that $v(0)=t_1$ and $v(1)=t_2$. The relevant comparison is between the numerical solution $y(\zeta) \equiv y[v(\zeta)]$ and the AdS and BTZ pieces of the matching computation, respectively given by $y_{\rm AdS}(\zeta) \equiv y_{\rm AdS}[v(\zeta)]$, $y_{\rm BTZ}(\zeta) \equiv y_{\rm BTZ}[v(\zeta)]$, and which can be immediately obtained from the expressions \eqref{AdS_matching}, \eqref{AdS_distance} and \eqref{BTZ_matching},  \eqref{BTZ_distance} in the main text. The result of this comparison for a thick-shell with $m_0=1$, $\gamma=16$ and boundary insertions $t_1=-0.75$, $t_2=0.75$ can be found in figure \ref{fig:comparison}, where we have set $\eta = 0.05$.  

As demonstrated in section \ref{sec:ads3_vaidya}, in the thin-shell Vaidya spacetime the junction of the AdS and the BTZ geodesics takes place at $v_s=0$. Taking this fact into account, we have chosen a path $v=v(\zeta)$ in the complex $v$-plane that passes close to $v=0$ when $\zeta = 0.5$. The expectation is that, for a sufficiently large $\gamma$, for $\zeta < 0.5$ the numerical solution $y(\zeta)$ is well-described by the AdS solution $y_{\rm AdS}(\zeta)$, and that for $\zeta \geq 0.5$ the numerical solution $y(\zeta)$ is well-described by the BTZ solution $y_{\rm BTZ}(\zeta)$. This expectation is confirmed by figure \ref{fig:comparison}, with excellent agreement between the real and imaginary parts, except for the immediate vicinity of $\zeta = 0.5$ for the latter. 

%%%%%%%%%%%%%%%%%%%%%%%%%%%%%%%%%%%%%%%%%%%
\section{Time ordering in four-point functions} \label{app:TO}

Here we briefly summarize how to implement  a prescribed  operator ordering when analytically continuing Euclidean correlation functions to Lorentzian signature. We refer to \cite{Hartman:2015lfa,Kundu:2025jsm} for more detailed discussions. 

The desired Lorentzian operator ordering can be implemented through the prescription
\begin{equation} \label{eq:opordering}
\langle O_1(t_1, x_1) O_2(t_2, x_2) \cdots O_n(t_n, x_n) \rangle
= \lim_{\varepsilon_i \to 0} \langle O_1(t_1 - i\varepsilon_1, x_1) \cdots O_n(t_n - i\varepsilon_n,  x_n) \rangle
\end{equation}
with the condition $ \varepsilon_1 > \varepsilon_2 > \cdots > \varepsilon_n > 0$. 
Notice that this prescription does not assume any ordering of the real Lorentzian times $t_i$. Instead, the ordering is encoded in the relative imaginary displacements of the insertion points.
For instance, one can immediately check that 
\be
\begin{aligned}
\langle O_1(t_1-i\varepsilon_1,x_1)\,
O_2(t_2-i\varepsilon_2,x_2)\rangle
&=
\langle O_1(t_1-i(\varepsilon_1-\varepsilon_2),x_1)\,
O_2(t_2,x_2)\rangle
\\
&=
\langle O_1(t_1,x_1)\,
O_2(t_2+i(\varepsilon_1-\varepsilon_2),x_2)\rangle
\end{aligned}
\ee
are equivalent and produce the same Lorentzian correlator.

When evaluating the entanglement entropy in a state $|\psi\rangle$, the operator ordering must be specified so that the Lorentzian correlator computes the time-ordered two-point function of replica twist operators in the state prepared by the Euclidean insertions of $\psi$.
For instance, on the cylinder  $w = i \phi + \tau  + i t$, where $\tau$ denotes Euclidean time and $t$ is obtained after analytic continuation, the desired Lorentzian correlator is 
\be
 \langle \psi |  \sigma_n( \Delta t, \Delta \phi ) \tilde \sigma_n (0)| \psi\rangle =  \langle  \psi(w_4 , \bar w_4  )  \sigma_n( \Delta t, \Delta \phi ) \tilde \sigma_n( 0) \psi(w_1, \bar w_1 )  \rangle 
\ee
for $\Delta t >0$, where the twist operators are understood to be time-ordered.

When the state is prepared by Euclidean insertions located at $w_1 = -\tau_{\infty}$ and $w_4 = \tau_{\infty}$, with $\tau_{\infty}>0$, comparison with \eqref{eq:opordering}  shows that the desired ordering is obtained through the analytic continuation
\be
\langle \psi |  \sigma_n( \Delta t, \Delta \phi ) \tilde \sigma_n (0)| \psi\rangle = \lim_{\veps \to 0} \langle  \psi( \tau_{\infty} ) \sigma_n(  w_2  =i \Delta \phi + i \Delta t +  \veps  ) \tilde \sigma_n(w_3= 0) \psi( -\tau_{\infty} )  \rangle 
\ee
while keeping $\tau_\infty>\varepsilon$ throughout. 
For Hamiltonian eigenstates one eventually takes the limit $\tau_\infty\to\infty$, so the condition $\tau_\infty>\varepsilon$ is automatically satisfied. For local operator quenches, on the other hand, one ultimately takes $\tau_\infty\to0$. The two limits therefore do not commute, and the condition $\tau_\infty>\varepsilon$ requires taking the $\varepsilon\to0$ limit only after the state preparation has been defined. This is precisely the prescription employed throughout the paper to determine the imaginary part of the TEE from analytically continued replica correlators.

%%%%%%%%%%%%%%%%%%%%%%%%%%%%%%%%%%%%%%%%%%%%%%
%%%%%%%%%%%%%%%%%%%%%%%%%%%%%%%%%%%%%%%%%%%%%%
\bibliographystyle{JHEP}
\bibliography{SingleIntervalTEE.bib}

\end{document}